\documentclass[11pt,a4paper]{report}
\usepackage{makeidx,latexsym,amssymb}
\usepackage{pst-all,bbm,slashed,fancyhdr}
\DeclareMathAlphabet{\boldmathe}{T1}{cmr}{bx}{it}
\DeclareMathAlphabet{\wipfa}{OT1}{pss}{m}{n}

\font\af=msbm10

\newcommand{\mbf}[1]{\boldmathe{#1}}
\newcommand{\mbfgr}[1]{\textit{\mbox{\boldmath$#1$}}}
\newcommand{\lrd}{\raise.3ex\hbox{$\stackrel{\leftrightarrow}{\partial }$}{}}
\newcommand{\delr}{\raise.3ex\hbox{$\stackrel{\leftarrow}{\delta }$}{}}

\def\chida{\chi^\dagger}

\def\varphib{\bar\varphi}
\def\varphid{\varphi_\alpha}
\def\varphiu{\varphi^\alpha}
\def\varphibu{\bar\varphi^{\dot\alpha}}
\def\varphibd{\bar\varphi_{\dot\alpha}}

\def\chiu{\chi^\alpha}
\def\chibu{\bar\chi^{\dot\alpha}}

\def\Cab{C^a_{\hskip4pt b}}

\def\num{\wipfa N}

\def\PL{P_{\wipfa L}}
\def\PR{P_{\wipfa R}}
\def\HB{H_{\wipfa B}}
\def\HF{H_{\wipfa F}}

\def\VB{V_{\wipfa B}}
\def\VF{V_{\wipfa F}}

\def\RB{R_{\wipfa B}}
\def\RF{R_{\wipfa F}}
\def\TB{T_{\wipfa B}}
\def\TF{T_{\wipfa F}}

\def\yuk{\textsc{Yukawa }}
\def\kk{\textsc{Kaluza-Klein }}
\def\ym{\textsc{Yang-Mills }}
\def\bog{\textsc{Bogomol'nyi }}
\def\ym{\textsc{Yang-Mills }}

\def\noe{\textsc{Noether }}
\def\wit{\textsc{Witten }}
\def\poin{\textsc{Poincar\'e }}
\def\ham{\textsc{Hamilton }}
\def\haman{\textsc{Hamilton}ian }
\def\lagan{\textsc{Lagrange}an }
\def\lor{\textsc{Lorentz }}
\def\lie{\textsc{Lie }}
\def\abel{\textsc{Abel}ian }
\def\mink{\textsc{Minkowski }}
\def\maj{\textsc{Majorana }}
\def\weyl{\textsc{Weyl }}

\def\noeth{\textsc{Noether }}

\def\hil{\textsc{Hilbert }}
\def\hils{\textsc{Hilbert} space }

\def\dirac{\textsc{Dirac }}
\def\schr{\textsc{Schr\"odinger }}

\def\cliff{\textsc{Clifford }}
\def\wz{\textsc{Wess-Zumino }}

\def\Ad{A^\dagger}

\def\Gam{\Gamma}

\def\psicc{\psi_{\rm c}}
\def\chicc{\chi_{\rm c}}
\def\psiL{\psi_{\wipfa L}}
\def\psiR{\psi_{\wipfa R}}
\def\Psib{\bar\Psi}

\def\tal{\tilde\alpha}

\def\ot{\otimes}
\def\detpr{{\det}^\prime}
\def\alda{\alpha^\dagger}
\def\psida{\psi^\dagger}
\def\thetab{\bar\theta}

\def\thetad{\theta_{\alpha}\,}

\def\thetabd{\bar\theta_{\dot\alpha}\,}
\def\zetau{\zeta^\alpha\,}

\def\zetabu{\bar\zeta^{\dot\alpha}\,}

\def\lamb{\bar\lambda}

\def\varphiald{\varphi^{\dot\alpha}}

\def\chiald{\bar\chi^{\dot\alpha}}

\def\chial{\chi_{\alpha}}

\def\chibed{\bar\chi^{\dot\beta}}

\def\bea{\begin{eqnarray}}

\def\eea{\end{eqnarray}}
\def\be{\begin{equation}}
\def\ee{\end{equation}}
\def\epsd{\varepsilon_{\alpha\beta}}

\def\dal{{\dot\alpha}}
\def\dbe{{\dot\beta}}

\def\zetab{\bar\zeta}

\def\de{{\dot 1}}

\def\fabc{f^a_{\;\;bc}}
\def\vac{\vert 0\rangle}
\def\bvac{\vert \hskip 1pt 0_{\wipfa B}\rangle}
\def\fvac{\vert \hskip 1pt 0_{\wipfa F}\rangle}
\def\kpsi{\vert\psi\ra}
\def\kpsibo{\vert\psi_{\wipfa B}\ra}
\def\kpsife{\vert\psi_{\wipfa F}\ra}
\def\psibo{\psi_{\wipfa B}}
\def\psife{\psi_{\wipfa F}}
\def\Z{\mbox{\af Z}}
\def\R{\mbox{\af R}}
\def\C{\mbox{\af C}}
\def\N{\mbox{\af N}}

\makeindex
\begin{document} 
\newtheorem{Lem}{Lemma}
\newcommand{\eqnn}[1]{\begin{eqnarray*}#1\end{eqnarray*}}
\newcommand{\eqngr}[2]{\begin{eqnarray*}#1\\#2\end{eqnarray*}}
\newcommand{\eqngrr}[3]{\begin{eqnarray*}#1\\#2\\#3\end{eqnarray*}}
\newcommand{\eqngrrr}[4]{\begin{eqnarray*}#1\\#2\\#3\\#4
\end{eqnarray*}}
\newcommand{\eqngrrrr}[5]{\begin{eqnarray*}#1\\#2\\#3\\#4\\#5
\end{eqnarray*}}
\newcommand{\eqnl}[2]{\begin{eqnarray}#1\label{#2}\end{eqnarray}}
\newcommand{\eqngrl}[3]{
\begin{eqnarray}#1\nonumber\\#2\label{#3}\end{eqnarray}}
\newcommand{\eqngrrl}[4]{
\begin{eqnarray}#1\nonumber\\#2\label{#4}\\
#3\nonumber\end{eqnarray}}
\newcommand{\eqngrrrl}[5]{
\begin{eqnarray}#1 \nonumber \\#2\nonumber \\
#3\label{#5}\\#4 \nonumber\end{eqnarray}}
\renewcommand{\arraystretch}{1.3}
\newsavebox{\uuunit}
\sbox{\uuunit}
    {\setlength{\unitlength}{0.825em}
     \begin{picture}(0.6,0.7)
        \thinlines
        \put(0,0){\line(1,0){0.5}}
        \put(0.15,0){\line(0,1){0.7}}
        \put(0.35,0){\line(0,1){0.8}}
       \multiput(0.3,0.8)(-0.04,-0.02){12}{\rule{0.5pt}{0.5pt}}
     \end {picture}}
\newcommand{\refs}[1]{(\ref{#1})}
\def\id{\mathbbm{1}}
\newcommand{\rmi}{{\rm i}}
\newcommand{\ft}[2]{{\textstyle\frac{#1}{#2}}}
\newcommand{\ome}[2]{\omega^{#1}_{\;\,#2}}
\def\Ca{C_{ab}^{\;\;\,c}}
\def\Ci{C_{ij}^{\;\;\;k}}
\def\Cia{C_{i\al}^{\;\;\,\beta}}
\def\Cab{C_{\al\beta}^{\;\;\,j}}
\def\spp{\;\;\;}
\def\veps{\varepsilon}
\def\hc{\hbox{ h.c.}}
\def\mod{{\rm mod ~}}
\def\tsigma{\tilde{\sigma}}
\def\dW{{\dot W}}
\def\da{{\dot A}}
\def\db{{\dot B}}
\def\dc{{\dot C}}
\def\dd{{\dot D}}
\def\de{{\dot 1}}
\def\dz{{\dot 2}}
\def\dv{{\dot v}}
\def\barpsi{\bar{\psi}}

\def\diag{{\rm diag}}
\def\ali{{\al_{(i)}}}
\def\alj{{\al_{(j)}}}
\def\mui{{\mu_{(i)}}}
\def\muj{{\mu_{(j)}}}
\def\alic{\al^\vee_{(i)}}
\def\aljc{\al^\vee_{(j)}}
\def\muc{\mu^\vee}
\def\muic{\mu^\vee_{(i)}}
\def\mujc{\mu^\vee_{(j)}}
\def\smui{\vert\mui\rangle}
\def\smuj{\vert\muj\rangle}
\def\smu{\vert\mu\rangle}
\def\Hali{H_{\ali}}
\def\cA{{\cal A}}
\def\cB{{\cal B}}
\def\cC{{\cal C}}
\def\cd{{\cal D}}
\def\cE{{\cal E}}
\def\cG{{\cal G}}
\def\cF{{\cal F}}
\def\cH{{\cal H}}
\def\cHB{\cH_{\wipfa B}}
\def\cHF{\cH_{\wipfa F}}
\def\cL{{\cal L}}
\def\cM{{\cal M}}
\def\cN{{\cal N}}
\def\cQ{{\cal Q}}
\def\cQd{{\cal Q}^\dagger}
\def\cP{{\cal P}}
\def\cR{{\cal R}}
\def\cS{{\cal S}}
\def\cZ{{\cal Z}}
\def\cmb{{\pa{\cal M}}}
\def\fd{{^{*\!}F}}
\def\intl{\int\limits}
\def\nup{\vert n_\uparrow\rangle}
\def\mup{\vert m_\uparrow\rangle}
\def\ndown{\vert n_\downarrow\rangle}
\def\psiup{\psi_\uparrow}
\def\psidown{\psi_\downarrow}
\def\hha{{\hbar\ov 2}}
\def\un{\underline}
\def\tsi{\tilde\sigma}
\def\hb{\hbar}
\def\dett{{\det}_\theta}
\def\di{\slashed{D}}
\def\as{\slashed{a}}
\def\As{\slashed{A}}
\def\psl{\slashed{p}}
\def\fdi{\slashed{\partial}}
\def\fsi{\slashed{\sigma}}
\def\fsig{\tilde\sigma\!\!\!\:\!\!\slash\,}
\def\ddx{d^dx}
\def\mbal{\mbfgr{\alpha}}
\def\mbpsi{\mbfgr{\psi}}
\def\mbphi{\mbfgr{\phi}}
\def\mbpi{\mbfgr{\pi}}
\def\mbsigma{\mbfgr{\sigma}}
\def\mbtau{\mbfgr{\tau}}
\def\mbb{\mbf{b}}
\def\mbe{\mbf{e}}
\def\mbg{\mbf{g}}
\def\mbff{\mbf{f}}
\def\mbn{\mbf{n}}
\def\mbv{\mbf{v}}
\def\mbk{\mbf{k}}
\def\mbm{\mbf{m}}
\def\mbp{\mbf{p}}
\def\mbr{\mbf{r}}
\def\mbx{\mbf{x}}
\def\mby{\mbf{y}}
\def\mb0{\mbf{0}}
\def\mbA{\mbf{A}}
\def\mbB{\mbf{B}}
\def\mbC{\mbf{C}}
\def\mbD{\mbf{D}}
\def\mbE{\mbf{E}}
\def\mbJ{\mbf{J}\,}
\def\mbK{\mbf{K}}
\def\mbL{\mbf{L}}
\def\mbM{\mbf{M}}
\def\mbN{\mbf{N}}
\def\mbP{\mbf{P}}
\def\mbS{\mbf{S}}
\def\mbV{\mbf{V}}
\def\mbW{\mbf{W}}
\def\val{\vec{\alpha}}
\def\vtheta{\vec{\theta}}
\def\vpi{\vec{\pi}}
\def\vsigma{\vec{\sigma}}
\def\vOmega{\vec{\Omega}}
\def\hi{{\hbar\ov i}}
\def\ih{{i\ov \hbar}}
\def\nablad{\nabla\cdot}
\def\tr{{\rm Tr}\hskip 1pt}
\def\str{{\rm S}\,{\rm Tr}\hskip 1pt}
\def\sdet{{\rm S}\,{\rm det}\hskip 1pt}
\def\bet{\bar\eta}
\def\etab{\bar\eta}
\def\bal{\bar\alpha}
\def\psib{\bar\psi}
\def\lamb{\bar\lam}
\def\chib{\bar\chi}
\def\epsb{\bar\eps}
\def\alb{\bar\alpha}
\def\vepsb{\,\bar\varepsilon}
\def\ald{\alpha^\dagger}
\def\betab{\bar\beta}
\def\betad{\beta^\dagger}
\def\thetab{\bar\theta}
\def\phib{\bar\phi}
\def\lambdab{\bar\lambda}
\def\gan{\gamma_n}
\def\gamf{\gamma_5}
\def\gams{\gamma_*}

\def\ra{\rangle}
\def\la{\langle}
\def\gff{\bar\gamma}
\def\pan{\par\noindent}
\def\cov{\bigtriangledown}
\def\pd{\psi^{\dagger}}
\def\phid{\phi^\dagger}
\def\Phid{\Phi^\dagger}
\def\mtxt#1{\quad\hbox{{#1}}\quad}
\def\pa{\partial}
\def\gam{\gamma}
\def\gammu{\gamma^\mu}
\def\eps{\epsilon}
\def\es{\!=\!}
\def\ms{\!-\!}
\def\ps{\!+\!}
\def\lapf{\triangle_f}
\def\lap{\triangle}
\def\olap{{1\ov\lap}}
\def\ov{\over}
\def\om{\omega}
\def\pamu{{\partial_\mu}}
\def\panu{{\partial_\nu}}
\def\al{\alpha}
\def\si{\sigma}
\def\sp{\hbox{Sp}\,}
\def\pr{\prime}
\def\ppr{{\,\prime}}
\def\lam{\lambda}
\def\ha{{1\over 2}}
\def\h{\ft12}
\title{\huge Non-perturbative Methods in\\
Supersymmetric Theories}
\author{Andreas Wipf\\
Theoretisch-Physikalisches-Institut\\
Friedrich-Schiller-Universit\"at Jena, Max Wien Platz 1\\
07743 Jena}
\date{\textbf{Abstract}\\[3ex]
These notes are based on Graduate Lectures
I gave over the past 5 years.\\ The aim of these
notes is to provide a short \emph{introduction} to
supersymmetric theories: supersymmetric
quantum mechanics, Wess-Zumino models and supersymmetric
gauge theories. A particular emphasis is put on
the underlying structures and non-perturbative
effects in $\cN=1,\,\cN=2$ and $\cN=4$
Yang-Mills theories.\\[5ex]
{\small Extended version of lectures given at the\\
\emph{TROISIEME CYCLE DE LA PHYSIQUE EN SUISSE ROMANDE}.}}
\maketitle
\tableofcontents

\chapter{Introduction} \label{intro}
Supersymmetric theories are highly symmetric and beautiful.
They unify fermions (matter) with bosons (carrier of forces),
either in flat or in curved space-time. 
Supergravity theories with local supersymmetries 
unify the gravitational with the other interactions. 
The energy at which gravity and quantum
effects become of \emph{comparable strength} can be estimated from
the only expression with the dimension of mass that can
be formed from the fundamental constants of nature $\hbar, c$ and $G$:
the \textsc{Planck} mass
\eqnn{
m_{\mathrm{Pl}}=\left(\frac{\hbar c}{G}\right)^{1/2}
\approx 10^{19}\,{\hbox{GeV}\ov c^2}.}
For a particle with this mass the \textsc{Schwarzschild}
radius, where its gravitational field becomes strong, is
just twice its \textsc{Compton} wavelength, which is the
minimal length to which it can be localized,
\eqnn{
r_{\rm S}={2G m_{\rm Pl}\ov c^2} =
\frac{\hbar}{2m_{\rm Pl}c}=
2\lam_c\quad\mtxt{for}m=m_{\rm Pl}.}
Supersymmetry (susy) transformations relate bosons to fermions,
\eqnn{
\cQ\,|\hbox{Boson}\ra\sim |\hbox{Fermion}\ra\mtxt{and}
\cQ\,|\hbox{Fermion}\ra\sim |\hbox{Boson}\ra,}
and hence relate particles with different
spins. The particles fall into multiplets and
the supersymmetry transforms different members of such 
a supermultiplet into each other. Each supermultiplet
must contain at least one boson and one fermion whose spins
differ by $1/2$ and all states in a multiplet (of unbroken
supersymmetry) have the same mass. \pan
So far no experimental 
observation has revealed particles or forces which manifestly 
show such a symmetry. Yet supersymmetry has excited great
enthusiasm in large parts of the community and more recently
in the context of superstring theories. It has even been 
said of the theory that it
\begin{center}
\textsc{is so beautiful it must be true.}
\end{center}
The first part of these lectures deals with \emph{supersymmetric
quantum mechanics}. There are at least three good reasons
to consider such systems,\\
$\bullet$ they contain the essential structures of susy theories,\\
$\bullet$ they appear as lattice versions of susy
field theories,\\
$\bullet$ they describe the infrared
dynamics of susy field theories in finite
volumes.\\
Some of these topics will be discussed in these lectures.

In the second part of these lectures we review some textbook
material, in particular the \textsc{Coleman-Mandula} theorem,
supersymmetry algebras, representation theory and
simple supersymmetric models.

The third part contains more recent results on supersymmetric
gauge theories with one, two and four supercharges, central 
charges, BPS-states and $\beta$-functions.\\

\textbf{Notation:}
\begin{center}
\begin{tabular}{|l|l|l|}\hline
symbols &range & meaning\\ \hline
$i,j,k,\dots$& $1,2,\dots ,d-1$ & space indices\\
$\mu,\nu,\rho,\sigma,\dots$&$ 0,1,\dots,d-1$& spacetime indices\\
$\al,\beta,\gam,\delta\dots$&$1,\dots 2^{[d/2]}$& Dirac-spinor indices\\
$\al,\beta,\dot{\al},\dot{\beta}\dots$&$1,\dots 2^{d/2-1}$ &
Weyl-spinor-indices (d even)\\
$A^\dagger,A^*,A^T$&$A$ matrix & adjoint, complex conjugate and transpose of
$A$\\ \hline \end{tabular}
\end{center}
The (anti)symmetrization of a tensor $A_{\mu_1\ldots\mu_n}$ is
\eqnn{
A_{(\mu_1\ldots\mu_n)}={1\ov n!}\sum_\sigma A_{\sigma(\mu_1)\ldots
\sigma(\mu_n)},\quad
A_{[\mu_1\ldots\mu_n]}={1\ov n!}\sum_\sigma
\hbox{sign}(\sigma)A_{\sigma(\mu_1)\ldots
\sigma(\mu_n)}.}
\paragraph{Reading:}
The introductory books and review articles
\cite{intwess}-\cite{intsachs} I found useful
when preparing these lecture notes.

\chapter{Supersymmetric QM}
In this chapter we examine simple $1\!+\!0$-dimensional supersymmetric
(susy) field theories.  In $1+0$ dimensions the \poin algebra
reduces to time translations generated by the
\haman $H$ and the hermitian field and momentum operators 
$\phi(t)$ and $\pi(t)$ may be viewed  as position 
and momentum operators of a point particle on the 
real line in the \textsc{Heisenberg}-picture.
Hence susy field theories in
$1+0$ dimensions are particular quantum mechanical 
systems \cite{witten81}. There are no 
technical difficulties hiding the essential 
structures. Such systems are interesting in their own right
since they describe the infrared-dynamics of susy
field theories in finite volumes \cite{witten82a}. 
This observation may be used to improve our understanding
of supersymmetric quantum field theories beyond perturbation 
theory. A susy quantum mechanics with $16$ supercharges 
also emerges in the matrix theory description of $M$ theory 
\cite{banks97}. In mathematical physics susy QM has 
been useful in proving index theorems for physically relevant
differential operators \cite{witten82b}. 
There exist several extensive texts on susy
quantum mechanics \cite{cooper,junker,kalka} in
which the one-dimensional systems are discussed
in detail. But some of the material presented in
these notes (in particular on higher-dimensional
systems) is not found in reviews.

\section{Pairing and ground states}
\index{pairing of states}
\index{ground states in SQM}
The \hils of a supersymmetric system is the sum of its
bosonic and fermionic subspaces, $\cH=\cHB\oplus \cHF$.
Let $A$ be a linear operator $\cHF\to \cHB$.
In most cases it is a first order differential operator. 
We shall use a block notation such that the vectors
in $\cHB$ have upper components and those in
$\cHF$ lower ones
\eqnn{
\kpsi=\pmatrix{\kpsibo\cr \kpsife}.}
Then the nilpotent
\emph{supercharge} and its adjoint take the form
\eqnl{
\cQ=\pmatrix{0&A\cr 0&0},\quad
\cQd=\pmatrix{0&0\cr \Ad&0}\Longrightarrow
\{\cQ,\cQ\}=0.}{sqm1}
The block-diagonal super-\haman
\eqnl{
H\equiv\{\cQ,\cQd\}=\pmatrix{A\Ad & 0\cr 0& \Ad A }=
\pmatrix{\HB&0\cr 0&\HF},}{sqm3}
commutes with the supercharge
\eqnl{
[\cQ,H]=0.}{sqm5}
It is useful to introduce the (fermion) number operator
\eqnl{
\num=\pmatrix{0&0\cr 0&1}}{sqm7}
which commutes with $H$. Bosonic states have number $\num=0$ 
and fermionic states $\num=1$.
The supercharge and its adjoint decrease and increase 
this number by one,
\eqnl{
[\num,\cQ]=-\cQ\mtxt{and}
[\num,\cQd]=\cQd.}{sqm9}
The energies of the partner-\textsc{Hamilton}ians
$\HB$ and $\HF$ in \refs{sqm3} are either zero or positive. 
A bosonic state in $\cHB$ has zero energy if and only
if it is annihilated by $\Ad$ and a fermionic state 
in $\cHF$ has zero energy if and only
if it is annihilated by $A$,
\eqngrl{
\HB\bvac=0&\Longleftrightarrow&\! \Ad\bvac=0}
{\HF\fvac=0&\Longleftrightarrow& A\,\fvac=0.}{sqm11}
The states with positive energies come in pairs.
Let $\kpsife$ be a fermionic eigenstate with 
\emph{positive energy},
\eqnn{
\HF\kpsife= \Ad A\kpsife=E\kpsife,\qquad E>0.}
It follows, that $A\kpsife$ is a bosonic eigenfunction
with the same energy,
\eqnn{
\HB\big(A\kpsife\big)=(A\Ad )A\kpsife=
A(\Ad A)\kpsife=A\HF\kpsife=E\big(A\kpsife\big).}
The fermionic state $\kpsife\in\cHF$ and its partner state
\eqnl{
\kpsibo={1\ov \sqrt{E}}\,A\kpsife\in\cHB}{sqm13}
have equal norms,
\eqnl{
\la \psibo\kpsibo={1\ov E}
\big\la \psife\vert \Ad A\vert\psife\big\ra
=\la\psife\kpsife}{sqm15}
and this proves, that the partner state of any
excited state is never the null-vector.
Likewise, the nontrivial partner state of any 
bosonic eigenstate $\kpsibo\in\cHB$ with positive
Energy $E$ is 
\eqnl{
\kpsife={1\ov \sqrt{E}}\,\Ad\kpsibo\in\cHF.}{sqm17}
This then proves that the partner 
\textsc{Hamilton}ians $\HB$ and
$\HF$ have identical spectra, up to possible 
zero-modes.

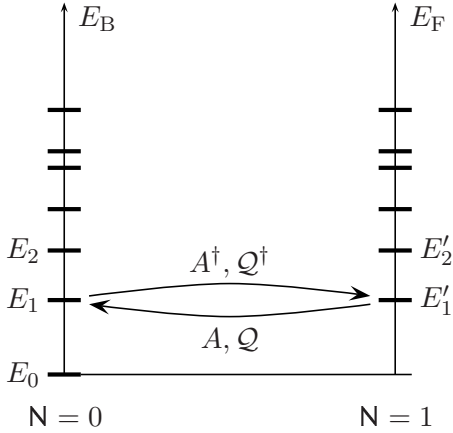
\begin{minipage}[t]{7.6cm}
\psset{unit=1.1mm,linewidth=0.6pt,dotsize=1mm}
\hskip3mm\label{figure1}
\begin{pspicture}(-30,-5)(30,50)
\psline(-18,0)(22,0)
\psline{->}(-20,0)(-20,45)
\rput(-16,43){$E_{\rm B}$}
\rput(24.0,43){$E_{\rm F}$}
\psline{->}(20,0)(20,45)
\rput(-20,-5){$\num=0$}
\rput(20,-5){$\num=1$}
\pscurve[arrowsize=2mm]{->}(-17,9.5)(0,11)(17,9.5)
\rput(0,13.6){$\Ad,\cQd$}
\pscurve[arrowsize=2mm]{->}(17,8.5)(0,7)(-17,8.5)
\rput(0,4){$A,\cQ$}
\psset{linewidth=1.6pt}
\psline(-22,0)(-18,0)
\psline(-22,9)(-18,9)\psline(18,9)(22,9)
\psline(-22,15)(-18,15)\psline(18,15)(22,15)
\psline(-22,20)(-18,20)\psline(18,20)(22,20)
\psline(-22,25)(-18,25)\psline(18,25)(22,25)
\psline(-22,27)(-18,27)\psline(18,27)(22,27)
\psline(-22,32)(-18,32)\psline(18,32)(22,32)
\rput(-25,0){$E_0$}
\rput(-25,9){$E_1$}
\rput(-25,15){$E_2$}
\rput(25,9){$E'_1$}
\rput(25,15){$E'_2$}
\end{pspicture}
\end{minipage}
\begin{minipage}[b]{6.2cm}
The pairing of the non-zero energies and
eigenfunctions in supersymmetric quantum
mechanics is depicted in the
figure on the left. The supercharge $\cQd$
containing $\Ad$ maps bosonic eigenfunctions into
fermion ones and $\cQ$ containing
$A$ maps fermionic
eigenfunctions into bosonic ones. For potentials
with scattering states there is a corresponding
relation between the transmission and reflection
coefficients of $\HB$ and $\HF$, see below.
\end{minipage}\\
\par
In SQM on $\R$ the operator $A$ and its adjoint read in
position space
\eqnl{
A=i\pa_x+iW(x)\quad,\quad A^\dagger=i\pa_x-iW(x)}{sps1}
and the partner \textsc{Hamilton}ians take the familiar
forms
\eqngrl{
\HB=p^2+\VB,&& \VB=W^2+W'}
{\HF=p^2+\VF,&& \VF=W^2-W'.}{sps3}
For such simple systems we can find the ground 
state(s) of the super-\haman explicitly. 
With \refs{sqm11} we must study the first order differential 
equations
\eqngrl{
\left(\pa_x-W(x)\right)\psibo(x)&=&0}
{\left(\pa_x+W(x)\right)\psife(x)&=&0.}
{sps5}
The solutions are
\eqnl{
\psibo(x)\propto e^{\chi(x)}\mtxt{and}
\psife(x)\propto e^{-\chi(x)}}{sps7}
where we have introduced the function
\eqnl{
\chi(x)=\int^x W(x')dx'}{sps9}
If at least one of the two solutions is normalizable
then susy is \emph{unbroken}. But since $\psibo(x)\cdot\psife(x)$ 
is constant, at most one of the two solutions can be
\emph{normalized}. For example, for  
\eqnn{
W=\lam x^p+o\left(x^p\right)\mtxt{and}
V_{\wipfa{B,F}}=\lam^2 x^{2p}+o\left(x^{2p}\right)}
supersymmetry is unbroken for odd $p$: For
positive $\lam$ the ground state is fermionic and for 
negative $\lam$ it is bosonic. Below we have depicted
the partner potentials and ground state wave function
for $W=x(1-x^2)$.
The corresponding partner \ham operators have the
same positive eigenvalues.
\par
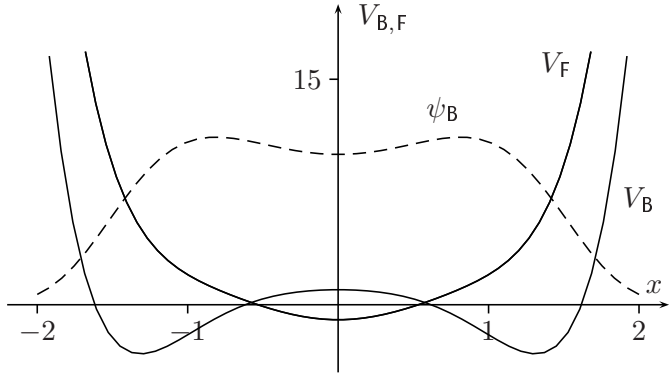
\psset{xunit=20mm,yunit=2mm,linewidth=0.6pt,dotsize=1mm}
\hskip25mm
\begin{pspicture}(-2.1,-5.5)(2.2,23)
\psaxes[Dy=15,tickstyle=bottom,labelsep=1mm]{->}(0,0)(-2.2,-4.5)(2.2,20)
\rput(2.1,1.2){$x$}
\rput(1.45,16){$\VF$}
\rput(2,7){$\VB$}
\rput(.3,19){$V_{\wipfa{B,F}}$}
\rput(.7,13){$\psibo$}
\psplot{-1.92}{1.92}{x x mul dup dup 1 sub dup mul mul exch 3 mul sub 1 add}
\psplot{-1.68}{1.68}{x x mul dup dup 1 sub dup mul mul exch 3 mul add 1 sub}
\psplot{-1.68}{1.68}{x x mul dup dup 1 sub dup mul mul exch 3 mul add 1 sub}
\psplot[linestyle=dashed]{-2}{2}{2.7183 x x mul dup 0.25 mul 0.33 sub mul neg exp 10 mul}
\end{pspicture}
\par
A long time ago, \schr asked the
following question \cite{schroed40}: Given a general
non-negative \haman
\eqnl{
H=p^2+V(x)\geq 0}{sps13}
in one dimension. Is there always a (in position space)
first order differential operator $A=i\pa_x+iW(x)$, such 
that $H=\HF=A^\dagger A$? This is the so-called 
\emph{factorization-problem}\index{factorization}.
In one dimension every non-negative $H$ can be
factorized\footnote{If $H$ is bounded from below
but has negative energies, then we add a big enough 
constant $c^2$ to $H$ and factorize $H+c^2$.}.

To construct the factorization we ompare \refs{sps13} with 
\refs{sps3} which leads 
to the nonlinear differential equation 
of \textsc{Ricatti},
\eqnl{
V(x)=
\VF(x)=W^2(x)-W'(x).}{fact25}
This equation is solved by the following well-known
trick: setting
\eqnl{
W(x)=-{\psi'_0(x)\ov\psi_0(x)},}{fact27}
the \textsc{Ricatti} equation transforms
into the linear \schr equation 
for $\psi_0$,
\eqnl{
-\psi''_0+V\psi_0=0.}{fact29}
Since $H\geq 0$ the solution $\psi_0$ has no node
and $W$ in \refs{fact27} is real and regular,
as required. If the ground state energy $E_0$
of $H$ is zero, then the transformation \refs{fact27}
is just the relation \refs{sps7} between
the superpotential and the ground state wave
function in the fermionic sector. If $E_0$
is positive, then the solution $\psi_0$
will not be square integrable.

As a simple example we consider a constant positive
potential $\VF=c^2$. The non-normalizable solution
of \refs{fact29} is $\psi_0=ae^{cx}+be^{-cx}$
and is used for the factorization,
\eqnl{
c^2=W^2-W'\Longrightarrow
W=-{d\ov dx}\log \psi_0.}{fact31}
The corresponding partner potential
\eqnl{
\VB=W^2+W'=c^2-2{d^2\ov dx^2}\log \psi_0}{fact33}
has exactly one zero-energy bound state
and scattering states with energies bigger
than $c^2$. For $a=b$
it is the reflectionless \textsc{P\"oschl-Teller}
potential
\eqnl{
\VB=c^2\left(1-{2\ov\cosh^2 cx}\right)}{fact35}
In the seminal paper by \textsc{Infeld} and \textsc{Hull}
\cite{infeld}
the factorization method for second order differential
equations has been worked out in great detail. It was
applied to six possible factorization types.
These types include the
\textsc{P\"oschl-Teller-, Morse-, Rosen-Morse-} and
radial \textsc{Coulomb} potential.

\section{SUSY breaking in SQM}
\index{susy-breaking in SQM}
The fact that susy has not been observed in
nature so far does not imply that there are no
practical uses for supersymmetric theories.
It could be that every occurring supersymmetry
is a broken one. We still would have a supercharge
and super-\haman obeying the super algebra.
But the symmetry could be \emph{spontaneously broken},
in which case there is no invariant ground state.

In order for supersymmetry to exist and be unbroken,
we require a ground state such that $\HB\vac=\HF\vac=0\vac$.
This means that the ground state is annihilated by
the generators $\cQ$ and $\cQd$ of supersymmetry.
Thus we have
\eqnn{
\hbox{susy unbroken}\Longleftrightarrow
\hbox{exist }\vac\in\cH\hbox{ with }
\cQ\vac=\cQd\vac=0.}
\par
\textbf{The Witten Index:}
\index{Witten index}
\wit defined an index to determine whether
supersymmetry is unbroken.
Formally this index is
\eqnl{
\Delta=\tr (-1)^{\num},}{wit1}
where $\num$ is the \emph{fermion number}.
\index{fermion number in SQM}
For simplicity we assume that the spectrum
of $H$ is discrete and use the energy eigenfunctions
to calculate $\Delta$. Let us first assume
that supersymmetry is broken, in which case
there is \emph{no} normalizable zero-energy state. Then
all eigenstates of $H$ have positive energies
and come in pairs: one bosonic state with $\num=0$
and one fermionic state with $\num=1$ having
the same energy. Their contribution to $\Delta$
cancel. Since all states with positive energy
are paired we obtain $\Delta=0$.
 
If there are $n_{\wipfa B}$ bosonic and $n_{\wipfa F}$ fermionic
ground states then their contribution to the \wit index
is $n_{\wipfa B}-n_{\wipfa F}$. Since the contributions of the excited
states cancel pairwise we obtain
\eqnl{
\Delta=n_{\wipfa B}-n_{\wipfa F}.}{wit3}
This yields an efficient method to determine
whether susy is broken,
\eqnl{
\Delta\neq 0\Longrightarrow \hbox{supersymmetry is unbroken}.}{wit5}
The converse need not be true. It could be that
susy is unbroken but the number of
zero-energy states in the bosonic and fermionic
sectors are equal so that $\Delta$ vanishes.
This does not happen in one-dimensional SQM, so
that
\eqnl{
\Delta\neq 0\Longleftrightarrow \hbox{supersymmetry is unbroken
in SQM}.}{wit7}
Already in SQM the operator $(-)^{\num}$ is  \emph{in general not 
trace class} and its trace must be regulated for the \wit index 
to be well defined. A natural definition is
\eqnl{
\Delta=\lim_{\al\searrow 0}\Delta(\al),\qquad
\Delta(\al)=\tr \left((-1)^{\num}e^{-\al H}\right).}{wit9}
In SQM with discrete spectrum $\Delta(\al)$ does not
depend on $\al$, since the contribution of all
super partners cancel in \refs{wit9}. The
contribution of the zero-energy states is still
$n_{\wipfa B}-n_{\wipfa F}$. In field theories the excited states
should still cancel in $\Delta(\al)$ in which case it
is $\al$-independent. Since $\Delta(\al)$
is constant, it may be evaluated for small $\al$.
But for $\al\searrow 0$ one can use the asymptotic
expansion for the \emph{heat kernel} of $\exp(-\al H)$
\index{heat kernel} to actually calculate the \wit index.
\section{Scattering states}
\index{scattering states in SQM}
Let us see, how supersymmetry relates
the \emph{transmission} and \emph{reflection coefficients}
\index{transmission coefficient}
\index{reflection coefficient}
of $\HB$ and $\HF$ for potentials supporting
scattering states \cite{cooperwipf}. Thus we assume that the 
superpotential tends to constant values for large 
$\vert x\vert$,
\eqnl{
\lim_{x\to\pm\infty}W(x)=W_\pm,\mtxt{such that}
\lim_{x\to\pm\infty}\VB(x)=\lim_{x\to\pm \infty}\VF(x)=W_\pm^2.
}{scat1}
We consider an \emph{incoming} plane wave from
the left. The asymptotic form of the wave function for 
scattering from the one-dimensional potential $\VB$
is given by
\eqnl{
\psibo(k,x)\longrightarrow
\cases{e^{ikx}+\RB e^{-ikx}&$x\to -\infty$\cr
               \TB\, e^{ik'x}&$x\to +\infty$},}{scat2}
where $\RB$ and $\TB$ are the reflection
and transmission coefficient in the
bosonic sector.
The properly normalized fermionic partner state has 
the asymptotic form (cp. \ref{sqm17})
\eqnn{
\psife(k,x)=-{1\ov k+iW_-}\Ad\psibo(x)\longrightarrow
\cases{e^{ikx}+\RF e^{-ikx}&$x\to -\infty$\cr
       \TF\, e^{ik'x}&$x\to+\infty$}} 
with the following \emph{reflection} and 
\emph{transmission coefficients},
\eqnl{
\RF={W_-+ik\ov W_--ik}\RB\mtxt{and}
\TF={W_+-ik'\ov W_--ik}\TB,}{scat3}
where $k$ and $k'$ are given by
\eqnl{
k=(E-W_-^2)^{1/2}\mtxt{and}
k'=(E-W_+^2)^{1/2}.}{scat4}
The scattering data for the supersymmetric
partners are not the same but they are related
in this simple way. For real $k,k'$ we have 
\eqnn{
\vert \RB\vert^2=\vert \RF\vert^2\mtxt{and}
\vert \TB\vert^2=\vert \TF\vert^2} 
and the partner systems
have identical reflection and transmission
probability.

The transmission coefficients have
\emph{physical poles} in the upper half of the complex
$k$-plane, their positions $k_j=i\kappa_j$ correspond 
to energies of bound states
\eqnl{
E_j=W_-^2-\kappa_j^2.}{scat5}
For negative $W_-$ and positive $W_+$ there is one more 
zero-energy bound state in $\cH_{\wipfa F}$
and for positive $W_-$ and negative $W_+$ 
one more zero-energy bound state in $\cH_{\wipfa B}$.

As an example we consider the kink
\eqnl{
W(x)=-\lam\tanh(x)\mtxt{with}
W_-=-W_+= \lam,}{scat6}
giving rise to the partner potentials
\eqnl{
\VB(\lam;x)=\lam^2-{\lam(\lam+1)\ov\cosh^2x}\mtxt{and}
\VF(\lam;x)=\lam^2-{\lam(\lam-1)\ov\cosh^2x}}{scat7}
Supersymmetry, together with the shift-property 
\eqnl{
\VF(\lam;x)=\VB(\lam-1;x)+2\lam-1}{scat9}
allows one to find the scattering coefficients for
an infinite tower of potentials. Let us assume that 
we know the coefficients $\RB(\lam)$ and $\TB(\lam)$
for a certain value of the parameter $\lam$. It follows that
\eqnl{
\TB(\lam-1)=\TF(\lam)\stackrel{\refs{scat3}}{=}
-{\lam+i k\ov \lam-i k}\,\TB(\lam)}{scat11}
and a similar formula for $\RB$ and $\RF$.
The iteration of this relation yields
\eqnl{
\TB(\lam)=(-)^N\prod_{n=0}^{N-1}
{\lam-n-ik\ov \lam-n+ik}\,\TB(\lam-N)}{scat13}
For $\lam=N$ the transmission coefficient $\TB$ on the right
hand side is $1$ so that
\eqnl{
\TB(N)=(-)^N\prod_{n=0}^{N-1}{N-n-ik\ov N-n+ik}}{scat15}
is the transmission coefficient for the
\textsc{P\"oschl-Teller} potentials with $\lam\in \N_0$,
\eqnl{
\VB=N^2-{N(N+1)\ov\cosh^2 x}.}{scat17}
The reflection coefficients for these potentials vanish
since $\RB(0)=0$.
The poles $k_n=i(N-n)$ of $\TB$
yield the energies of the bound states,
\eqnl{
E_n=N^2-(N-n)^2\qquad n=0,\dots,N-1.}{scat19}
Supersymmetry is unbroken, since the ground state has 
energy zero.
\section{Shape invariance}
\index{shape invariance}
Shape invariance is a property that arises when there
is an additional relationship between the partner
\textsc{Hamilton}ians $\HB$ and $\HF$. Suppose that these
\textsc{Hamilton}ians are linked by the condition
\eqnl{
\HF(\lam)=
A(\lam)\,A^\dagger(\lam)=A^\dagger\big(f(\lam)\big)\,A\big(f(\lam)\big)+c(\lam)
=\HB\big(f(\lam)\big)+c(\lam),}{sinv1}
where $f$ is a mapping from the space of coupling
constants into itself and $c(\lam)$ is a real-valued function.
When this condition holds, then
the \ham $\HB$ is said to be \emph{shape invariant} 
\cite{gendenshtein}.
For example, the partner potentials \refs{scat7} define
a shape invariant system with $f(\lam)=\lam-1$ and $c(\lam)=2\lam-1$. 

One can readily derive recursion relations for
the energies and scattering coefficients 
of a shape-invariant \haman on $\R$.\index{shape invariance}
As indicated in the figure on
page~\pageref{figure1} we denote the energy levels of 
$\HB$ by $E_n$ and those of $\HF$ by $E'_n$.
Then \refs{sinv1} implies 
\eqnn{
E'_{n+1}(\lam)=E_n\big(f(\lam)\big)+c(\lam)\mtxt{for}n\in \N_0,}
while supersymmetry implies $E_n(\lam)=E'_{n}(\lam)$ for $n\in \N$.
Combining these two properties yields 
\eqnl{
E_{n+1}(\lam)=E'_{n+1}(\lam)=E_n\big(f(\lam)\big)+c(\lam).}{sinv3}
Iterating this useful relation leads to
\eqnl{
E_N(\lam)=E_0\left(f_{\circ N}(\lam)\right)+\sum_{n=0}^{N-1} 
c\left(f_{\circ n}(\lam)\right),}{sinv5}
where $f_{\circ n}$ is the $n$-times iterated map $f$.
This result yields an explicit formula for the energies $E_N$
in case $\HB\left(f_{\circ N}(\lam)\right)$ admits a
zero energy bound state.

The shape invariance \refs{sinv1} also implies that
the \emph{scattering coefficients} of $\HF(\lam)$ and $\HB(f(\lam))$
are the same. Together with the supersymmetric relations
\refs{scat3} we obtain the recursion relations
\eqngrl{
\TB(\lam)=\prod_{n=0}^{N-1}{W_-\left(f_{\circ n}(\lam)\right)-ik\ov
{W_+\left(f_{\circ n}(\lam)\right)-ik'}}\,\TB\left(f_{\circ N}(\lam)\right)}
{\RB(\lam)=\prod_{n=0}^{N-1}{W_-\left(f_{\circ n}(\lam)\right)-ik\ov
{W_-\left(f_{\circ n}(\lam)\right)+ik}}\,\RB\left(f_{\circ N}(\lam)\right).}
{sinv17}
For the kink in \refs{scat6} with $W_-=\lam$ and $f(\lam)=\lam-1$ this
simplifies to the formula \refs{scat13} for the transmission
coefficient of the \textsc{P\"oschl-Teller} potential.
\subsection{Hydrogen atom in Einstein universe}
\index{hydrogen atom!in Einstein universe}
As an application we follow \schr \cite{schrhatom}
and consider a hydrogen atom in a closed \textsc{Einstein} 
universe with spatial line element on a $3$-dimensional 
sphere $S^3$,
\eqnn{
ds^2=R^2\left(d\xi^2+\sin^2\xi(d\theta^2+\sin^2\theta\,
d\varphi^2)\right).}
The radial-type coordinate $R\xi$ takes its values in the
finite interval $[0,\pi R]$. For a radially symmetric
potential $V(\xi)$ the angular momenta on the equatorial
$2$-spheres commute with the \schr operator and can be
diagonalized. The radial \schr equation takes the form ($\hbar=1$)
\eqnl{
-\pa_\xi(\sin^2\xi\pa_\xi f)+\ell(\ell+1)f+
\kappa\sin^2\xi \left(V(\xi)-E\right)f=0,\quad \kappa=2mR^2,}{sinv7}
where $\ell\in \N_0$ denotes the angular momentum.

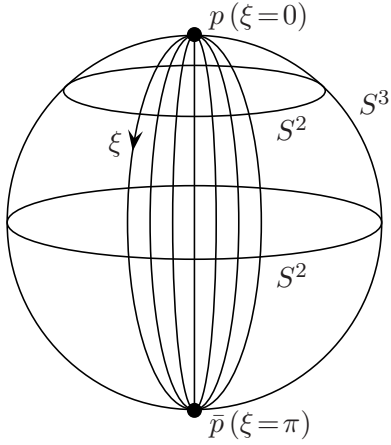
\begin{minipage}[t]{7.2cm}
\psset{unit=1mm,linewidth=0.6pt,dotsize=1mm}
\hskip3mm\label{figure2}
\begin{pspicture}(-30,-30)(30,32)
\pscircle(0,0){25}
\psellipse(0,0)(25,5)
\psellipse(0,17.5)(17.5,3.5)
\psellipse(0,0)(9,25)
\psellipse(0,0)(6,25)
\psellipse(0,0)(3,25)
\psline(0,-25)(0,25)
\psdot[dotsize=2mm](0,25)
\rput[l](2,27.3){$p\,(\xi\!=\!0)$}
\psline[arrowsize=2mm]{->}(-8,11)(-8.2,9.5)
\rput(-10.5,10.5){$\xi$}
\psset{dotsep=1pt}
\psdot[dotsize=2mm](0,-25)
\rput[l](2,-27){$\bar p\,(\xi\!=\!\pi)$}
\rput(24,16){$S^3$}
\rput(13,-7){$S^2$}
\rput(13,12.5){$S^2$}
\end{pspicture}
\end{minipage}
\begin{minipage}[b]{6.7cm}
The \textsc{Coulomb}-type potential on the spatial
section $S^3$ of the \textsc{Einstein} universe 
reads\index{Coulomb potential!in Einstein universe}
\eqnn{
V=-{e^2\ov R}\cot\xi}
and belongs to a proton at $\xi=0$ and 
an anti-proton at the opposite side of the 
universe, as depicted in the figure on the left.
As for any closed space without boundary there is
overall charge neutrality.\\
\end{minipage}
Setting $\psi=\sin\xi\,f$, the radial 
\schr equation for $\psi$ becomes
\eqnl{
-{d^2\psi\ov d\xi^2}+\VF\psi=\lam \psi
\,,\mtxt{where}\lam=1+\kappa E+a(\ell),\quad
a(\ell)={\nu^2\ov\ell^2}-\ell^2}{sinv9}
and we introduced $\nu=mRe^2$. The potential has the form
\eqnl{
\VF=
{\ell(\ell+1)\ov \sin^2\xi}-2\nu\cot\xi+a(\ell)=W^2-\frac{dW}{d\xi}\mtxt{with}
W=\ell\cot \xi-\frac{\nu}{\ell}.}{sinv11}
Actually this system is \emph{shape invariant} with intertwining relation
\eqnl{
\VF(\ell)=\VB(\ell+1)+c(\ell),\mtxt{where}
c(\ell)=a(\ell)-a(\ell+1).}{sinv13}
There is one (non-normalizable) fermionic zero-mode, and
the general formula \refs{sinv5} yields the eigenvalues
$\lam_N=a(\ell)-a(\ell+N)$. Setting $\ell+N=n$ we end up
with the following energies for hydrogen in an
Einstein universe,
\eqnl{
E_n=E'_n={n^2\!-1\ov 2mR^2}-{mc^2\ov 2}\left({\al\ov n}\right)^2
,\qquad n=1,2,\dots.}{sinv15}
The $E_n$ have no upper limit and all eigenvalues
are discrete. With the exception of the ground state energy 
all energy levels will be shifted as a
result of the interaction of the atom with the
curvature of space. \par
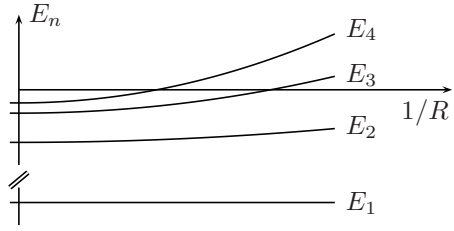
\begin{minipage}[t]{7.6cm}
\psset{xunit=1.2mm,yunit=1mm,linewidth=0.6pt,dotsize=1mm}
\hskip3mm\label{figure3}
\begin{pspicture}(-5,-16)(55,15)
\psline{->}(0,-10)(0,10)
\psline{->}(0,0)(48,0)
\psline(-1,-15)(35,-15)
\psline(0,-18)(0,-13)
\psline[doubleline=true](-1,-13)(1,-11)
\psplot{-1}{35}{x dup mul 0.0015 mul 7 sub}
\psplot{-1}{35}{x dup mul 0.004 mul 3.11 sub}
\psplot{-1}{35}{x dup mul 0.0075 mul 1.75 sub}
\small
\rput(38,-15){$E_1$}
\rput(38,-5){$E_2$}
\rput(38,2){$E_3$}
\rput(38,8){$E_4$}
\rput(3,10){$E_n$}
\rput(45,-3){$1/R$}
\end{pspicture}
\end{minipage}
\begin{minipage}[b]{6.2cm}
The effect differs from the usual gravitational
and Doppler shifts in that it perturbs each energy level
to a different extend. As expected, for $R\to\infty$ we
recover the energy levels of the
hydrogen atom in flat space.
\end{minipage}

\section{Isospectral deformations}
\index{isospectral deformation}
Let us assume that $\VF$ supports $n$ bound
states. By using supersymmetry one can easily 
construct an $n$-parameter family of potentials 
$V(\lam;x),\;\lam=(\lam_1,\dots,\lam_n)$, for which the
\haman has the same energies and scattering coefficients
as $H=p^2+\VF$. The existence of such
families of isospectral potentials has been
known for a long time from the inverse scattering
approach \cite{chadan}
which is technically more involved than the
method based on supersymmetry. We show how a
\emph{one-parameter} isospectral family of potentials is
obtained by first deleting and then re-inserting the
ground state of $\VF$ using the \textsc{Darboux}-procedure
\cite{darboux}. The generalization to an $n$-parameter
family is described in \cite{keung}.

Suppose that $\psife(x)$ is a normalizable zero-energy ground state
of the \haman with potential $\VF=W^2-W'$.
Its explicit form in position space is
\eqnl{
\psife(x)\propto e^{-\chi(x)},\qquad \chi(x)=\int^x W(x')dx'.}{isod1}
Suppose further that the partner potential $\VB=W^2+W'$ is
kept fixed. A natural question is whether there
are other superpotentials leading to the \emph{same}
potential $\VB$. A second solution $\hat W=W+\phi$ gives
rise to the same $\VB$ if
\eqnl{
\big(\hat W^2+\hat W'\big)-\big(W^2+W'\big)
=\phi^2+2W\phi+\phi'=0.}{isod3}
The transformation $\phi=(\log F)'$ leads to a linear
differential equation for $F'$
\eqnl{F''+2WF'=0}{isdef5}
with solution
\eqnl{
F'(x)=\exp\left(-2\int^x W(x')dx')\right)=\psife^2(x).}{isdef7}
The integration constant is just the lower bound of the
integral which determines the norm
of the fermionic ground state. A further integration yields
$F$ and hence $\phi$ and introduces another constant $\lam$
which is identified with the deformation parameter,
\eqnl{
\phi(x,\lam)={d\ov dx}\log\big(I(x)+\lam\big),\qquad
I(x)=\int\limits_{-\infty}^x \psife^2(x')dx'}
{isdef9}
In this formula for $\phi$ we
could change the lower integration bound
or multiply $I$ with any non-vanishing
constant. This is equivalent to a redefinition
of the constant $\lam$.
\par
By construction $W$ and $\hat W=W+\phi$ lead to the same
$\VB$. But the corresponding partner potentials
are different,
\eqnl{
\hat W^2-\hat W'=
W^2-W'+\phi^2+2W\phi-\phi'
\stackrel{\refs{isod3}}{=}\VF-2\phi'.}
{isdef11}
Thus the fermionic \ham operators with superpotentials
$W$ and $W+\phi$ are unequal.
But since they share the same partner
\haman $\HB$ they must have the same
spectrum, up to possible zero modes. This
then proves that all \ham operators of the
one-parameter family
\begin{eqnarray}
\HF(\lam)&=&-{d^2\ov dx^2}+\VF(\lam;x),\label{isdef13a}\\
\VF(\lam;x)\!\!&=&\VF(x)-2{d^2\ov dx^2}
\log\big(I(x)+\lam\big)\label{isdef13b}
\end{eqnarray}
have the same spectrum, up to possible
zero modes. The isospectral deformation \refs{isdef13a}
depends via the function $I(x)$
on the ground state wave function of the undeformed operator $\HF$.
\par
\paragraph{Deformation of the harmonic oscillator:}
Let us see how the deformation looks like for
the harmonic oscillator with potential
$\VF=\om^2x^2-\om$ and ground state wave function
$\psife(x)\propto\exp(-\om x^2/2)$.
We obtain
\eqnl{
\phi(\lam,x)=2\sqrt{\om\ov\pi}\,{e^{-\om x^2}\ov
\hbox{erf}\big(\sqrt{\om}\,x\big)+\lam},\mtxt{where}
\hbox{erf}(y)={2\ov\sqrt{\pi}}\int_0^y e^{-t^2}dt}{isdef15}
is the error function, and this leads to the following
isospectral deformation
\eqnl{
\VF(\lam;x)=\VF(x)+4\om x\,\phi(\lam,x)+2\phi^2(\lam,x)
=\VF(-\lam;-x).}{isdef17}
\par
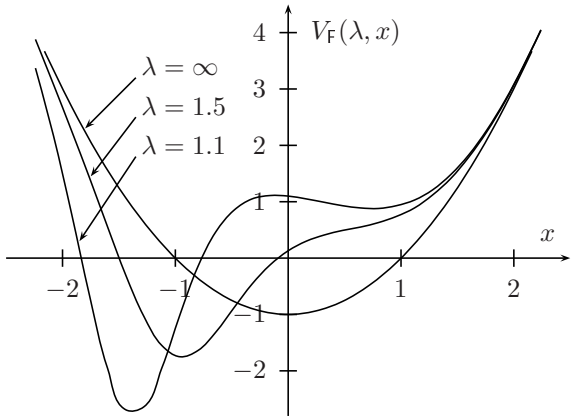
\begin{minipage}[t]{8.6cm}
\psset{xunit=15mm,yunit=7.5mm,linewidth=0.6pt,dotsize=1mm}
\begin{pspicture}(-2.5,-3)(2.5,4.5)
\small
\psaxes{->}(0,0)(-2.5,-2.8)(2.5,4.5)
\pscurve
(-2.16,3.67)(-1.92,2.69)(-1.68,1.82)(-1.44,1.07)
(-1.20,0.44)(-0.96,-0.08)(-0.72,-0.48)(-0.48,-0.77)(-0.24,-0.94)
(0.00,-1.00)(0.24,-0.94)(0.48,-0.77)(0.72,-0.48)(0.96,-0.08)
(1.20,0.44)(1.44,1.07)(1.68,1.82)(1.92,2.69)(2.16,3.67)
\pscurve
(-2.24,3.88)(-2.08,3.08)(-1.92,2.26)(-1.76,1.42)
(-1.60,0.55)(-1.44,-0.30)(-1.28,-1.03)(-1.12,-1.55)(-0.96,-1.75)
(-0.80,-1.64)(-0.64,-1.31)(-0.48,-0.88)(-0.32,-0.46)(-0.16,-0.12)
(0.00,0.13)(0.16,0.30)(0.32,0.41)(0.48,0.49)(0.64,0.55)
(0.80,0.63)(0.96,0.74)(1.12,0.89)(1.28,1.12)(1.44,1.41)
(1.60,1.79)(1.76,2.24)(1.92,2.77)(2.08,3.38)(2.24,4.04)
\pscurve
(-2.24,3.37)(-2.08,2.17)(-1.92,0.79)(-1.76,-0.67)
(-1.60,-1.96)(-1.44,-2.69)(-1.28,-2.64)(-1.12,-1.95)(-0.96,-1.01)
(-0.80,-0.15)(-0.64,0.47)(-0.48,0.85)(-0.32,1.05)(-0.16,1.11)
(0.00,1.10)(0.16,1.05)(0.32,0.99)(0.48,0.93)(0.64,0.89)
(0.80,0.88)(0.96,0.93)(1.12,1.03)(1.28,1.21)(1.44,1.48)
(1.60,1.83)(1.76,2.27)(1.92,2.79)(2.08,3.39)(2.24,4.05)
\rput(2.3,0.4){$x$}
\rput(.6,4){$\VF(\lam,x)$}
\rput[l](-1.3,3.4){$\lam=\infty$}
\psline{->}(-1.35,3.2)(-1.8,2.3)
\rput[l](-1.3,2.7){$\lam=1.5$}
\psline{->}(-1.35,2.5)(-1.75,1.45)
\rput[l](-1.3,2){$\lam=1.1$}
\psline{->}(-1.35,1.8)(-1.85,.3)
\end{pspicture}
\end{minipage}
\begin{minipage}[b]{5.3cm}
In the figure on the left we have plotted
the potential of the harmonic oscillator
and two deformations with parameters
$\lam=1.5$ and $\lam=1.1$. We have set
$\om=1$. For the deformed
potential to be regular we must assume
$\vert\lam\vert>1$. For $\lam\to\pm\infty$ the
potential tends to the potential
of the harmonic oscillator. For
$\vert\lam\vert\downarrow 1$ the deviation
from the oscillator potential become
significant near the origin. 
\end{minipage}
\par
\paragraph{Deformation of reflectionless P\"oschl-Teller
potentials:}
We deform the reflectionless \textsc{P\"oschl-Teller} Potential
\eqnl{
\VF=1-2\cosh^{-2} x,}{isdef21}
with just \emph{one supersymmetric bound state},
$\psife(x)=1/\cosh x$.
Since $\int^x \psife^2=\tanh x$ we obtain
\eqnl{
\phi(x)={1\ov (\cosh x)^2}\,{1\ov \tanh x+\lam}}{isdef25}
giving rise to the following isospectral deformation of $\VF$,
\eqnl{
\VF(\lam,x)=\VF(x)
+4\tanh(x)\phi(x)+2\phi^2(x)=\VF(-\lam,-x).}{isdef27}
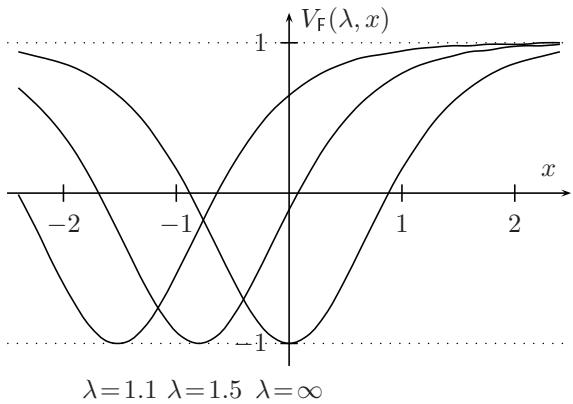
\begin{minipage}[t]{8.6cm}
\psset{xunit=15mm,yunit=20mm,linewidth=0.6pt,dotsize=1mm}
\hskip4mm
\begin{pspicture}(-2.5,-1.7)(2.5,1.4)
\small
\psaxes{->}(0,0)(-2.5,-1.15)(2.5,1.2)
\psline[linestyle=dotted](-2.5,1)(2.5,1)
\psline[linestyle=dotted](-2.5,-1)(2.5,-1)
\pscurve
(-2.40,0.94)(-2.24,0.91)(-2.08,0.88)(-1.92,0.84)(-1.76,0.78)
(-1.60,0.70)(-1.44,0.60)(-1.28,0.47)(-1.12,0.30)(-0.96,0.11)
(-0.80,-0.12)(-0.64,-0.36)(-0.48,-0.60)(-0.32,-0.81)(-0.16,-0.95)
(0.00,-1.00)(0.16,-0.95)(0.32,-0.81)(0.48,-0.60)(0.64,-0.36)
(0.80,-0.12)(0.96,0.11)(1.12,0.30)(1.28,0.47)(1.44,0.60)
(1.60,0.70)(1.76,0.78)(1.92,0.84)(2.08,0.88)(2.24,0.91)
(2.40,0.94)
\pscurve
(-2.40,0.70)(-2.24,0.59)(-2.08,0.46)(-1.92,0.30)(-1.76,0.10)
(-1.60,-0.13)(-1.44,-0.37)(-1.28,-0.61)(-1.12,-0.81)(-0.96,-0.95)
(-0.80,-1.00)(-0.64,-0.95)(-0.48,-0.80)(-0.32,-0.59)(-0.16,-0.35)
(0.00,-0.11)(0.16,0.11)(0.32,0.31)(0.48,0.47)(0.64,0.60)
(0.80,0.70)(0.96,0.78)(1.12,0.84)(1.28,0.88)(1.44,0.91)
(1.60,0.94)(1.76,0.95)(1.92,0.97)(2.08,0.98)(2.24,0.98)
(2.40,0.99)
\pscurve
(-2.40,-0.01)(-2.24,-0.24)(-2.08,-0.49)(-1.92,-0.71)(-1.76,-0.89)
(-1.60,-0.99)(-1.44,-0.99)(-1.28,-0.89)(-1.12,-0.71)(-0.96,-0.48)
(-0.80,-0.24)(-0.64,0.00)(-0.48,0.21)(-0.32,0.39)(-0.16,0.54)
(0.00,0.65)(0.16,0.74)(0.32,0.81)(0.48,0.86)(0.64,0.90)
(0.80,0.92)(0.96,0.94)(1.12,0.96)(1.28,0.97)(1.44,0.98)
(1.60,0.98)(1.76,0.99)(1.92,0.99)(2.08,0.99)(2.24,1.00)
(2.40,1.00)
\small
\rput(2.3,0.15){$x$}
\rput(.5,1.15){$\VF(\lam,x)$}
\rput(0,-1.3){$\lam\!=\!\infty$}
\rput(-.75,-1.3){$\lam\!=\!1.5$}
\rput(-1.5,-1.3){$\lam\!=\!1.1$}
\end{pspicture}
\end{minipage}
\begin{minipage}[b]{5.3cm}
In the figure to the left we have plotted
the reflectionless \textsc{P\"oschl-Teller}
potential with one bound state 
and two of its isospectral deformations with
parameters $\lam=1.5$ and $1.1$. For the deformed
potential to be regular we must assume
$\vert\lam\vert>1$. For $\lam\to\pm\infty$ the
potential tends to the
\textsc{P\"oschl-Teller} potential.
For $\lam\downarrow 1$ the minimum
of the potential tends to $-\infty$ and
for $\lam\uparrow -1$ to $\infty$.
\end{minipage}
\par

The potential $\VF$ may be viewed as a \emph{soliton}\index{soliton}
with its center at the minimum.
For $\lam=1$ the soliton is at $x=-\infty$ and moves
to the origin for $\lam\to\infty$.
For $\lam=-1$ the soliton is centered
at $\infty$ and moves with decreasing
$\lam$ to the left. For $\lam=-\infty$
it reaches the origin. Actually one can show
that for $\lam=-\coth(4t+c)$ the function $u(t,x)=\VF(\lam(t),x)-1$ 
solves the wellknown \textsc{Korteweg-deVries} equation,
\index{Korteweg-deVries equation}
\eqnl{
u_t+u_{xxx}-6uu_x=0.}{isdef29}
For a generalization of this construction to $n$-soliton
solutions I refer to the review of F. Cooper et al. \cite{cooper}.

\section{SQM in higher dimensions}
\index{SQM in higher dimensions}
Supersymmetric quantum mechanical systems 
also exist in higher dimensions \cite{andrianov}. 
The construction is motivated by the following 
rewriting of the supercharge \refs{sqm1}:
\eqnl{
\cQ=\psi\otimes A\mtxt{and}\cQd=\psi^\dagger\otimes \Ad}{hdi1}
containing the \emph{fermionic} operators
\eqnl{
\psi=\pmatrix{0&1\cr 0&0}\mtxt{and}
\psi^\dagger=\pmatrix{0&0\cr 1&0}}{hdi2}
with anti-commutation relations
\eqnl{
\{\psi,\psi\}=\{\psi^\dagger,\psi^\dagger\}=0\mtxt{and}
\{\psi,\psi^\dagger\}=\id.}{hdi3}
For the choice
$A=i\pa_x+iW$ as in \refs{sps1} the super-\haman \refs{sqm3} reads
\eqngrl{
H&=&\left(p^2+W^2\right)\{\psi,\psi^\dagger\}+W'[\psi,\psi^\dagger]}
{&=&\HB-2W'\psida\psi=\HF+2W'\psi\psida,}{hdi5}
where we skipped the tensor product symbol.
In \cite{andrianov} this construction has been
generalized to $d>1$ dimensions. Then one
has $d$ fermionic annihilation operators $\psi_k$ 
and $d$ creation operators $\psi^\dagger_k$,
\eqnl{
\{\psi_k,\psi_\ell\}=\{\psida_k,\psida_\ell\}=0\mtxt{and}
\{\psi_k,\psida_\ell\}=\delta_{k\ell},\quad k,\ell=1,\dots,d.}{hdi7}
For the supercharge we make the ansatz
\eqnn{
\cQ=i\sum \psi_k \left(\pa_k+W_k(\mbx)\right)=
i\mbpsi\cdot(\nabla+\mbW),}
where $\mbpsi$ denotes the $d$-tupel $(\psi_1,\dots,\psi_d)^T$.
The supercharge is \emph{nilpotent} if and only if $\pa_kW_\ell-\pa_\ell W_k=0$ 
holds true. Locally this integrability condition is 
equivalent to the existence of a potential
$\chi(\mbx)$ with $\mbW=\nabla\chi$. Thus we are lead to
the following \emph{nilpotent} supercharge and its adjoint,
\eqngrl{
\cQ&=&i\mbpsi\cdot\left(\nabla+\nabla\chi\right)
=e^{-\chi}\cQ_0\,e^\chi\;\mtxt{with}
\cQ_0=i\mbpsi\cdot\nabla}
{\cQd\!\!&=&i\mbpsi^\dagger\!\cdot \left(\nabla-\nabla\chi\right)
=e^\chi \cQd_0\,e^{-\chi}\mtxt{with}\cQd_0=i\mbpsi^\dagger\cdot\nabla.}{hdi9}
The super \haman takes the simple form
\eqngrl{
H=\{\cQ,\cQd\}&=&
H_0\otimes \id_{2^d}
-2\sum\psida_k \psi_\ell\,\pa_k\pa_\ell\chi}
{&=&
H_d\otimes \id_{2^d}
+2\sum\psi_k \psida_\ell\,\pa_k\pa_\ell\chi,}
{hdi11}
where $H_0$ and $H_d$ are the \schr operators in
the extreme sectors,
\eqnl{
H_0=-\triangle+(\nabla\chi,\nabla\chi)+\triangle \chi
\quad,\quad
H_d=-\triangle+(\nabla\chi,\nabla\chi)-\triangle \chi}{hdi12}
This super \ham generalizes the \textsc{Nicolai-Witten}
operator \refs{hdi5} in one dimension to $d$ dimensions.

Again there exists a (fermion) \emph{number operator}
\index{number operator}
\eqnl{
\num=\sum\psida_k\psi_k}{hdi13}
and the $\psi_k$ decrease it by one unit
whereas the $\psida_k$ increase it by one unit.
The same is then true for the supercharge and its
adjoint,
\eqnl{
[\num,\cQ]=-\cQ\mtxt{and}[\num,\cQd]=\cQd.}{hdi14} 
A direct way to find a representation
for the fermionic operators makes use of the \textsc{Fock} 
construction over a  'vacuum'-state $\vac$ which is 
annihilated by all $\psi_k$,\index{Fock construction}
\index{fermionic operators in SQM}
\eqnl{
\psi_k\vac=0,\quad k=1,\dots,d.}{hdi15}
Acting with the $d$ raising operators on $\vac$ yields the states 
\eqnn{
\vert k\ra =\psida_k\vac}
with $\num=1$. When counting the states with higher fermion
number we should recall that the raising operators
anticommute, such that
\eqnl{
\vert k_1\dots k_n\ra=\psida_{k_1}\cdots\psida_{k_n}\vac}{hdi17}
is antisymmetric in $k_1,\dots,k_n$. The states and their
corresponding  eigenvalues of $\num$ together with their 
degeneracies are listed in the following table: 
\eqnn{
\begin{array}{|c|c|c|c|c|c|}\hline 
\hbox{states:}&\vac&\vert k\ra&\vert k,\ell\ra&\cdots&
\vert 1,2,\dots,d\ra \\ \hline
\num&0&1&2&\dots&d\\ \hline
\#\hbox{ of states}&{d\choose 0}=1&{d \choose 1}=d&{d\choose 2}&\cdots&
{d\choose d}=1 \\ \hline
\end{array}}
The total number of independent states is $2^d$
and thus we obtain a $2^d$-dimensional irreducible
representation of the fermionic algebra \refs{hdi7}.
The states with even $\num$ are called
\emph{bosonic}, those with odd $\num$ \emph{fermionic}.
The number of bosonic states equals the
number of fermionic states.

With the help of \refs{hdi7} and \refs{hdi15} one
may calculate the matrix elements of $\psi_k$ between
any two \textsc{Fock} states \refs{hdi17}. In \emph{one dimension}
there is one bosonic and one fermionic state and for
the orthonormal basis
\eqnn{
e_1=\vac\mtxt{and}e_2=\psida\vac}
we recover the annihilation operator $\psi$ in \refs{hdi2}.
In \emph{two dimensions} there are two bosonic and
two fermionic states and with respect to the orthonormal basis
\eqnl{
\{e_1,e_2,e_3,e_4\}=
\{\vac,\vert 1\ra,\vert 2\ra,\vert 12\ra\}}{hdi19}
the annihilation operators are represented by
\eqnl{
\psi_1=\pmatrix{0&1&0&0\cr 0&0&0&0\cr0&0&0&1\cr0&0&0&0}\mtxt{and}
\psi_2=\pmatrix{0&0&1&0\cr 0&0&0&-1\cr0&0&0&0\cr0&0&0&0}.}{hdi21}
Taking into account the $\mbx$-dependency of the states,
the \hils of supersymmetric quantum mechanics in $d$ dimensions is
\eqnl{
\cH=L_2\big(\R^d\big)\otimes \C^{2^d}}{hdi23}
and decomposes into sectors with different fermion numbers,
\eqnl{
\cH=
\cH_0\oplus \cH_1\oplus\dots\oplus \cH_d\mtxt{with}
\num\big\vert_{\cH_p}=p\id.}{hdi25}
An arbitrary element in $\cH$ has the expansion
\eqnl{
\psi(x)=
f(\mbx)\vac+
f_k(\mbx)\vert k\ra+
\ha f_{k\ell}(\mbx)\vert k\ell\ra+
{1\ov 3!}f_{k\ell m}(\mbx)\vert k\ell m\ra+\dots.}{hdi27}
Recalling that $\cQ$ lowers and $\cQd$ raises
$\num$ by one unit it follows at once that
the super-\haman \refs{hdi11} commutes with $\num$.
In a basis adapted  to the decomposition \refs{hdi25}
the number operator is block-diagonal,
\eqnl{
\num=\pmatrix{0\cdot \id_{d\choose 0}&&&\cr &1\cdot\id_{d\choose 1}&&\cr &&
\ddots&\cr &&&d\cdot \id_{d\choose d}}}{hdi29}
and so is the super-\haman
\eqnl{
H=\pmatrix{H_0&&&\cr &H_1&&\cr
&&\ddots&\cr &&&H_d}\qquad
\matrix{H_0=p^2+(\nabla\chi^2)+\triangle \chi
\cr \cr
H_d=p^2+(\nabla\chi^2)-\triangle \chi\,.}}{hdi31}
Note that in the extremal sectors with $\num=0$ and
$\num=d$ the super \haman reduces to ordinary
\schr operators.

SQM in higher dimensions with a nilpotent supercharge
defines a \emph{complex} of the following structure:
\index{complex!formed by superscharge}
\par
\psset{unit=1mm,linewidth=.6pt}
\hskip8mm
\begin{pspicture}(0,-4)(135,15)
\rput(5,5){$\cH_0$}
\psline{->}(9,6)(21,6)
\psline{<-}(9,4)(21,4)
\rput(15,9){$\cQd$}
\rput(15,1){$\cQ$}
\rput(25,5){$\cH_1$}
\psline{->}(29,6)(41,6)
\psline{<-}(29,4)(41,4)
\rput(35,9){$\cQd$}
\rput(35,1){$\cQ$}
\rput(45,5){$\cH_2$}
\psline{->}(49,6)(61,6)
\psline{<-}(49,4)(61,4)
\rput(55,9){$\cQd$}
\rput(55,1){$\cQ$}
\multirput(64,5)(4,0){4}{\psdot}
\psline{->}(79,6)(89.5,6)
\psline{<-}(79,4)(89.5,4)
\rput(85,9){$\cQd$}
\rput(85,1){$\cQ$}
\rput(95,5){$\cH_{d-1}$}
\psline{->}(100,6)(111,6)
\psline{<-}(100,4)(111,4)
\rput(105,9){$\cQd$}
\rput(105,1){$\cQ$}
\rput(115,5){$\cH_d$}
\end{pspicture}
\par
For the free supercharge $\cQ_0$ this complex is isomorphic
to the \textsc{de Rham} complex for differential forms.
The nilpotent charge $\cQd_0$ is identified with
the \emph{exterior differential} $d$ and $\cQ_0$ with the
co-differential $\delta$. The super \haman
$H=\{\cQ_0,\cQd_0\}$ corresponds to the \textsc{Laplace-Beltrami}
operator $-\triangle$.

The nilpotent supercharge gives
rise to the following  \textsc{Hodge}-type decomposition
of the \textsc{Hilbert} space,
\eqnl{
\cH=\cQ\cH \oplus \cQd\cH \oplus \cH_0,\qquad
\cH_0=\hbox{Ker}\,H\;,}{hdi33}
where the finite dimensional subspace $\cH_0$
is spanned by the zero-modes of $H$. Indeed,
on the orthogonal complement of $\cH_0$ we may 
invert $H$ and write (using $[H,\cQ]=0$),
\eqnn{
\cH_0^\perp=(\cQ\cQd+\cQd \cQ)H^{-1}\cH_0^\perp
=\cQ\left(H^{-1}\cQd\cH_0^\perp\right)
+\cQd \left(H^{-1}\cQ\cH_0^\perp\right),}
which proves \refs{hdi33}.

Before we study two relevant systems we rewrite
the superalgebra in terms of the hermitian supercharges
$\cQ_1$ and $\cQ_2$ in the decompositions
\eqnl{
\cQ=\ft12(\cQ_1+i\cQ_2)\mtxt{and}
\cQd=\ft12(\cQ_1-i\cQ_2).}{hdi34}
$\cQ_1$ and $\cQ_2$ anticommute
and both are roots of the super-\textsc{Hamilton}ian,
\eqnl{
\{\cQ_i,\cQ_j\}=2\delta_{ij}H.}{hdi35}
This means that the systems with super-\haman \refs{hdi11}
actually possess an \emph{extended supersymmetry} with two
real supercharges. More generally, the non-negative
super-\haman of a SQM with $\cN$ supersymmetries can
be written as\index{extended supersymmetry}
\index{supersymmetry, extended}
\eqnl{
\delta_{ij}H=\ft12 \left\{\cQ_i,\cQ_j \right\},
\qquad i,j=1,\ldots,\cN,}{hdi37}
with \emph{hermitian} supercharges $\cQ_i$ anticommuting with an
involutary operator $\Gamma$,
\eqnl{
\{\cQ_i ,\Gamma \}= 0, \qquad \Gamma^\dagger=\Gamma,\qquad
\Gamma^2 =\id . }{hdi39}
In our case $\Gamma$ is just the number operator
modulo $2$.

There exist other definitions for SQM in the
literature, for a recent discussion,
in particular concerning the role of the
grading operator $\Gamma$, we refer to \cite{Combescure:2004ey}.
One may also relax the condition for the left-hand side
of \refs{hdi37}, see for example \cite{Hull:1999ng}, but
in these lectures we will not consider such generalizations.

\subsection{The $2d$ supersymmetric anharmonic oscillator}
\index{anharmonic oscillator!$2d$, supersymmetric}
As an application we consider susy
oscillators in $\R^2$ with polar coordinates,
\eqnl{
z=x_1+ix_2=r e^{i\varphi}.}{sanh1}
They emerge in the \emph{strong coupling limit}
of certain \textsc{Wess-Zumino}-models
on space lattices and for this reason
we are interested in their vacuum structure. We choose the
harmonic superpotential,
\eqnl{
\chi(\mbx)=\frac{\lam}{p}\,r^p\cos(p\varphi),}{sanh3}
and obtain the following super-\haman in the 
basis \refs{hdi19}, 
\eqnl{
H=\{\cQ,\cQd\}=\pmatrix{H_0&0&0\cr 0&H_1&0\cr 0&0&H_2},}{sanh5}
with the well-studied anharmonic oscillator in the extremal
sectors $\cH_0$ and $\cH_2$,
\eqnl{
H_0=H_2=-\triangle +(\nabla\chi)^2=-\triangle+\lam^2 (z\bar z)^{p-1}.}{sanh7}
and a matrix \schr operator in $\cH_1$,
 \eqnl{
H_1=H_0\cdot\id_2+2\lam (p-1)\pmatrix{-\Re z^{p-2}
&\Im z^{p-2}\cr \;\Im z^{p-2}&\Re z^{p-2}}}{sanh9}
The ground states of $H$ are known
\cite{Elitzur:1983nj,Bruckmann:1997,wipfkirch3},
in contrast to the ground state(s) of
non-supersymmetric anharmonic oscillator $H_0$.
To construct these states we observe that
the 'angular momentum'
\eqnl{
J=L-s\pmatrix{0&0&0\cr 0&\sigma_2&0\cr 0&0&0},
\qquad L={1\ov i}\pa_\varphi,\quad s={p-2\ov 2},}{sanh11}
is conserved and that the ground states must
reside in $\cH_1$, since in the
other sectors $H=H_0$ is positive.
Diagonalizing $J$ on $\cH_1$
leads to the ansatz
\eqnl{
\psi_j(\mbx)=e^{ij\varphi}\left(f_j(r)\,e^{-is\varphi}
\sigma_3
+g_j(r)\,e^{is\varphi}\right)\pmatrix{1\cr i},}{sanh13}
where the eigenvalues $j$ of $J$ are integers for
even $p$ and half-integers for odd $p$.
The zero-energy conditions $\cQ\psi_j= \cQd\psi_j= 0$
yields a coupled system of first order differential
 equations for the radial functions $f_j$ and $g_j$.
The square integrable solutions are just \textsc{Bessel}
functions,
\eqnl{
f_j (r)=c\, r^{p-1}\,
\mbox{K}_{\ha+\frac{j}{p}}
\left(\textstyle\frac{\lam}{p}r^p\right)\mtxt{and}
g_j (r)=c\, r^{p-1}\,
\mbox{K}_{\ha-\frac{j}{p}}
\left(\textstyle\frac{\lam}{p}r^p\right),}{sanh15}
where $j\in\{-s,-s+1,\dots,s-1,s\}$. The number of
supersymmetric ground states of the oscillator
with potential $\propto r^{2p-2}$ is 
just $p-1$. For example, the supersymmetric
anharmonic oscillator with $r^4$ potential
has $2$ normalizable zero modes.

\subsection{The supersymmetric hydrogen atom}
\index{hydrogen atom!supersymmetric}
For a closed system of two non-relativistic point
masses interacting via a central force the angular
momentum $\mbL$ of the relative motion is conserved
and the motion is always in the plane perpendicular to
$\mbL$. If the force is derived from the \textsc{Coulomb}
potential, there is an additional conserved quantity: the 
\textsc{Laplace-Runge-Lenz}\footnote{A more suitable name for
this constant of motion would be 
\textsc{Hermann-Bernoulli-Laplace}
vector, see \cite{goldstein}.} vector.
This vector is perpendicular to $\mbL$ and
points in the direction of the semi-major axis.
For the hydrogen atom the corresponding hermitian vector
operator has the form
\eqnl{
\mbC={1\ov 2m}\big(\mbp\times\mbL-\mbL\times\mbp\big)-{e^2\ov r}\mbx
.}{hydro1}
with reduced mass $m$ of the proton-electron
system.
By exploiting the existence of this conserved vector operator,
\textsc{Pauli} calculated the spectrum of the hydrogen atom by purely
algebraic means \cite{pauli}. He noticed that 
the angular momentum $\mbL$ together with the vector
\eqnl{
\mbK=\sqrt{-{m\ov 2H}}\,\mbC \;,}{hydro3}
which is well-defined and hermitian on bound states with
negative energies,
generate a hidden $SO(4)$ symmetry algebra,
\eqnl{
[L_i,L_j]=[K_i,K_j]=i\hbar\eps_{ijk}L_k \;,\quad
[L_i,K_j]=i\hbar\eps_{ijk}K_k \;,}{hydro5}
and that the \haman
can be expressed in terms of $\mbK^2+\mbL^2$, one of the
two second-order \textsc{Casimir} operators of this algebra,
as follows
\eqnl{
H=-{me^4\ov 2}{1\ov \mbK^2+\mbL^2+\hbar^2} \;.}{hydro7}
One also notices that the second \textsc{Casimir}
operator  $\mbK\cdot\mbL$ vanishes and arrives
at the bound state energies by purely group
theoretical methods. The existence
of the conserved vector $\mbK$ also explains the 
accidental degeneracy of the hydrogen spectrum.

In a recent publication with \textsc{A. Kirchberg} and
\textsc{D. Lange} we 'supersymmetrized' this construction 
and showed that the \emph{supersymmetric hydrogen atom}
admits generalizations of the angular momentum
and \textsc{Laplace-Runge-Lenz} vector \cite{wipfkirch2}. 
Similarly as for the ordinary \textsc{Coulomb} problem the 
hidden $SO(4)$-symmetry generated by these two 
vector operators allows for a purely algebraic 
solution of the supersymmetric system.

To find the supersymmetrized hydrogen
atom we choose $\chi=-\lam r$ in \refs{hdi11} and
obtain in $3$ dimensions the super-\haman
\cite{wipfkirch2}
\eqnl{
H=(-\triangle+\lam^2)\id_8-{2\lam\ov r}B,\qquad
B=\id-\num+S^\dagger S,\quad S=\hat\mbx\cdot \mbpsi}{hydro9}
on the \hils
\eqnl{
\cH=L_2(\R^3)\times \C^8=\cH_0\oplus \cH_1\oplus\cH_2\oplus\cH_3.}{hydro11}
We defined the triplet $\mbpsi$ containing the $3$
annihilation operators $\psi_1,\psi_2,\psi_3$.
States in $\cH_0$ are annihilated by $S$
and states in $\cH_3$ by $S^\dagger$. With
$\{S^\dagger,S\}=\id$ we obtain the following
\ham operators in these subspaces,
\eqnl{
H_0=-\triangle+\lam^2-{2\lam\ov r}\mtxt{and}
H_3=-\triangle+\lam^2+{2\lam\ov r}.}{hydro13}
Hence, the \schr operators for both the electron-proton 
$H_0$ and positron-proton systems $H_3$ are part of $H$.
The conserved angular momentum contains a spin-type term,
\eqnl{
\mbJ=\mbL+\mbS=\mbx\wedge\mbp-i\mbpsi^\dagger\wedge\mbpsi,}
{hydro15}
and the operators $S$ and $B$ in \refs{hydro9} commute
with this total angular momentum, since
$\mbx$ and $\mbpsi$ are both vector operators.
To find the susy extension of the \textsc{Runge-Lenz} 
vector is less simple. It reads
\cite{wipfkirch2}
\eqnl{
\mbC=\mbp\wedge\mbJ-\mbJ\wedge\mbp-2\lam\, \hat\mbx B}{hydro17}
with $\mbJ$ from \refs{hydro15}. The properly normalized vector
\eqnl{
\mbK=\ha{\mbC\ov \sqrt{\lam^2-H}}}{hydro19}
together with $\mbJ$ form an $SO(4)$ symmetry algebra
on the subspace of bound states for
which $H<\lam^2$.

Finally we would like to find a relation
similar to \refs{hydro7} to solve for the
spectrum. However, one soon realizes that
there is no algebraic relation between the
conserved operators $\id, \num,\mbJ^2,
\mbK^2$ and $H$. However, we can prove
the equation
\eqnl{
\lam^2\cC_{(2)}=\mbK^2 H
+\left(\mbJ^2 + (1-\num)^2 \right) \cQ\cQd
+\left(\mbJ^2+(2-\hbox{N})^2 \right) \cQd \cQ \;,}
{hydro21}
where $\cC_{(2)}$ is the second-order \textsc{Casimir}
of the dynamical symmetry group $SO(4)$,
\eqnl{
\cC_{(2)} =\mbJ^2+\mbK^2}{hydro23}
and this relation is sufficient to obtain the
energies of the supersymmetric $H$-atom.

Each of the three subspaces in the decomposition
\refs{hdi33} is left invariant by $H$
and thus we may diagonalize it on each subspace separately.
Since $H\vert_{\cQ\cH}=\cQ\cQd$ 
and $H\vert_{\cQd\cH}=\cQd \cQ$ we can solve 
\refs{hydro21} for $H$ in both subspaces,
\eqnl{
H\big\vert_{\cQ\cH}=\lam^2\,{\cC_{(2)}\ov (1-\num)^2+\cC_{(2)}}
\mtxt{and}
H\big\vert_{\cQd\cH}=
\lam^2\,{\cC_{(2)}\ov (2-\num)^2+\cC_{(2)}}.}{hydro25}
The states with zero energy are annihilated by both
$\cQ$ and $\cQd$, and according to \refs{hydro21} 
the second-order \textsc{Casimir} must vanish on these modes,
such that
\[
\cC_{(2)}\big\vert_{\hbox{Ker}\,H}=0\;.\]
We conclude that every supersymmetric ground state of $H$
is an $SO(4)$ singlet.\\

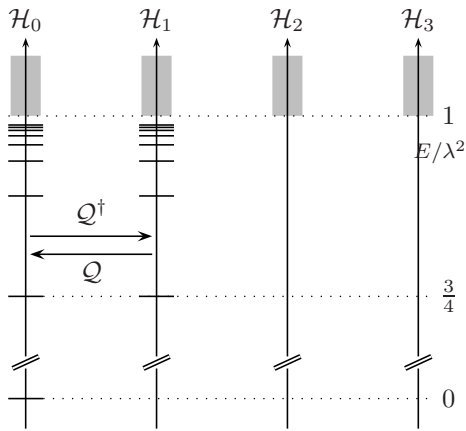
\begin{minipage}[t]{7.1cm}
\psset{xunit=.58cm,yunit=.8cm,linewidth=0.6pt,dotsize=1mm}
\begin{pspicture}(0,1)(11,8.5)
\multirput(1,1)(3,0){4}{\psline(0,0)(0,1)}
\psline(.6,1.5)(1.4,1.5)
\psline[linestyle=dotted](1.4,1.5)(10.4,1.5)
\multirput(1,2.1)(3,0){4}{\psline[doubleline=true](-.3,-.1)(.3,.1)}
\multirput(1,6.2)(3,0){4}{
\psline[linewidth=4,linecolor=lightgray](0,0)(0,1)}
\multirput(1,2.2)(3,0){4}{\psline{->}(0,0)(0,5.3)}
\psline[linestyle=dotted](1.4,3.2)(10.4,3.2)
\psline[linestyle=dotted](.6,6.2)(10.4,6.2)
\multirput(0,0)(3,0){2}{
\psline(0.6,3.2)(1.4,3.2)
\psline(0.6,4.87)(1.4,4.87)
\psline(0.6,5.45)(1.4,5.45)
\psline(0.6,5.72)(1.4,5.72)
\psline(0.6,5.87)(1.4,5.87)
\psline(0.6,5.96)(1.4,5.96)
\psline(0.6,6.01)(1.4,6.01)
\psline(0.6,6.05)(1.4,6.05)}
\small
\rput(10.7,1.5){$0$}
\rput(10.7,3.2){$\ft{3}{4}$}
\rput(10.7,6.2){$1$}
\rput(10.5,5.6){\scriptsize$E/\lam^2$}
\rput(1,7.8){$\cH_0$}
\rput(4,7.8){$\cH_1$}
\rput(7,7.8){$\cH_2$}
\rput(10,7.8){$\cH_3$}
\psline[arrowsize=1.4mm]{->}(1.1,4.2)(3.9,4.2)
\rput(2.5,4.6){$\cQd$}
\psline[arrowsize=1.4mm]{->}(3.9,3.9)(1.1,3.9)
\rput(2.5,3.6){$\cQ$}
\end{pspicture}
\end{minipage}
\begin{minipage}[b]{6.7cm}
In the figure on the left we have plotted
the spectrum of the supersymmetric
$H$-atom. The bound states are in the sectors
with fermion number $0$ and $1$. The allowed
$SO(4)$-representations, energies, degeneracies
and wave functions have been
calculated in \cite{wipfkirch2} with
group theoretical methods. Actually the
\haman \refs{hdi11} with $\chi(\mbx)=-\lam/r$
can be diagonalized in arbitrary dimensions.
Again there exist a generalized angular momentum
and \textsc{Runge-Lenz} vector which generate
a symmetry $SO(d+1)$.
\end{minipage}
Much work went into investigating
supersymmetric QM since the pioneering
papers of \textsc{Infeld} and \textsc{Hull} \cite{infeld},
\wit \cite{witten81} and \textsc{Gendenshtein}
\cite{gendenshtein}. It is impossible 
to cover all topics in one chapter and I
presented those which I personally find
most interesting and/or to which we contributed. 
For example, I omitted the formulation
of SQM in superspace and the interesting
interrelation with index theorems.
I omitted the applications of SQM to atomic,
nuclear and statistical physics. Also there is
a close relation between 'shape-invariant susy' 
and group theory. Most of these aspects are
covered in the reviews \cite{cooper,junker,kalka}.
Very recently the following question has been
answered: Given a \textsc{Dirac} operator
$i\di$ for charged particles in a (\textsc{Euclid}ean)
curved space. What are the conditions on the
background gauge field $A_\mu$ and
metric $g_{\mu\nu}$ such that
$-\di^2$ possesses $\cN>1$ first order
hermitian differential operator $\cQ_i$ as 
square roots and that these roots form an
extended superalgebra
\eqnl{
\{\cQ_i,\cQ_j\}=2\delta_{ij}H,\qquad i,j=1,\dots,\cN.}{hydro27}
For example, one finds $\cN=2$ hermitian supercharges
in \textsc{K\"ahler} spaces and for particular
gauge fields and $\cN=4$ supercharges in 
hyper-\textsc{K\"ahler} spaces and (in $4$ dimensions)
selfdual or anti-selfdual gauge fields.
The solutions for arbitrary $\cN$ and space dimensions
can be found in \cite{wipfkirch1}. 

\chapter{Symmetries and Spinors}\label{chap:3}

Some time ago the idea came up that perhaps the
approximate $SU(3)$ symmetry of strong interaction
is part of a larger $SU(6)$ symmetry and that
mesons (or baryons) with different spins
belong to \emph{one multiplet} of this bigger
symmetry group. Various attempts were made
to generalize this symmetry of the 
non-relativistic quark model to a fully relativistic
QFT. These attempts failed, and several
authors proved no-go theorems showing that in fact this is 
impossible. Well-known is the \textsc{Coleman-Mandula}
theorem  \cite{Coleman}
which states that in a theory with \emph{nontrivial
scattering} in more than $1+1$ dimensions, the only
possible conserved quantities that transform
as tensors under the \lor group (i.e. without
spinor indexes) are the $4$-momentum
$P_\mu$ generating spacetime translations,
the generalized angular momenta $J_{\mu\nu}$
generating \lor transformations and possible
'internal' symmetry charges $B_k$ which commute
with $P_\mu$ and $J_{\mu\nu}$. The $(P_\mu,J_{\mu\nu})$
generate the \poin \emph{algebra} $\cP$,
\index{Poincare algebra}
\begin{eqnarray}
\,[J_{\mu\nu},J_{\rho\sigma}]&=&
i\big(\eta_{\mu\rho}J_{\nu\sigma}+\eta_{\nu\sigma}J_{\mu\rho}
-\eta_{\mu\sigma}J_{\nu\rho}-\eta_{\nu\rho}J_{\mu\sigma}\big)\label{lotr19a}\\
\,[J_{\mu\nu},P_\rho ]
&=&i\big(\eta_{\mu\rho}P_\nu-\eta_{\nu\rho}P_\mu\big)
\label{lotr19b}\\
\,[P_\mu,P_\nu]&=&0\label{lotr19c}
\end{eqnarray}
with $(\eta_{\mu\nu})=\hbox{diag}(1,-1,\dots,-1)$.
It is the symmetry algebra of any relativistic
field theory. There exists an extension
of the \textsc{Coleman-Mandula} theorem for
\emph{massless particles} which allows for
the generators of conformal transformations.
\section{Coleman-Mandula theorem}
\index{Coleman-Mandula theorem}
In order to understand the
\textsc{Coleman-Mandula} theorem better we consider
the theory for two free real scalar fields 
with \lagan density
\eqnl{
\cL_0'=\ft12 \pa_\mu\phi_1\pa^\mu\phi_1
+\ft12 \pa_\mu\phi_2\pa^\mu\phi_2}{cman1}
and linear \textsc{Euler-Lagrange} equations
\eqnl{
\Box\phi_i=0,\qquad i=1,2.}{cman3}
Besides the conserved energy-momentum
and 'angular momentum' tensors
\eqnl{
T_{\mu\nu}=\sum_i \pa_\mu\phi_i\pa_\nu\phi_i-\eta_{\mu\nu}\cL_0'
\mtxt{and}
J_{\mu\rho\sigma}=\h x_\sigma T_{\mu\rho}-\h x_\rho T_{\mu\sigma}}{cman5}
such a model has infinitely many conserved currents.
For example it follows immediately that the
series of currents
\eqnl{
J_{\mu\rho_1\dots\rho_n}=\phi_1\pa_\mu
\left(\pa_{\rho_1}\dots\pa_{\rho_n}\phi_2\right)
-\pa_\mu\phi_1
\left(\pa_{\rho_1}\dots\pa_{\rho_n}\phi_2\right)
}{cman7}
are conserved for solutions of the field equations.
One says they are conserved \emph{on shell}, since
for solutions the $4$-momentum lies on the mass shell.
The corresponding conserved charges
\eqnl{
Q_{\rho_1\dots\rho_n}=
\int d\mbx J_{0\rho_1\dots\rho_n},}{cman9}
where one integrates with $d\mbx$ over space,
are tensorial charges of higher rank. 
According to the \textsc{Coleman-Mandula}
theorem these conservation laws can not be 
extended to the interacting case in $d>2$
dimensions \cite{Coleman,Witten}, since
all additional conserved charges must be \lor
scalars.

The theorem does not apply to spinorial charges,
though. Let us add a \maj spinor in $4$
dimensions to the above system and consider the \lagan
\eqnl{
\cL_0=\cL_0'+\ft{i}{2}\psib\fdi\psi}{cman11}
The equations of motion are \refs{cman3}
supplemented by the free \dirac equation $\fdi\psi=0$.
Now there is an infinite number of
conserved currents with \emph{spinor indexes}, e.g.
\eqngrl{
S_{\mu\al}&=&\pa_\rho(\phi_1-i\phi_2)(\gam^\rho\gam_\mu\psi)_\al}
{S_{\mu\nu\al}&=&\pa_\rho(\phi_1-i\phi_2)(\gam^\rho\gam_\mu\pa_\nu
\psi)_\al.}{cman13}
For example, the first current is conserved since in
\eqnn{
\pa_\mu S^\mu_{\;\al}=\pa_\mu\pa_\rho(\phi_1-i\phi_2)
(\gam^\rho\gam^\mu\psi)_\al+\pa_\rho(\phi_1-i\phi_2)(\gam^\rho
\fdi\psi)_\al.}
the first term is proportional to $\Box(\phi_1-i\phi_2)$
and the second term to $\fdi\psi$.
Adding the interaction
\eqnl{
\cL_1=
-g\psib\left(\phi_1-i\gam_5\phi_2\right)\psi-\h
g^2\left(\phi_1^2+\phi_2^2\right)^2}{cman15}
to the \lagan $\cL_0$ of the free model, $S_{\mu\al}$
(with corrections proportional to $g$) remains
conserved. But a current with more indexes
can never be deformed to a conserved current of
the interacting theory.

After these preliminaries we formulate the theorem
more precisely. The \hil space of scattering theory,
$\cH$ is the infinite sum of $n-$particle subspaces
\eqnl{
\cH=\cH^{(1)}\oplus \cH^{(2)} \oplus\cH^{(3)}\oplus \dots,}{cman17}
where $\cH^{(n)}$ is the (properly symmetrized) subspace
of the tensor product of one-particle \hil spaces.
The scattering matrix is then a unitary operator
on $\cH$ describing all possible scattering processes
of the underlying theory.\index{S-matrix}\index{scattering matrix}
It is usually written as
\eqnl{
S=\id-i(2\pi)^4\delta^4(P-P')T,}{cman19}
\\[1ex]
\begin{minipage}[t]{6.6cm}
\psset{unit=1mm,linewidth=0.6pt,dotsize=1mm}
\hskip3mm\label{figure4}
\begin{pspicture}(0,10)(50,32)
\psellipse[fillstyle=solid,fillcolor=lightgray](25,20)(4,6)
\psline{->}(0,10)(20,16)
\psline{->}(0,20)(20,20)
\psline{->}(0,30)(20,24)
\psline{->}(30,20)(50,20)
\psline{->}(30,16)(50,10)
\psline{->}(30,24)(50,30)
\small
\rput(3,13){$p_1$}
\rput(3,23){$p_2$}
\rput(3,33){$p_3$}
\rput(47,14){$p_1'$}
\rput(47,23){$p_2'$}
\rput(47,33){$p_3'$}
\end{pspicture}
\end{minipage}
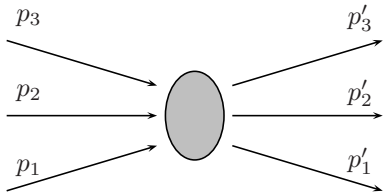
\begin{minipage}[b]{7.2cm}
where the delta distribution ensures the conservation
of the total energy and momentum during the
scattering process. In \refs{cman19} $P=\sum p$ denotes the
sum of the ingoing $4$-momenta and $P'=\sum p'$ the sum
of the outgoing momenta.
\end{minipage}
\\[4ex]
A unitary operator $U$ on $\cH$ is a
\emph{symmetry transformation}\index{symmetry transformation}
of the $S$-matrix if
\begin{itemize}
\renewcommand{\itemsep}{0pt}
\item
it maps $1$-particle states into $1$-particle states;
\item 
it acts on many-particle states as if they were tensor products
of one-particle states;
\item
$U$ commutes with $S$.
\end{itemize}
\emph{Internal symmetries} of the scattering matrix
are symmetries which do not act on space-time
coordinates, prominent examples being
gauge symmetries. The question arises whether
the \poin group $\cP$ can be combined in a nontrivial
way with internal symmetries of the S-matrix.

In what follows we restrict ourselves to theories
for which the scattering states are in positive
mass representations of $\cP$. Such
representations are characterized by the mass $M$
and the spin. We shall further assume that
for any finite $M$ there is only a finite
number of particle types with masses less than $M$. 
\begin{Lem}[The Coleman-Mandula-Theorem \cite{Coleman}]
Let a \lie group $G$ be a symmetry
group of the S-matrix which contains the \poin group
and which connects a finite number of particles in
a multiplet. Assume furthermore that
\begin{itemize}
\item Elastic-scattering amplitudes are analytic functions of
center-of-mass energy squared $s$ and invariant momentum transfer
squared $t$, in some neighborhood of the physical region, except
at normal thresholds.
\item For $\vert p,q\rangle$ one
has $S\vert p,q\rangle\neq \vert p,q\rangle$ for
almost all $s$.
\end{itemize}
Then G is locally isomorphic to the direct product of an
internal symmetry group and the \poin group $\cP$.
\end{Lem}

The theorem of \textsc{Coleman} and \textsc{Mandula}
implies that the most general symmetry algebra
of the S-matrix contains the $4$-momentum
$P_\mu$, the \lor generators $J_{\mu\nu}$
and a finite number of \lor scalars $B_k$,
\eqnl{
[P_\mu,B_k]=0\quad,\quad [J_{\mu\nu},B_k]=0,}{cman21}
where the $B_k$ constitute a \textsc{Lie}-algebra
with structure constants $c_{k\ell}^{\;\;\;m}$:
\eqnl{
[B_k,B_\ell]=ic_{k\ell}^{\;\;\;m}B_m.}{cman23}
It follows that the \textsc{Casimir} operators of
the \poin algebra, $P^2$ and $W^2$,
where
\eqnl{
W_\mu =\ft12 \eps_{\mu\nu\rho\sigma}P^\nu J^{\rho\sigma},\qquad
\eps_{0123}=1.}{cman25}
is the \textsc{Pauli-Ljubanski} polarization vector,
commute with \emph{all} generators of $\cG$, and in
particular with the generators of the internal symmetry
group,
\eqnl{
[B_\ell,P^2]=0\mtxt{and}[B_\ell,W^2]=0.}{cman27}
The first equation says that all members of an irreducible
multiplet of the internal symmetry group have the same
mass. This is known as \textsc{O'Raifeartaigh's}
\emph{theorem} \cite{loch1}.
\index{O'Raifeartaigh theorem}
The second equation says that they must also have the same spin.
For \emph{massless states} with discrete helicities we have
\eqnl{
W_\mu=\lam P_\mu,\qquad \lam\in \left\{0,\ft12,1,\dots\right\}}{cman29}
and no generator $B_\ell$ can change the helicity since 
$[B_\ell,\lam]=0$.

To see the arguments leading to the \textsc{Coleman-Mandula}
theorem consider a forbidden tensorial charge
$\cQ_{\mu\nu}$ which for simplicity we shall
assume to be traceless, $\cQ^\mu_\mu=0$. Assume a
scalar particle of mass $m$, carrying the charge
$\cQ_{\mu\nu}$ appears in the theory and let
$\vert p\rangle$ be a corresponding one-particle state.
Then
\eqnl{
\la p\vert\cQ_{\mu\nu}\vert p\rangle=\left(p_\mu p_\nu-
\frac{p^2}{d}\eta_{\mu\nu}
\right)C,\quad C\neq 0.}{cman33}
We consider a $2\to 2$ scattering process. The incoming particles
with momenta $p_1,p_2$ scatter and then go out with final momenta
$p'_1$ and $p'_2$. The conservation law of $\cQ$ applied
between asymptotic incoming and outgoing states requires
\eqnl{
C\left(p_{1\mu}p_{1\nu}+p_{2\mu}p_{2\nu}-{1\ov d}\eta_{\mu\nu}(m^2+m^2)\right)
=C\left(p_{1\mu},p_{2\mu}\to p'_{1\mu},p'_{2\mu}\right).}{cman35}  
If $C\neq 0$, these equations imply that the scattering
must proceed either in the forward or backward direction
whereas in all other directions there is no scattering.
This conflicts the analyticity properties of scattering 
amplitudes in more then $2$ dimensions.
No interacting theory can carry the charge $\cQ_{\mu\nu}$.
Similar arguments can be used for amplitudes of
non-identical particles to prove this 'no-go' theorem.

The \textsc{Coleman-Mandula} theorem shows the impossibility of
nontrivial symmetries that connect particles of different
spins, if all the particles have integer spin or if
all the particles have half-odd-integer spins.
\section{Noether theorem}
For \lagan field theories the field equations
are the \textsc{Euler-Lagrange} equation of a \poin
invariant action integral $S$. In a local field 
theory the action is the space-time integral of 
a local \lagan density $\cL(x)$,
\index{action}
\index{Lagrangean}
\eqnl{
S=\int d^dx \;\cL(x)\equiv\int dt d\mbx \cL(t,\mbx),}{symm1}
where $\cL$ depends on the fields and their derivatives.
Here we consider theories for which $\cL$ does not depend
on second or higher derivatives, $\cL=\cL(\phi,\pa_\mu\phi)$.
The dimension of spacetime is left open. The volume 
element of space is denoted by $d\mbx$.

The action is assumed to be invariant under
\poin \emph{transformations},
\eqnl{
\tilde x=\Lambda x+a,\qquad \Lambda\in L_+^{\uparrow},\;
a\in \R^d,}{symm3}
and the fields transform covariantly under the \poin
transformations:
\eqnl{
\phi'(x)=S(A)\phi
\left(\Lambda^{-1}(A)x+a\right)\sim \phi(x)+\delta_\xi\phi(x),}{symm5}
where $A$ is from the the universal covering group
(the \emph{spin group})\index{spin group}
of the restricted \textsc{Lorentz} group.
Here $A\to S(A)$ a finite-dimensional
representations of the  spin group and $\Lambda(A)$
the \lor transformation belonging to $A$. For simplicity,
in the following the 'label' $A$ in $\Lambda(A)$ will
not always be spelled out. 

In addition the action may be invariant under
\emph{global gauge transformations}\index{gauge transformation!global}
\eqnl{
\phi'(x)=U\phi(x)\sim \phi+
\delta_\xi\phi.}{symm7}
They are called global since the transformation matrix
$U$ it the same for all space-time points.
\subsection{Noether theorem for internal symmetries}
\index{Noether theorem!for internal symmetries}
According to the first theorem of \textsc{Emmy Noether},
to each parameter of the symmetry
group there corresponds a conserved
current. The global gauge transformations \refs{symm7}
leave the \lagan density invariant so that
\eqnl{
0=\delta_\xi \cL={\delta\cL\ov\delta(\pamu\phi)}\;
\pamu\left(\delta_\xi \phi\right) 
+{\delta\cL\ov\delta \phi}\,\delta_\xi\phi=
\pamu\left({\delta\cL\ov\delta(\pamu\phi)}\delta_\xi\phi\right),}
{noeth1}
where we used the \textsc{Euler-Lagrange} equation 
(field equation, equation of motion)
\index{Euler-Lagrange equation}
\eqnl{
\pamu\left({\delta\cL\ov \delta(\pamu\phi)}\right)-
{\delta\cL\ov\delta\phi}=0}{noeth2}
in the last step. Thus the conserved \noe 
\emph{current} for an internal symmetry takes the form
\index{Noether current!for internal symmetry}
\eqnl{
J_\xi^\mu={\delta\cL\ov \delta(\pamu\phi)}\delta_\xi\phi,\qquad
\pamu J_\xi^\mu=0.}{noeth3}
Integrating the last equation over the 
space-time region $[t_0,t]\times \R^{d-1}$
and converting the volume- into a surface integral, 
shows that the \noe \emph{charge}\index{Noether charge}
\eqnl{
Q_\xi=\int_{x^0} d\mbx J^0_\xi=
\int_{x^0} d\mbx\,\pi(x)\,\delta_\xi\phi(x),}{noeth5}
is time-independent. We introduced
the momentum density conjugate to $\phi$,
\eqnl{
\pi(x)={\pa\cL\ov\pa \dot\phi(x)}.}{noeth7}
To every internal symmetry there is one
conserved \noe charge. The dimension of the 
symmetry group equals the number of independent
vector fields $\xi$ in \refs{noeth5} and hence
equals the number of independent \noe charges.

The fundamental \textsc{Poisson} bracket between 
field and momentum density is
\eqnl{
\{\phi(x),\pi(y)\}_{x^0=y^0}=\delta(\mbx-\mby),}{noeth9}
and can be used to calculate the \textsc{Poisson} brackets
between the conserved charges and the field,
\eqnl{
\{\phi(x),Q_\xi\}=\int d\mby\,\{\phi(x),\pi(y)\,\delta_\xi\phi(y)\}
=\delta_\xi\phi(x),}{noeth11}
where we assumed that $\delta_\xi\phi$ contains
no time-derivatives of the field.
We used that $Q_\xi$ is conserved and set
$y^0=x^0$ when calculating  \refs{noeth11}. For
the quantized field the corresponding result reads
\eqnl{
[iQ_\xi,\phi_(x)]=\delta_\xi\phi(x).}{noeth12}
\noe charges generates
the symmetries from which they have been derived!
\subsection{Noether theorem for space-time symmetries}
\index{Noether theorem!for spacetime symmetries}
Under space-time translations a field changes
into $\phi(x+a)\sim \phi(x)+\delta_a\phi(x)$
and the infinitesimal variation is a total divergence,
\eqnl{
\delta_a\phi=\pa_\mu V^\mu_a\mtxt{with} V_a^\mu=a^\mu\phi(x).}{noeth15}
Under \lor transformations a general field transforms into
\eqnl{
S(A)\,\phi\left(\Lambda^{-1}(A)x\right)=
e^{\frac{i}{2}\,\om^{\mu\nu}S_{\mu\nu}}
\phi\left(e^{-\om}x\right)\sim\phi(x)+\delta_\om\phi(x)}{noeth17}
where the $S_{\mu\nu}$ form a representation of the
\lor algebra \refs{lotr19a} and
\eqnl{ 
\delta_\om\phi=\ft{i}{2}\om^{\mu\nu}
\left(L_{\mu\nu}+S_{\mu\nu}\right)\phi,\qquad
L_{\mu\nu}=\ft{1}{i}(x_\mu\pa_\nu-x_\nu\pa_\mu).
}{noeth19}
For a scalar field $S_{\mu\nu}$ is absent
and the variation becomes a total 
divergence\footnote{Exercise: prove this}
\eqnl{
\delta_\om\phi=\pa_\mu V^\mu_\om,\qquad
V_\om^\mu=-\om^{\mu\rho}x_\rho \phi(x)\quad (\phi\hbox{ scalar}).}
{noeth21}
The \lagan density of a \poin invariant theory
is a scalar and its variation under
\emph{all} small spacetime symmetries is a
total divergence, $\delta \cL=\pa_\mu V_\xi^\mu$,
where $\xi$ characterizes the type of transformation.
The action stays invariant and this is sufficient
for the field equations to be covariant.

Since the \lagan transforms into a total
divergence, the zero on the left in \refs{noeth1}
is replaced by $\pa_\mu V^\mu_\xi$ and
we obtain the conserved \noeth current
\eqnl{
J^\mu_\xi={\delta\cL\ov\delta(\pamu\phi)}
\delta_\xi \phi-V^\mu_\xi.}{noeth23}
The corresponding conserved charge takes the form
\eqnl{
Q_\xi=\int d\mbx\,\left(\pi(x)\,
\delta_\xi\phi(x)-V^0_\xi(x)\right)}{noeth25}
Now we discuss the currents for translations
and \lor transformations in turn.
\paragraph{Translations:}
The \noeth current belonging to the translations
defines the \emph{canonical energy-momentum} tensor
\index{Noether current!for translations}
\index{energy-momentum tensor}
\eqnl{
J^\mu_a=a^\nu T^\mu_{\;\nu},\qquad T_{\mu\nu}=
{\delta\cL\ov\delta(\pa^\mu\phi)}\,\pa_\nu\phi-
\eta_{\mu\nu}\cL.}{noeth27}
The conserved charges are the total momentum of the field,
\eqnl{
P^\mu=\int_{x^0}d\mbx\,T^{0\mu},\qquad
\dot P^\mu=0.}{noeth29}
Its components, the \emph{energy} and \emph{momentum},
have the explicit forms
\eqnl{
P^0\equiv H=\int_{x^0}d\mbx \,\big(\pi\dot\phi-\cL\big)
\quad,\quad
P^i=\int_{x^0} d\mbx\,\pi\,\pa^i\phi,}{noeth31}
and they generate infinitesimal spacetime translations,
\eqnl{
\{\phi,P_\mu\}=\pa_\mu\phi\mtxt{or after quantization}
[iP_\mu,\phi]=\pa_\mu\phi.}{noeth32}
\paragraph{Lorentz transformations:}
Under infinitesimal \lor transformations
\eqnl{
\delta_\om\cL=
\pamu V_\om^\mu\mtxt{with}V_\om^\mu=-\om^{\mu\rho}x_\rho\cL.}{noeth33}
Inserting $\delta_\xi\phi$ from \refs{noeth19}
in the general formula \refs{noeth23}
and subtracting $V_\om^\mu$ yields the following \noe
current for \lor transformations,
\index{Noether current!for Lorentz transformations}
\eqnl{
J_\om^\mu=\om_{\rho\sigma}M^{\mu\rho\sigma},\quad
M^{\mu\rho\sigma}=
\ft12 x^\rho T^{\mu\sigma}-\ft12 x^\sigma T^{\mu\rho}
+\ft{i}{2}{\delta\cL\ov\delta(\pamu\phi)}\,S^{\rho\sigma}\phi.}{noeth35}
The corresponding conserved \noeth charges read
\eqnl{
J^{\rho\sigma}=-J^{\sigma\rho}=-2i
\int_{x^0}d\mbx\,M^{0\rho\sigma},}{noeth37} 
and they generate infinitesimal \lor transformations,
\eqnl{
\{\phi,J^{\mu\nu}\}=\left(L_{\mu\nu}+S_{\mu\nu}\right)
\phi\mtxt{or}
[iJ^{\mu\nu},\phi]=\left(L_{\mu\nu}+S_{\mu\nu}\right)\phi}
{noeth39}
In passing we note, that for theories with
non-scalar fields things may become tricky. The 
canonical energy-momentum tensor is generically 
not symmetric and must be improved.
It is possible to correct the non-symmetric $T_{\mu\nu}$
through the \textsc{Belinfante} symmetrization 
procedure \cite{belinfante}.\index{Belinfante symmetrization}

For bosonic fields the most efficient way to do 
this is to couple the fields to gravity and vary 
the resulting \lagan with respect to the 
metric\footnote{One may even allow for a
non-minimal coupling to gravity to
obtain an improved $T_{\mu\nu}$.}. 
However, when coupling fermions to gravity one needs
a vielbein. When one varies the action with
respect to the vielbein one again gets a conserved
but not necessarily symmetric $T_{\mu\nu}$ which
needs further improvement. 
\section{Spinors}\label{sect:spinors}
\index{spinors}
In this section we study the transformation property of spinors
under '\textsc{Lorentz}'-transformations and the
properties of \textsc{Majorana-, Weyl-} and
\dirac spinors.
\subsection{Clifford algebras}
\index{$\gamma$-matrices!in $d$ dimensions}
\index{Clifford algebra}
The \cliff algebra is the free algebra generated
by the $d$ elements $\gam_0,\dots,\gam_{d-1}$, 
modulo the quadratic relation
\eqnl{
\gam_\mu \gam_\nu + \gam_\nu \gam_\mu
\equiv\{\gam_\mu,\gam_\nu\}=
2\eta_{\mu\nu}\,.}{gam3}
For even $d$ there exists \emph{one} irreducible
representation of dimension $2^{d/2}$ and for odd
$d$ \emph{two} inequivalent irreducible
representations of dimension $2^{(d-1)/2}$. Thus
\eqnl{
\gammu\in \hbox{GL}\left(2^{[d/2]},\C\right),}{gam5}
where $[a]$ is the biggest integer less or equal to $a$.
\par
For applications the
following observations are relevant:
\begin{itemize}
\item[$\bullet$] In \emph{even dimensions} a complete
set of $2^{d/2}$-dimensional matrices is provided by
the antisymmetrized products 
of gamma-matrices,
\eqnl{
\gam_{\mu_1\ldots \mu_n}\equiv
\gam_{[\mu_1}\gam_{\mu_2}\ldots \gam_{\mu_n]}
}{gam7}
The antisymmetrized product of all $\gam$'s is
proportional to
\eqnl{
\gam_*=i^{1+d/2}\gam_0\ldots \gam_{d-1}\mtxt{with}
\gam_* \gam_* =1.}{gam9}
In addition, we have
\eqnl{
\gam_{\mu_1\ldots \mu_n}=i^{1+d/2}\frac{1}{(d-n)!}
\eps_{\mu_1\ldots \mu_d}\,
\gam_*\gam^{\mu_d\ldots \mu_{n+1}},\quad
\eps_{01\dots d-1}=1.}{gam11}
$\gam_*$ anti-commutes with all $\gam_\mu$ and thus 
can be viewed as $\gam_d$ in $d+1$ dimensions.
\item[$\bullet$] In \emph{odd dimensions} the product of
all $\gam$-matrices is a multiple of the identity. 
A basis for the $2^{[d/2]}$-dimensional matrices
is formed by antisymmetrized products with $n=0,1,\ldots ,
\h(d-1)$.
\end{itemize}
The formula for the expansion of an arbitrary
$\Delta=2^{[d/2]}$-dimensional matrix $M$ is
\eqnl{
M= {1\ov\Delta}\sum_{n=0}^{D}\frac{1}{n!}\,(-)^{n(n-1)/2}
\gam_{\mu_1\ldots \mu_n}\,\tr
\left( \gam^{\mu_1\ldots \mu_n}M\right)\,,}{gam13}
where $D=d$ in even dimensions and $D=\h(d-1)$
in odd dimensions.
\subsection{Spin transformations}\label{ltr}
\index{spin transformation}
To study the transformation property of spinors
under '\textsc{Lorentz}'-transformations we introduce
the $d(d-1)/2$ matrices
\eqnl{
\Sigma^{\mu\nu}=-\Sigma^{\nu\mu}=\ft{1}{2i}\gam^{\mu\nu}=
\ft{1}{4i}(\gammu\gam^\nu-\gam^\nu\gammu).}{spint5}
They  furnish a $2^{[d/2]}$-dimensional
\emph{representation} of the \lor algebra \refs{lotr19a}
and generate \emph{spin-transformations}\index{spin-representation},
\eqnl{
S=e^{\frac{i}{2}(\om,\Sigma)}=\id+\ft{i}{2}(\om,\Sigma)+\dots .}{spint7}
The mapping $S\to \Lambda(S)$ defined by 
\eqnl{
S^{-1}\gam^\rho S=\Lambda^\rho_{\;\sigma}\gam^\sigma\mtxt{with}
\Lambda=e^{\om}}{spint9}
defines a representation of the spin group as \lor 
transformations. The infinitesimal \lor transformations 
are  $(\delta^\mu_{\;\;\nu}+\om^\mu_{\;\;\nu})$ with
antisymmetric $\om_{\mu\nu}$. A \dirac 
spinor transforms with
$S$ in \refs{spint7} such that
\eqnl{
\delta_\om\psi(x)=\ft{i}{2}\om_{\mu\nu}\left(L^{\mu\nu}
+\Sigma^{\mu\nu}\right)\psi(x),}{spint15}
where the $\Sigma_{\mu\nu}$ generate spin rotations and
the $L_{\mu\nu}$ orbital transformations.
Both satisfy the commutation relations of the
\lor algebra and so does their sum.
To construct bilinear \emph{tensor fields}
we use that $\gam^0$ conjugates the $\gam$ and
$\Sigma$-matrices into their adjoints.
It follows that $\gam^0$ conjugates the adjoint of $S$
into the inverse,
\eqnl{
\gam^0 S^\dagger \gam^0=S^{-1},}
{spint25}
such that the \emph{conjugate spinor},
\index{Dirac conjugate}
$\psib\equiv \psi^\dagger \gam^0$
transforms with the inverse spin rotation,
\eqnl{
\psib(x)\longrightarrow
\psib\left(\Lambda^{-1}x\right) S^{-1}.}{spint29}
With the help of \refs{spint9} it is now easy to
prove that the bilinear objects\index{antisymmetric tensor fields}
\index{fermionic bilinears}
\eqnl{
A^{\mu_1\dots\mu_n}=
\psib \gam^{\mu_1\dots\mu_n}\psi}{spint31}
are \emph{antisymmetric tensor fields}. The transformation
of these objects follow from that of $\psi$ and $\psib$,
\eqnn{
A^{\mu_1\dots\mu_n}(x)
\longrightarrow
\Lambda^{\mu_1}_{\;\,\nu_1}\cdots \Lambda^{\mu_n}_{\;\,\nu_n}\,
A^{\nu_1\dots\nu_n}\left(\Lambda^{-1}x\right).}
In $4$ dimensions there are $5$ tensor fields
\eqnn{
\psib\psi\quad,\quad \psib\gams\psi\quad,\quad\psib\gammu\psi,\quad,\quad
\psib\gams\gammu\psi\quad,\quad \psib \gam^{\mu\nu}\psi,}
a scalar, pseudo-scalar, vector, pseudo-vector and antisymmetric
2-tensor field.
\subsection{Charge conjugation}
\index{charge conjugation}
Assume that there exists a \emph{charge
conjugation matrix} $\cC$ which fulfills
\index{charge conjugation matrix}
\eqnl{
\cC\gam_\mu^T \cC^{-1}=\eta\, \gam^\mu,\mtxt{with}\eta=\pm 1,}{chc5}
in which case we define the \emph{charge conjugated spinor}
\index{charge conjugated spinor}
\eqnl{
\psicc=\cC\psib^{\,T}=\cC\gam_0^T \psi^*.}{chc7}
Now we multiply the \dirac equation
\eqnl{
i\gam^\mu D_\mu(e)\psi-m\psi=0,\qquad
D_\mu(e)=\pa_\mu-ieA_\mu}{chc1}
with $\gam^0$, complex conjugate and
multiply with $\cC$ from the left and obtain
\eqnl{
i\gam^\mu D_\mu(-e)\psicc+\eta \bar m\psicc=0.}{chc9}
For a vanishing mass $\psicc$ fulfills 
the \dirac equation with reversed electric charge and 
this justifies its name charge conjugated 
spinor\footnote{In the presence of a real mass
we would want $\eta=-1$.}.
If there exists  a \maj representation with \emph{real} or
\emph{imaginary} $\gam$'s then we may choose in 
$\cC=\gam_0^T$ in \refs{chc5},
\index{Majorana representation}
\eqnl{
\gam_\mu^*=\eta\gam_\mu:\quad
\cC=\gam_0^T,\quad \psicc=\psi^*}{chc13}
A \emph{spinor} invariant under charge conjugations 
is called \maj spinor\index{Majorana spinor}.
In a \maj representation such spinors are real.

Now one can prove that there always exist a
\emph{symmetric} or \emph{antisymmetric} charge
conjugation matrix $\cC$ \cite{Scherk,KugoTown}.
In the following table we have summarized the
results. The symbols $S$ and $A$ denote symmetric
and antisymmetric matrices. For example, in
$3$ dimensions there exists an antisymmetric
solution with $\eta=-1$ but no solution with $\eta=1$.
Since the results are identical in
$d$ and $d+8n$ dimensions, it is sufficient to give
the results for $d=1,\dots,8$:
\eqnl{
\begin{array}{|c||c|c|c|c|c|c|c|c|}\hline
d&1&2&3&4&5&6&7&8\\ \hline\hline
\eta=+1&S&S&&A&A&A&&S\\ \hline
\eta=-1&&A&A&A&&S&S&S\\ \hline
\end{array}}{chc23}
\subsection{Irreducible spinors}
\index{spinor!irreducible}
We do not know whether the spinor representations
\refs{spint7} are \emph{irreducible}. In general
they are not and there are two types of projections
onto invariant subspaces that one can envisage:
\par 
$\bullet$ The first exists in even dimensions only, 
where one has a $\gam_*$ anti-commuting
with all $\gam$-matrices. $\gam_*$ can be used
to define \emph{left} and \emph{right}-handed (chiral) 
spinors,
\index{spinor!left-handed}
\index{spinor!right-handed}
\eqnl{
\psiL=\ft12\left(1-\gam_*\right) \psi\equiv \PL\psi \,,\qquad
\psiR=\ft12\left(1+\gam_*\right) \psi\equiv \PR\psi\,.}{irs1}
Since $\gams$ commutes with the
generators $\Sigma_{\mu\nu}$ a left(right)-handed
spinors is left(right)-handed in any inertial system.

$\bullet$ 
The second possible projection is a \maj
\emph{reality condition}.
\index{reality condition!for spinor} 
\eqnl{
\psi= \psicc=\cC\psib^T.}{irs7}
This is a \lor invariant condition since
$\psicc$ transforms the same way as $\psi$, as
follows from $S\cC S^T=\cC$. Since $\psi^{**}=\psi$
we must demand
\eqnl{
\big(\cC\gam_0^T\big)^*\big(\cC \gam_0^T\big)=
\cC^*\gam_0\cC\gam_0^T\stackrel{\refs{chc5}}{=}\eta\,\cC^*\cC
\stackrel{!}{=}\id\mtxt{or}
\cC^*=\eta\,\cC^{-1}=\eta\,\cC^\dagger.}{irs9}
Since $\cC$ is unitary this condition is equivalent to
$\cC^T=\eta\,\cC$. Comparing with the above table leads to 
the solutions given in the following table: 
\eqnl{
\begin{array}{|c||c|c|c|c|c|c|c|c|}\hline
d&1&2&3&4&5&6&7&8\\ \hline\hline
\eta=+1&S&S&&&&&&S\\ \hline
\eta=-1&&A&A&A&&&&\\ \hline
&&MW&&&&SM&&\\ \hline
\end{array}}{irs11}
In cases there exist no \maj spinor one may
still try do define \textit{symplectic} \maj \emph {spinors}
\index{symplectic Majorana spinor} in theories
with extended supersymmetry. They obey
\eqnl{
\psi_i=\cC\gam_0^T\, \Omega_{ij}\psi_j^*\,,}{irs13}
where $\Omega$ is some antisymmetric matrix, with $\Omega\Omega ^*=-1$.
Symplectic \maj spinors exist in $6,14,\dots$
dimensions, as indicated in the above table.

Having two projections, to chiral and to \maj spinors,
one may ask whether one can define a
reality condition respecting the chiral projection. 
This is indeed possible in $2+8n$ dimensions, where
such \textsc{Majorana-Weyl} fermions exist.
\subsection{Fierz identities}
\index{Fierz identities}
We take two spinors $\psi$ and $\chi$ whose components
anticommute and choose
\eqnl{
M=\psi\bar\chi,\mtxt{such that}
\tr\big(\gam^{\mu_1\ldots\mu_n}M\big)=
-\bar\chi \gam^{\mu_1\ldots\mu_n}\psi,}{fierz1}
where the minus sign originates from the anticommuting
nature of the spinor components.
The expansion \refs{gam13} becomes the general
\textsc{Fierz}-identity\index{Fierz identity!general}
\eqnl{
\psi\chib=-{1\ov\Delta}\sum_n \frac{1}{n!}\,(-)^{n(n-1)/2}
\,\gam_{\mu_1\ldots\mu_n}\;\big(\bar\chi
\gam^{\mu_1\ldots\mu_n}\psi\big),}{fierz3}
which allows us to write the matrix $\psi\chib$
as linear combination the antisymmetrized products
of $\gam$-matrices. For example, in $4$ dimensions
the general identity reads\index{Fierz identity!in
$4$ dimensions}
\eqnl{
4\psi\bar\chi=-(\bar\chi\psi)-\gam_{\mu}(\bar\chi\gam^\mu\psi)
+\ft12 \gam_{\mu\nu}(\bar\chi\gam^{\mu\nu}\psi)
+\gamf\gam_\mu(\bar\chi\gamf\gam^\mu\psi)
-\gamf(\bar\chi\gamf\psi),}{fierz5}
where $\gamf=-i\gam_0\gam_1\gam_2\gam_3$.

The \textsc{Dirac}-conjugate of the charge
conjugated spinor $\psicc$ is
$\bar\psicc=\eta\psi^T\cC^{-1}$
and is used to compute the bilinears for
the charge conjugated fields,
\index{fermionic bilinears!for charge conjugated fields}
\eqnl{
\bar\psicc\gam_{\mu_1\dots\mu_n}\chicc
=-\eta^{1+n}(-1)^{n(n-1)/2}\,\chib\gam_{\mu_1\dots\mu_n}\psi
\stackrel{\rm Majorana}{=}\bar\psi\gam_{\mu_1\dots\mu_n}\chi.}{fierz7}
In $4$ dimensions with $\eta=-1$ we find
for \maj spinors
\eqnl{
\psib M\chi=\pm\chib M\psi,\quad
+:\;M\in\{\id,\gamf,\gamf\gam_\mu\},\quad
-:\;M\in\{\gam_\mu,\gam_{\mu\nu}\}}{fierz9}
such that $\psib\gam_\mu\psi=\psib\gam_{\mu\nu}\psi=0$.
All \textsc{Fierz} identities you find in
the literature can be derived from the general
identity \refs{fierz3} and the symmetry
relations \refs{fierz7}. For example, setting $\chi=\psi$
in \refs{fierz5} we obtain for \maj spinors the identity
\eqnl{
\left(\psi\psib+\gamf\psi\psib\gamf\right)\psi=
\psi(\psib\psi)+\gamf\psi (\psib\gamf\psi)=0.}{fierz11}
\subsection{Hermitian conjugation}
\index{hermite conjugate!of bilinears}
We \emph{define} the hermitian conjugate as
if the spinor-components are operators in a
\textsc{Hilbert}-space (which they are
in quantum field theory). For example 
\eqnl{
(\psib\chi)^\dagger\equiv \chi^\dagger \psib^\dagger.}{herm7}
One can show that for fermionic bilinears 
\eqnl{
(\psib M\chi)^\dagger=
-\eta\, \bar\psicc M_{\rm c}\chicc\mtxt{with}
M_{\rm c}=\cC\gam_0^T M^*\gam_0^T\cC^{-1}.}{herm9}
In a \maj representation with $\cC=\gam_0^T$ this
simplifies to
\eqnl{
(\psib M\chi)^\dagger=-\eta\,\bar\psicc M^*\chicc}{herm11}
and with $\gam_\mu^*=\eta\gam_\mu$ (see \refs{chc13})
we obtain for \maj spinors
\eqnl{
\left(\psib\gam_{\mu_1\dots\mu_n}\chi\right)^\dagger=
(-1)^n\psib \gam_{\mu_1\dots\mu_n}\chi.}{herm13}
This property is useful when one enumerates all
possible terms in a real action or in supersymmetry
transformations.
\subsection{Chiral spinors in 4 dimensions}
\label{chiral}
\index{chiral spinors}
\index{spinor!chiral}
In theories with extended supersymmetry
one often uses left- and right-handed spinors. Then
it is convenient to use the \emph{chiral representation}
\index{chiral representation}
\eqnl{
\gam_\mu=\pmatrix{0&\sigma_\mu\cr \tilde\sigma_\mu&0}\mtxt{with}
\sigma_\mu=(\sigma_0,-\tau_i)\mtxt{and}
\tilde\sigma_\mu=(\sigma_0,\tau_i),}{chirr7}
where $\sigma_0=\id_2$ and $\tau_1,\tau_2,\tau_3$ are the
\textsc{Pauli} matrices.
Since $\gamf=\hbox{diag}(-\sigma_0,\sigma_0)$ the chiral
projectors take the simple form
\eqnl{
\PL=\pmatrix{\sigma_0&0\cr 0&0}\mtxt{and}
\PR=\pmatrix{0&0\cr 0&\sigma_0},\quad}{chirr9}
such that a lefthanded spinor has upper
and a right-handed spinor lower components.
The infinitesimal spin-rotations are block-diagonal,
\eqnl{
\gam_{\mu\nu}=\pmatrix{\sigma_{\mu\nu}&0\cr 0&\tsigma_{\mu\nu}},
\quad \sigma_{\mu\nu}=\h(\sigma_\mu\tsigma_\nu-\sigma_\nu\tsigma_\mu),
\quad \tsigma_{\mu\nu}=-\sigma_{\mu\nu}^\dagger}{chirr11}
and so are the spin rotations generated by them,
\eqnl{
S=\pmatrix{A&0\cr 0&A^{\dagger -1}},\quad
A=e^{\frac{1}{4}\om^{\mu\nu}\sigma_{\mu\nu}}\in SL(2,\C).}{chirr13}
The left- and right-handed parts of a \dirac spinor
\eqnl{
\psi=\pmatrix{\varphid\cr \chiald}}{chirr15}
transform with the two inequivalent irreducible
representations $A$ and $A^{\dagger -1}$ of $SL(2,\C)$,
\eqnl{
\varphid\longrightarrow A_\al^{\;\,\beta}\varphi_\beta\quad,\quad
\chiald\longrightarrow (A^{\dagger -1})^{\dot\al}_{\;\,\dot\beta}\,\chibed
}{chirr17}
Next we define the $\veps$-tensors 
\eqnl{
(\veps_{\al\beta})=(\veps_{\dal\dbe})
=-(\veps^{\al\beta})=-(\veps^{\dal\dbe})=
\pmatrix{0&-1\cr 1&0}\equiv \veps,}
{chirr19}
which obey the relations,
\eqnl{
\epsd\veps^{\beta\gamma}=\delta _{\al}^{\gam}\quad,\quad
\epsd\veps^{\delta\gamma}=\delta_{\al}^{\gam}\delta_{\beta}^{\delta} -
\delta_{\al}^{\delta}\delta_{\beta}^{\gam}.}{chirr21}
Because $A^T\veps A=\veps$ for any matrix $A$ 
with determinant one, the bilinears 
\eqnl{
\varphi\chi=\varphiu\chial\mtxt{and}
\varphib\chib=\varphib_{\dal}\chiald}{chirr23}
are \lor invariant,
where the raising and lowering of indexes are done with $\veps$,
\eqnl{
\varphiu=\veps^{\al\beta}\,\varphi_\beta,\quad 
\varphid=\veps_{\al\beta}\varphi^\beta\mtxt{and}
\varphibu=\veps^{\dal\dbe}\varphib_\dbe,\quad 
\varphib_\dal=\eps_{\dal\dbe}\bar\varphi^{\dot\beta}.}{chirr25}
The condition $A^T\veps A=\veps$ translates into
\eqnl{
\big(A^{T-1}\big)^\al_{\;\;\beta}=\veps^{\al\rho}A_\rho^{\;\;\sigma}
\veps_{\sigma\beta}\mtxt{and}
\big(A^{\dagger -1}\big)^{\dal}_{\;\;\dbe}=\veps^{\dal\dot\rho}
\bar A_{\dot\rho}^{\;\;\dot\sigma}\veps_{\dot\sigma\dbe}}{chirr27}
which means, that $A$ can be conjugated into $A^{T-1}$ 
and $\bar A$ into $A^{\dagger -1}$. Using these
formulas one can show that the spin-transformation in 
\refs{chirr17} are equivalent to 
\eqnl{
\varphi^\al\longrightarrow \big(A^{T-1}\big)^\al_{\;\;\beta}\varphi^\beta\quad,\quad
\chib_\dal\longrightarrow \bar A_\dal^{\;\;\dbe}\chib_\dbe.}{chirr29}
We conclude that the components $\varphi^\al$ transform as the complex
conjugate of the components $\chiald$ and the components $\chib_\dal$ 
as the complex conjugate of the components $\varphi_\al$. The index 
structure of the \textsc{Dirac}-conjugate spinor is
\eqnl{
\psib=(\chi^\al,\varphib_{\dal})\mtxt{such that}
\psib\psi=\chi\varphi+\varphib\chib.}{chirr31}
There is no mass term for a left- or for a right-handed
spinor. Since $\sigma_\mu$ maps right- into left-handed
spinors and $\tsigma_\mu$ does the opposite,
they have the index structure
\eqnl{
(\sigma_\mu)_{\al\dbe}\mtxt{and}
(\tsigma_\mu)^{\dal \beta}\,.}{chirr33}
The generator $\sigma_{\mu\nu}$ and $\tsigma_{\mu\nu}$ 
preserve chirality  and have the index structure
\eqnl{
(\sigma_{\mu\nu})_\al^{\;\;\beta}\mtxt{and}
(\tsigma_{\mu\nu})^\dal_{\;\;\dbe}\,.}{chirr35}
Spinor components are \textsc{Grassmann}
variables such that
\eqngrl{
\varphi\chi =\varphiu\chi_\al=-\chial\varphiu =\chi^{\al}\varphid=\chi\varphi}
{\varphib\chib=\varphib_{\dal}\chiald=-\chiald\varphib_{\dal} =
\chib_{\dal}\varphiald = \chib\varphib .}{chirr37}
The vector current, on the other hand, can be written as
\eqnl{
\psib\gammu\psi=\varphib_\dal\tsigma^{\mu\dal\beta}\varphid+
\chi^\al\sigma^\mu_{\al\dbe}\chib^{\dbe}
=\varphib\,\tsigma^\mu\varphi+\chi\sigma^\mu\chib,}{chirr39}
so that the 'kinetic term' for fermions take the form
\eqnl{
\psib\fdi\psi=
\varphib\,\tsigma^\mu\pa_\mu  \varphi+
\chi \sigma^\mu\pa_\mu\chib.}{chirr41}
In particular in $2$ and $4$ dimensions the 
\textsc{Fierz} identities take a simpler form 
when one uses chiral spinors. The explicit
formulas can be found in text books on
supersymmetry, for examples the ones cited
in the introduction.

\chapter{The Wess-Zumino Model}\label{chapWZ}
\index{Wess-Zumino model!free model}
In a supersymmetric model we expect
an equal number of bosonic and fermionic
states of equal mass. For example, a \maj fermion 
has $2$ polarization states and could
be accompanied by two neutral scalar particles.
Here we study the most simple of these models in 
four spacetime dimensions. It has been constructed by 
\textsc{Wess} and \textsc{Zumino} \cite{WZ} when they extended 
the $2$-dimensional supersymmetric string-model of \textsc{Gervais} 
and \textsc{Sakita}\footnote{which was based on earlier work of \textsc{Ramond} 
\cite{Ramond} and \textsc{Neveu} and \textsc{Schwartz} \cite{Neveu}.} 
\cite{Sakita} to four dimensions. The model contains a super-multiplet with
\begin{itemize}
\itemsep=1pt
\item a single \maj field $\psi$
\item  a pair of real scalar and pseudo-scalar
bosonic fields $\phi_1$ and $\phi_2$
\end{itemize}
In the off-shell formulations it also contains a pair of 
real scalar and pseudoscalar bosonic auxiliary fields 
$\cF_1$ and $\cF_2$. In passing we note that there 
was some earlier work on supersymmetric field theories in
four dimensions. A version of supersymmetric quantum
electrodynamics (QED) was found by
\textsc{Golfand} and \textsc{Likhtman} in 1970 and
published in 1971 \cite{likhtman}. This massive super-QED contains
a massive photon and photino, a charged \dirac
spinor and two charged scalars. Subsequently \textsc{Akulov}
and \textsc{Volkov} tried to associate the massless
fermion - appearing due to spontaneous supersymmetry breaking -
with the neutrino \cite{volkov}.

\section{The free massless Wess-Zumino model}
Without interaction the \lagan of the massless model
takes the form \refs{cman11}
\eqnl{
\cL_0=\ft12 (\pamu \phi_1)^2
+\ft12 (\pamu \phi_2)^2+\ft{i}{2}\psib\fdi\psi.}{wzm3}
Besides the well-known space-time symmetries the action 
$S_0=\int d^4x\,\cL_0$ is invariant under
the following supersymmetry transformations
\begin{eqnarray}
\delta_\veps \phi_1=\vepsb \psi &,&\delta_\veps \phi_2=i\vepsb\gamf\psi
\label{wzm5}\\
\delta_\veps\psi = -i\fdi \phi\veps &,&\phi=\phi_1+i\gamf\phi_2
\label{wzm7}
\end{eqnarray}
where $\veps$ is an arbitrary constant anticommuting \maj 
parameter with dimensions $[\veps]=L^{1/2}$. 
Clearly, these transformations map bosons into fermions and
vice versa. They are very restricted by the 
following requirements:
\begin{itemize}
\itemsep=1pt
\item
\lor covariance,
\item
dimensions of fields:
$[\delta_\veps\phi_i]=L^{-1},\quad [\delta_\veps \psi]=L^{-3/2}$,
\item
hermiticity:
$\psi,\phi_i$ real.
\end{itemize}
For example, with $\vepsb\psi$ the variation 
$\delta_\veps\phi_1$ must be a real scalar field 
and  with $i\vepsb\gamf\psi$ the variation 
$\delta_\veps\phi_2$ must be a real pseudo-scalar 
field. We conclude that $\phi_1$ is a scalar
and $\phi_2$ a pseudo-scalar field. Using
$\gam^0\gam_\mu\gam^0=\gam_\mu^\dagger$  the variation of the 
$\psib$ reads
\eqnl{
\delta_\veps\psib =(\delta_\veps\psi)^\dagger\gam^0=
i\vepsb \pamu \phi\gam^\mu.}{wzm9}
Not unexpected the \lagan density is invariant only
up to a total divergence. For newcomers to supersymmetry
the proof is enlightening and since it is simple
for the model \refs{wzm3} we shall give it. Clearly, 
\eqngrl{
\h\delta_\veps\left(\sum\pamu \phi_i\pa^\mu \phi_i\right)
&=&\vepsb \pa^\mu\phi \pamu\psi}
{\ft{i}{2}\delta_\veps(\psib\fdi\psi)&=&-\h\vepsb \fdi \gam^\mu\phi
\pamu\psi+\h\psib \Box\phi\,\veps,}{wzm11}
where we may interchange $\psi$ and $\veps$
in the last term since $\psib\veps=\vepsb\psi$
and $\psib\gamf\veps=\vepsb\gamf\psi$
hold true. The sum of the two terms is a divergence
\eqnl{
\delta_\veps \cL_0=\pamu (\vepsb V_0^\mu),\quad
V_0^\mu=\h\gammu\gam^\nu\pa_\nu\phi\,\psi}{wzm13}
which already proves the invariance of $S_0$. 
To find the \noeth current we calculate
\eqnl{
{\pa\cL\ov\pa(\pamu\phi)}\delta_\veps\phi=\vepsb K^\mu,\qquad
K^\mu=\left(\h\gammu\gam^\nu+\gam^\nu\gammu\right)\pa_\nu\phi\,\psi.}
{wzm15}
and subtract  $V_0^\mu$ in \refs{wzm13}. This yields 
the \emph{conserved current}\index{Noether current!for free WZ model}
\eqnl{
\vepsb J^\mu=\vepsb\pa_\nu\phi\gam^\nu\gammu\psi
=-i\delta_\veps\psib\gammu\psi=i\psib\gammu\delta_\veps\psi}{wzm17}
and conserved \noeth charge $\vepsb\cQ=\int d\mbx \vepsb J^0$ with
\emph{spinorial supercharge}
\index{supercharge!for free WZ model}
\eqnl{
\cQ=\int_{x^0}d\mbx\left(\pi_1+i\gamf \pi_2
-\al^i\pa\phi\right)\psi,\quad
\al^i=\gam^0\gam^i,}{wzm19}
where $\pi_i$ is the momentum field conjugate to $\phi_i$
and $\phi=\phi_1+i\gamf\phi_2$.
The fundamental equal-time (anti)commutators of the
field operators $\{\phi_i(\mbx),\pi_i(\mbx),\psi(\mbx)\}$ 
are\index{commutators!of field operators}
\eqnl{
[\phi_i(x),\pi_j(y)]_{x^0=y^0}=i\delta_{ij}\delta(\mbx-\mby)\;\;,\;\;
\{\psi_\al(x),\psi_\beta(y)\}_{x^0=y^0}=
\delta_{\al\beta}\delta(\mbx-\mby),}{wzm21}
where the anti-commutators holds in a \maj representation.
In an arbitrary representation $\delta_{\al\beta}$ is
replaced by $-(\gam^0\cC)_{\al\beta}$.
Now its not difficult to prove that the supercharge generate
the supersymmetry,
\eqnl{
i[\vepsb \cQ,\phi_i]=\delta_\veps\phi_i\mtxt{and}
i[\vepsb \cQ,\psi]=\delta_\veps\psi,}{wzm25}
similarly as $P_\mu$ and $J_{\mu\nu}$ generate
translations and \lor transformations.
\subsection{Superalgebra}
The commutator of two successive symmetry transformations
must itself be a symmetry transformation.
This way we can identify the algebra of group generators. 
Let us see what the  commutator
of two susy transformations looks like. For example,
\eqngrl{
[\delta_{\veps_1},\delta_{\veps_2}]\phi_1
&=&\delta_{\veps_1}(\vepsb_2\psi)-\delta_{\veps_2}(\vepsb_1\psi)=
-i\vepsb_2\fdi\phi\veps_1-
(1\leftrightarrow 2)}
{&=&-2i\vepsb_2\gammu \veps_1\pamu \phi_1.}{salg1}
Similarly one finds for the pseudo-scalar field
\eqnn{
[\delta_{\veps_1},\delta_{\veps_2}]\phi_2
=-2i\vepsb_2\gammu\veps_1\pamu \phi_2,}
Calculating the commutator on $\psi$ is more
involved. Upon using the \textsc{Fierz} rearrangement formula
\eqnl{
\veps_2\vepsb_1-\veps_1\vepsb_2=-\ft12 \gam_{\rho}(\vepsb_1\gam^\rho\veps_2)
+\gam_{\rho\sigma}(\vepsb_1\gam^{\rho\sigma}\veps_2)}{salg3}
one obtains
\eqnn{
[\delta_{\veps_1},\delta_{\veps_2}]\psi=
-2(\vepsb_2\gam^\rho\veps_1)\pa_\rho\psi+
i\gam_\rho(\vepsb_2\gam^\rho\veps_1)
\fdi\psi.}
Only if we impose the \emph{field equation} for $\psi$
do we get the expected commutator,
\eqnl{
[\delta_{\veps_1},\delta_{\veps_2}]\psi=
-2i\big(\vepsb_2\gammu\veps_1\big)\pamu\psi.}{salg5}
On \emph{all fields} the commutator of two
transformations yield a translation such that
\eqnl{
[\delta_{\veps_1},\delta_{\veps_2}]=
-2i\big(\vepsb_2\gammu\veps_1\big)\pamu.}{salg7}
Since $\vepsb\cQ$ generates the supersymmetry transformations
$\delta_\veps$, the result \refs{salg7} means that the 
commutator of two $\vepsb\cQ$ generates infinitesimal
translations.
Using
\eqnl{
[\vepsb_1 \cQ,\vepsb_2 \cQ]=[\bar\cQ\,\veps_1,\vepsb_2\cQ]
= \veps_{1\beta}\vepsb_2^\al\{\cQ_\al,\bar \cQ^\beta\}}{salg9}
and that the translations are generated by
$P_\mu$, see \refs{noeth32},  we obtain
\eqnl{
\bullet\;\;\{\cQ_\alpha,\bar\cQ^\beta\}=2(\gam^\mu)_\al^{\;\beta}P_\mu}{salg11}
which may be rewritten as
\eqnl{
\{\cQ_\al,\cQ_\beta\}=-2(\gammu\cC)_{\al\beta}P_\mu.}{salg13}
On the other hand the supersymmetry
commutes with the translations,
\eqnl{
[\delta_\veps,\delta_a]=0,\mtxt{where} \delta_a=a^\mu\pamu,}{salg15}
and we conclude that the supercharge commutes with
the momentum,
\eqnl{\bullet\;\;
[\cQ_\al,P_\mu]=0.}{salg17}
Let us finally calculate the commutator of \lor
and susy transformations. For the scalar field $\phi_1$
we find
\eqnl{
[\delta_\veps,\delta_\om]\phi_1=\ft{i}{2}\delta_\veps(\om,L)\phi_1
-\delta_\om \vepsb\psi
\stackrel{\refs{spint15}}{=}
-\ft{i}{2} \om_{\mu\nu}\vepsb\, \Sigma^{\mu\nu}\psi,}{salg19}
and similarly for the pseudo-scalar and spinor.
Finally, using
\eqnn{
[\delta_\veps,\delta_\om]\phi_1=
-\ft{i}{2}\left[\vepsb\cQ,[\om^{\mu\nu}J_{\mu\nu},\phi_1]\right]
\mtxt{and}i[\vepsb\cQ,\phi_1]=\vepsb\psi}
we conclude, that on $\phi_1$ we have
$[\cQ_\al,J_{\mu\nu}]=-(\Sigma_{\mu\nu}\cQ)_\al.$
The same holds true for the other fields, such that
\eqnl{
\bullet\;\;[J_{\mu\nu},\cQ_\al]=(\Sigma_{\mu\nu}\cQ)_\al.}{salg21}
As expected, the spinorial supercharge 
transforms as spin-$\ft12$ object.
\section{The off-shell formulation with interaction}
\index{Wess-Zumino model!off shell formulation}
\index{Wess-Zumino model!interacting model}
We continue and allow for mass terms and/or
interactions between the fermions and scalars.
If we proceed as we did for the free massless
model, then we would find that the supersymmetry
transformations (\ref{wzm5},\ref{wzm7}) were deformed
and would
\begin{itemize}
\itemsep=1pt
\item only close if $\psi$ fulfills the field equation (closes on-shell), 
\item become non-linear in the fields,
\item depend on the masses and coupling constants of the model.
\end{itemize}
However, if we introduce \emph{auxiliary} \lagan
multiplier fields, then the transformations become
linear, close off-shell and are independent
of the details of the model. Off-shell a \maj
spinor contains $4$ real fields, whereas $\phi$
only contains $2$. Thus we need \emph{two additional scalars}, 
denoted by $\cF_1$ and $\cF_2$, for the
degrees of freedom to match off-shell. Similarly
as for $\phi_1,\phi_2$ we introduce the field
$\cF=\cF_1+i\gamf \cF_2$.

The idea is to deform the susy transformation
of the \textsc{Fermi} field in \refs{wzm7} 
\eqnl{
\delta_\veps\psi=-i\fdi\phi+\cF\veps}{wzi1}
such that the algebra closes off-shell,
\eqngrl{
[\delta_{\veps_1},\delta_{\veps_2}]&=&
-2(\vepsb_2\gam^\rho\veps_1)\pa_\rho\psi+
i\gam_\rho(\vepsb_2\gam^\rho\veps_1)\fdi\psi+
(\delta_{\veps_1}\cF)\veps_2-(\delta_{\veps_2}\cF)\veps_1}
{&\stackrel{!}{=}&-2(\vepsb_2\gam^\rho\veps_1)\pa_\rho\psi}{wzi3}
The transformation law \refs{wzi1} reveals that $\cF_1$ is
a scalar and $\cF_2$ a pseudoscalar and that both fields 
have dimensions $L^{-2}$. Hence their variations can only 
be proportional to the hermitian objects 
$i\vepsb\fdi\psi$ and $\vepsb\gamf\fdi\psi$,
respectively. Indeed, setting
\eqnl{
\delta_\veps \cF_1=-i\vepsb \fdi\psi\mtxt{and}
\delta_\veps \cF_2=\vepsb\gamf \fdi\psi}{wzi5}
we find with $M=\veps_1\vepsb_2-\vepsb_2\veps_1$ the result
\eqnn{
(\delta_{\veps_1}\cF)\veps_2-(\delta_{\veps_2}\cF)\veps_1=
i\left(M-\gamf M\gamf\right)\fdi \psi=-i\gam_\rho (\vepsb_2\gam^\rho\veps_1)
\fdi\psi,}
where in the last step we used a \textsc{Fierz} identity.
In a perturbatively renormalizable
local field theory the \lagan density must
not contain any derivatives of the multiplier fields 
and their elimination should
yield the on-shell density \refs{wzm3}. This way one
is lead to the following ansatz,
\eqnl{
\cL_0=\ft12 \pamu \phi_1\pa^\mu \phi_1+\ft12 \pamu \phi_2\pa^\mu \phi_2+
\ft{i}{2}\psib\fdi\psi+\ft12 \left(\cF_1^2+\cF_2^2\right)}{wzi7}
and as required, this hermitian density transforms into a divergence,
\eqnl{
\delta_\veps\cL_0=\pamu\big(\vepsb V^\mu_0\big),\qquad V_0^\mu=
\ft12\left(\gammu\gam^\nu\pa_\nu\phi-i\cF\gammu\right)\psi.}{wzi9}
Note that the field-equations for the auxiliary fields
are simply $\cF_1=\cF_2=0$. 

We can add a \emph{mass term}\index{mass term!in WZ-model}
\eqnl{
\cL_m=m\left(\cF_1 \phi_1+\cF_2 \phi_2-\ft{1}{2}\psib\psi\right)}{wzi11}
to $\cL_0$, as well as 
\emph{interaction terms},\index{interaction term!in WZ-model}
for example 
\eqnl{
\cL_g=g\left(\cF_1\left(\phi_1^2-\phi_2^2\right)+2\cF_2\phi_1\phi_2
-\psib \left(\phi_1-i\gamf \phi_2\right)\psi\right).}{wzi13}
Each term of the resulting \lagan density
$\cL=\cL_0+\cL_m+\cL_g$ transforms into
a divergence,
\eqngrl{
\delta_\veps\cL_m=\pa_\mu(\vepsb V^\mu_m),&&V^\mu_m=-im\phi\gammu\psi}
{\delta_\veps\cL_g=\pa_\mu(\vepsb V^\mu_g),&&V^\mu_g=-ig\phi^2\gam^\mu\psi.}
{wzi15}
This then gives rise to the following \noeth current
\index{Noether current!for interacting WZ model}
\eqnl{
\vepsb J^\mu=\vepsb\left(\gam^\nu\gam^\mu\pa_\nu\phi\psi
+im\phi\gam^\mu\psi
+ig\phi\phi\gammu\right)\psi
=-i\delta_\veps\psib\gammu\psi.}{wzi17}
The corresponding spinorial \emph{supercharge}
\index{supercharge!for interacting WZ model}
has the form
\eqnl{
\cQ=\int_{x^0}d\mbx\left(\pi
-\al^i\pa_i\phi+im\phi\gam^0+ig\phi\phi\gam^0
\right)\psi,}{wzi19}
where we have introduced $\pi=\pi_1+i\gamf \pi_2$.
It generates the \emph{on-shell} supersymmetry 
transformations, which are just the off-shell
transformation (\ref{wzm5},\ref{wzi1})
in which one eliminates the auxiliary fields through 
their equations of motion
\eqnl{
\cF=-m\phi-g\phi\phi.}{wzi21}
This yields the following 'on-shell' \lagan density
\eqngrl{
\cL&=&\ft12 (\pamu \mbphi,\pa^\mu \mbphi\,)
-\ft12 m^2(\mbphi,\mbphi\,)+\ft{i}{2}\psib\fdi\psi-\ft12 m\psib\psi}
{&&-mg\phi_1(\mbphi,\mbphi\,)-\ft12 g^2(\mbphi,\mbphi\,)^2
-g\psib(\phi_1-i\gamf \phi_2)\psi,}{wzi23}
where $\mbphi$ denotes the doublet $(\phi_1,\phi_2)^T$,
and the following on-shell transformations
\index{Wess-Zumino model!on-shell transformations}
\index{susy transformations!for WZ-model}
\eqnl{
\delta_\veps \phi_1=\vepsb \psi,\quad\delta_\veps \phi_2=i\vepsb\gamf\psi
\mtxt{and}
\delta_\veps\psi=-i\fdi\phi\veps-(m\phi+g\phi\phi)\veps.}{wzi25}
They are generated by the supercharge
\refs{wzi19}.

\chapter{Representations of supersymmetric algebras}
\index{supersymmetry algebra!representations}
In a supersymmetric field theory the fields fall
into supermultiplets which transform according 
to some representation of the superalgebra. 
In this chapter we consider representations which
may occur in perturbatively renormalizable field theories. 
We consider the superalgebra in $4$ dimensions in the 
\weyl basis and use the conventions introduced in section
\refs{chiral}. Since the charge conjugation
matrix in the chiral representation \refs{chirr7} is
\eqnl{
\cC=\pmatrix{\veps&0\cr 0&-\veps},\mtxt{such that}
\cC\gam_0^T=\pmatrix{0&\veps\cr -\veps&0},}{reps1}
the spinorial supercharge $\cQ$ is \maj if and only if
\eqnl{
\cQ=\pmatrix{\cQ_\al\cr \bar \cQ^\dal}=\cQ_{\rm c}
\Longleftrightarrow
\bar \cQ_\dal=\cQ_\al^\dagger,\quad \bar \cQ^\dal
=(\cQ^\al)^\dagger.}{reps3}
The nontrivial anticommutator \refs{salg11} takes the form
\eqnn{
\{\cQ_\al,\cQ^\beta\}=0\mtxt{and}
\{\cQ_\al,\bar\cQ_\dbe\}=2\sigma^\mu_{\al\dbe}P_\mu,}
and for convenience we recall the index structure
of the relevant matrices,
\eqnl{
(\sigma_\mu)_{\al\dal},\quad
(\tsigma_\mu)^{\dal \al},\quad
(\sigma_{\mu\nu})_\al^{\;\;\beta},\quad
(\tsigma_{\mu\nu})^\dal_{\;\;\dbe}.}{reps5}
Under \lor transformations the supercharges 
$\cQ_\al$ and $\bar \cQ^\dal$ transform 
with the chiral representations
of the spin group. The hermitian adjoint of a 
left(right)-handed operator is a linear combination 
of right(left)-handed operators.

In a classic paper  \textsc{Haag, Lopuszanki} and \textsc{Sohnius}
re-examined the result of \textsc{Coleman} and \textsc{Mandula}
by redefining the very notion of symmetry to encompass
\textsc{Lie} superalgebras. They characterized the most 
general symmetry \textsc{Lie} superalgebra of an
$S$-matrix \cite{Haag}.
The \textsc{Coleman-Mandula} theorem applies to
the bosonic sector of this algebra, which is 
a \lie algebra, so that the bosonic subalgebra is the 
direct product of  the \poin algebra and an internal 
symmetry \lie algebra.
The bosonic generators
$B_k, P_\mu$ and $J_{\mu\nu}$  belong to the $(0,0), (\h,\h)$
and  $(0,1)\oplus(0,1)$ representations of $SL(2,\C)$.
The novelty lies in the fermionic sector, which is generated
by spinorial charges $\cQ^i_\al$ in the $(\h,0)$ representation
and their hermitian adjoints $\bar \cQ^i_\dal=(\cQ^i_\al)^\dagger$
in the $(0,\h)$ representation. Here $i$ runs from $1$ to 
some positive integer $\cN$. 
The \textsc{Haag-Lopuszanski-Sohnius} 
theorem\footnote{A proof can be found in the textbook of Weinberg
\cite{intweinberg}.}
states in part that the fermion symmetry generators
can only belong to these representations. 
\section{Extended superalgebras}
\index{extended superalgebras}
\index{extended supersymmetry}
\index{central charge}
Let us now assume that there are $\cN$ such spinorial $(\h,0)$
supercharges $\left\{\cQ^1,\dots,\cQ^\cN\right\}$.
They must commute with translations,
\eqnl{
[\cQ^i_\al,P_\mu]=0,\quad i=1,\dots,\cN,}{reps7}
and have spin $\ft12$ which fixes
their commutators with the \lor generators,
\eqnl{
[J_{\mu\nu},\cQ^i_\al]=\ft{1}{2i}
(\sigma_{\mu\nu})_\al^{\;\;\beta}\cQ^i_\beta}
{reps9}
Thus the $\cN$-extended superalgebra consists of the 
generators of the \poin algebra plus $\cN$ 
spinorial supercharges with anticommutators
\eqnl{
\{\cQ^i_\al,\bar \cQ^j_\dbe\}=2\delta^{ij}(\sigma^\mu)_{\al\dbe}P_\mu .}
{reps11}
The missing supercommutators $\{\cQ^i_\al,\cQ^j_\beta\}$
will be discussed soon.
The sign in the anticommutator \refs{reps11} is 
determined by the requirement that in a relativistic
quantum field theory the energy should be
a non-negative operator: we get for each value
of the index $i$
\eqnl{
\sum_{\al=1}^2 \{\cQ^i_\al,\cQ^{i\dagger}_\al\}
\stackrel{\refs{reps3}}{=}
\sum_{\al=1}^2 \{\cQ^i_\al,\bar\cQ^i_\al\}=
2\,\tr\sigma^\mu\, P_\mu=4P_0
\mtxt{(no sum over i.)}}{reps13}
The left hand side is manifestly non-negative, since
each term has this property,
\eqnn{
\la\psi\vert\{\cQ^i_\al,\cQ^{i\dagger}_\al\}\psi\ra=
||\cQ^i_\al\psi||^2+||\cQ^{i\dagger}_\al\psi||^2\geq 0,}
and it follows that\\[2ex]
$\bullet$ the \emph{spectrum of $H=P_0$ in a theory
with supersymmetry contains no negative eigenvalues.}
\\[2ex]
We denote the state (or family of states) with the lowest
energy by $\vac$ and call it \emph{vacuum state}.\index{vacuum state} 
The vacuum will have zero energy, $H\vac=0$, if and only if
\eqnl{
\cQ^i_\al\vac=0\mtxt{and} \cQ^{i\dagger}_\al\vac=0\qquad
\forall \;\al,i.}{reps17}
Any state with positive energy cannot be invariant under
supersymmetry transformations. It follows in particular
that every one-particle state $\vert 1\ra$ must have
super partner states $\cQ^i_\al\vert 1\ra$ or $\cQ^{i\dagger}_\al\vert 1\ra$.
The spin of these partners will differ by $\ft12$ from that
of $\vert 1\ra$. Thus\\[1ex]
$\bullet$ \emph{each supermultiplet must contain
at least one boson and one fermion whose spins differ by
$\ft12$}.\\[1ex]
The translation invariance of $\cQ$ implies that
$\cQ$ does not change energy-momentum
\eqnl{
P_\mu \vert p\ra=p_\mu\vert p\rangle\Longrightarrow
P_\mu \cQ^i_\al\vert p\ra=p_\mu \cQ^i_\al\vert p\ra\;\;,\;\;
P_\mu \cQ^{i\dagger}_\al\vert p\ra=p_\mu \cQ^{i\dagger}_\al\vert p\ra,}
{reos19}
and therefore\\[1ex]
$\bullet$ \emph{all states in a multiplet of unbroken
supersymmetry have the same mass}.
\\[1ex]
Supersymmetry is
spontaneously broken if the ground state will not
be invariant under all supersymmetry transformations,
\eqnl{
\cQ^i_\al\vac\neq 0\mtxt{or}\cQ^{i\dagger}_\al\vac\neq 0}{reps21}
for same $\al$ and $i$. We conclude that\\[1ex]
$\bullet$
\emph{supersymmetry is spontaneously broken if and
only if the energy of the lowest lying state is not exactly
zero.}\\[1ex]
In \emph{extended supersymmetry} the $\cQ^i$ may
carry a representation of some internal symmetry,
\eqnl{
[T_r,\cQ^i_\al]=(t_r)^i_{\;j}\cQ^j_\al.}{reps23}
Since we assume this so-called \emph{R-symmetry}\index{R-symmetry}
to be compact, the representation matrices $t$ can be chosen 
Hermitian, $t_r=t_r^\dagger$.

Now let us consider the anticommutator
$\{\cQ,\cQ\}$. It must be a linear combination
of the bosonic operators in the representation $(0,0)$ and
$(1,0)$ of the \lor group. The only three-dimensional 
$(1,0)$ representation in the bosonic sector is the (anti)selfdual
part of $J_{\mu\nu}$. Such a term would not
commute with the $4$-momentum, whereas $\{\cQ,\cQ\}$ does. Thus
we are left with
\eqnl{
\{\cQ^i_\al,\cQ^j_\beta \}=2\veps_{\al\beta}Z^{ij},}{reps25}
were $Z^{ij}$ commutes with the space-time symmetries and
hence is some linear combination of the internal symmetry 
generators,
\eqnl{
Z^{ij}=\al^{rij}T_r.}{reps27}
Using the super-\textsc{Jacobi} identity
one shows that the $Z^{ij}$ commute with
\emph{all} generators of the superalgebra, including
the $T_r$ and thus with themselves,
\eqnl{
[T_r,Z^{ij}]=0\Longrightarrow
[Z^{ij},Z^{kl}]=0.}{reps29}
For this reason they are called
\emph{central charges}.\index{central charge} The
internal symmetries are generated by
$\{T_r\}=\{T'_\ell,Z^{ij}\}$. In what follows we shall
skip the prime at the non-central generators
$T'_\ell$.
\paragraph{Summary:}
It may be useful to collect the relevant (anti)commutation
relations of the $\cN$-extended superalgebra in
$4$ dimensions containing the \poin algebra
(\ref{lotr19a}-\ref{lotr19c}) as a subalgebra. 
The supercharges commute with translations \refs{reps7},
transform as spinors under \lor transformations \refs{reps9}
and transform under the $R$-symmetry as 
\eqnl{
[T_r,\cQ^i_\al]=(t_r)^i_{\;j}\cQ^j_\al\quad,\quad
[T_r,\bar \cQ^i_\dal]=-\bar \cQ^j_{\dal}(t_r)_j^{\;\,i}.}{reps31}
The generators of the $R$-symmetry generate
a \lie subalgebra,
\eqnl{
[T_r,T_s]=iC_{rs}^{\;\;\,t}T_t}{reps33}
and commute with the bosonic generators of the
\poin algebra,
\eqnl{
[T_r,J_{\mu\nu}]=[T_r,P_\mu]=0.}{reps35}
The supercharges fulfill the following anticommutation
relations
\begin{eqnarray}
\{\cQ_{\al}^i,\cQ_{\beta}^j\}&=&2\veps_{\al\beta}Z^{ij}\label{reps37}\\
\{\bar \cQ_{\dal}^i,\bar \cQ_{\dbe}^j\}&=&2\veps_{\dal\dbe}\bar Z^{ij}
\label{reps38}\\
\{\cQ_{\al}^i,\bar
\cQ_{\dal}^j\}&=&2\delta^{ij}\sigma^{\mu}_{\al\dal}P_{\mu}\,,
\label{reps39}
\end{eqnarray}
where $(Z^{ij})$ is the antisymmetric \emph{central charge matrix}.
The central charges $Z^{ij}$ commute with all
generators of the super-\poin algebra. 
According to the \textsc{Haag-Lopuszanski-Sohnius}
theorem \cite{Haag} the operators $\{P_\mu,J_{\mu\nu},
T_r,Z^{ij},\cQ^i_\al\}$ generate the \emph{most general} 
\textsc{Lie} superalgebra of an $S$-matrix. The bosonic 
ones generate the direct product of the \poin and internal
symmetry group. The remaining (anti)commutators are given
in (\ref{reps7}-\ref{reps9}), \refs{reps31} and 
(\ref{reps37}-\ref{reps39}).
\section{Representations}
\index{representation of superalgebras}
Let us discuss the representation theory
of $\cN$-extended supersymmetry in  four dimensions.
First we shall assume that the \emph{central charges are zero}.
\subsection{Massive representations without central charges}
\index{superalgebra!massive representations}
For vanishing central charges the supercharges anticommute,
\eqnl{
\{\cQ^i_{\al},\cQ^j_{\beta}\}=\{\bar \cQ^i_{\dal},\bar
\cQ^j_{\dbe}\}=0.}{mrep1}
For a massive particle we may choose the rest frame in which 
$P\sim(M,\mb0)$. Then the relations \refs{reps37} simplify as 
follows:
\eqnl{
\{\cQ_{\al}^i,\bar \cQ^j_{\dal}\}=
2M\delta^{ij}(\sigma_0)_{\al\dal}=
2M\delta_{\al\dal}\delta^{ij}.}{mrep3}
$\cQ$ is a tensor operator of spin $\ft12$, as follows from
\eqnn{
[\cQ^i,\mbJ]=\ft12 (\mbsigma \cQ^i)}
and in particular
\eqnl{
[J_3,\cQ^i_1]=-\h \cQ^i_1\mtxt{and}
[J_3,\cQ^i_2]=\h \cQ^i_2.}{mrep5}
Therefore, the result of the action of $\cQ$ on a state
with spin $s$ will be a linear combination of
states with spin $s\pm \ft12$.
Now we define the $2\cN$ fermionic creation and 
annihilation operators
\eqnl{
A^i_{\al}={1\over \sqrt{2M}}\,\cQ^i_{\al}\mtxt{and}
\bar A^i_{\dal}={1\over \sqrt{2M}}\,\bar \cQ^i_{\dal}}{mrep7}
which satisfy the simple algebra
\eqnl{
\{A^i_\al,\bar A^j_\dal\}=\delta^{ij}\delta_{\al\dal}.}{mrep9}
Building irreducible representations is now straightforward. 
We introduce a \textsc{Clifford} \emph{vacuum} $|\Omega\ra$, 
which is annihilated  by the $A^i_{\al}$ and generate
the \textsc{Fock} states by acting with the creation 
operators $\bar A^i_\dal$. A typical state would be
\eqnl{
\bar A^{i_1}_{\dal_1}\cdots \bar A^{i_n}_{\dal_n}\;\vert\Omega\ra.}{mrep11}
It is antisymmetric under interchange of index-pairs
$(\dal,i)\leftrightarrow (\dbe,j)$.
Since there are $2\cN$ creation operators
there are ${2\cN}\choose{n}$ states at the $n$'th oscillator level.
The total number of states built on \emph{one} vacuum
state is 
\eqnl{
\sum_{n=0}^{2\cN}{2\cN\choose n}=(1+1)^{2\cN}=4^\cN,}{mrep13}
half of them being bosonic and half fermionic. 
If the vacuum sector is degenerate, which happens
if the \cliff vacuum $\vert\Omega\ra$ is a member of 
a spin multiplet, then the number of states is
\begin{center}
\textsc{
number of states = $4^\cN\cdot$ dimension of vacuum sector.}
\end{center}
The operators $\bar A^i_{\dz}$ increase $J_3$ by $1/2$, 
such that the \emph{maximal spin} is equal to the 
spin $s_0$ of the ground-state plus $\cN/2$,
\eqnn{
s_{\rm max}=s_0+\cN/2.}
The \emph{minimal spin} is $0$ if $\cN/2\geq s_0$ or $s_0-\cN/2$
otherwise. 

Since renormalizability requires \emph{massive matter} to have spin
$\leq \ft12$. We conclude from the above
expression for $s_{\rm max}$ that we must have\\[1ex]
$\bullet\; \cN=1$ \emph{for renormalizable coupling of massive matter.}
\\[1ex]
We see that in the absence of central charges, the only
relevant massive multiplet is that of the massive
\wz model which has $\cN=1$ and $s_0=0$. It contains
a scalar, a pseudo-scalar and the two spin states 
of a massive \maj spinor
\eqnl{
\renewcommand{\arraystretch}{1.5}
\begin{array}{|l|l|lll|}\hline
\cN=1 & s^P: & 0^+ & \ft12 & 0^- \\ \hline
& \hbox{states:} & 1
 & 2 & 1\\ \hline
\end{array}}{mrep15}
For example, the spin-zero states are $\vert\Omega\ra$ and 
$\bar A_\de \bar A_{\dot 2}\vert\Omega\ra$.
\subsection{Massless representations} 
\index{superalgebra!massless representations}
For massless particles we can choose an
inertial frame in which $P_\mu=(E,0,0,E)$.
Then the nontrivial anti-commutation relations 
have the form
\eqnl{
\{\cQ^i_{\al},\bar \cQ^{j}_{\dal}\}=2E\delta^{ij}(\sigma_0+\tau_3)_{\al\dal}
=\pmatrix{4E&0\cr 0&0\cr} \delta^{ij}
\,,}{mlrep1}
the other anticommutators being zero. Since
\eqnl{
\{\cQ^i_2,\bar \cQ^i_{\dot 2}\}=\cQ^i_2\cQ^{i\dagger}_2+
\cQ^{i\dagger}_2\cQ^i_2=0,}
{mlrep3}
the $\cQ^i_2$ are represented by zero in a unitary theory.
Thus we only have $\cN$ non-trivial creation and annihilation operators
\eqnl{
A^i={1\ov 2\sqrt{E}}\,\cQ_1^i\mtxt{and}
\bar A^i={1\ov 2\sqrt{E}}\,\bar \cQ_{\dot 1}^i,}{mlrep5}
and the \textsc{Fock}-representation is $2^\cN$-dimensional.
It is much shorter than the massive one which contains
$4^\cN$ states.

The following \lor generators commute with 
$P_\mu=(E,0,0,E)$:
\eqnl{
J_1=J_{10}+J_{13},\quad J_2=J_{20}+J_{23}\mtxt{and} J_3=-J_{12}.}{mlrep7}
Observe that $J_3$ is just the \emph{helicity operator}
\index{helicity}
$\lam$ for a massless particle moving in 
the $3$-direction. The $J_i$ generate the \emph{non-compact} 
little group $E_2$ of translations and rotations in the $2$-plane,
\eqnl{
[J_1,J_3]=iJ_2,\quad [J_2,J_3]=-iJ_1\mtxt{and}
[J_1,J_2]=0.}{mlrep9}
In any \emph{finite dimensional} unitary representation 
the generators $J_1,J_2$ are trivially represented.
Note that $A^i$ increases the helicity by $\ft12$ and $\bar A^i$
decreases it by $\ft12$,
\eqnl{
[\lam,A^i]=\ft12 A^i\mtxt{and}
[\lam,\bar A^i]=-\ft12 \bar A^i.}{mlrep11}
We introduce the \cliff vacuum $\vert\Omega\ra$
with maximal helicity $\lam$ which is annihilated by all
$A^i$. The states in an
irreducible representation are gotten by acting with the
creation operators on this state. For example,
\eqnl{
\bar A^{i_1}\cdots \bar A^{i_n}\vert\Omega\ra}{mlrep13}
has helicity $\lam-n/2$. This way we get the
following states
\eqnl{
\renewcommand{\arraystretch}{1.5}
\begin{array}{|l|cccc|}\hline
\hbox{helicity:}& \lam & \lam-\ha & \dots & \lam-\frac{\cN}{2}\\ \hline
\hbox{multiplicity:}& 1 & {\cN\choose 1} & \dots & {\cN\choose \cN}\\ \hline
\end{array}
}{mlrep14}
The total number of states in an irreducible massless
representation is
\eqnl{
\sum_{k=0}^\cN {\cN\choose k}=2^\cN.}{mlrep15}
According to the CPT theorem a physical massless state
contains both helicities $\lam$ and $-\lam$.
A \emph{single} supermultiplet can contain all massless states
for $\lam =\cN/4$. Otherwise the states must be doubled
starting with a second \cliff vacuum with
helicity $\lam^\pr=\cN/2-\lam$.
Here we will describe the important examples with 
helicities up to one.

\begin{enumerate}
\item $\cN=1$ \textbf{supersymmetry:}
The number of \textsc{Clifford}-states in an irreducible multiplet is
just $1+1$ and we need at least two \cliff vacua to built
a CPT invariant model.
\begin{itemize}
\item For the \textit{chiral multiplet} $\lam=\ft12$ and $\lam^\pr=0$
and we have the following states\index{chiral multiplet}
\begin{center}
\renewcommand{\arraystretch}{1.1}
\begin{tabular}{lrl}
helicity:&$1/2$&1 \maj spinor\\
&$0$&2 real scalars\\
&$-1/2$&1 \maj spinor
\end{tabular}
\end{center}
These are the fields of the massless \wz model.
\item The \textit{vector multiplet}\index{vector multiplet}
 with $\lam=1$ and $\lam^\pr=-1/2$
consists of 
\begin{center}
\renewcommand{\arraystretch}{1.1}
\begin{tabular}{lrl}
helicity:&$1$&1 gauge field\\
&$1/2$& 1 \maj spinor\\
&$-1/2$&1 \maj spinor\\
&$-1$&1 gauge field
\end{tabular}
\end{center}
\end{itemize}
\item
\textbf{$\cN=2$ supersymmetry:}
A irreducible representation of the $\cN=2$-extended
superalgebra contains $3$ different helicities and 
$1+2+1$ 'states'.  The relevant multiplets are the hyper- 
and vector multiplet.
\begin{itemize}
\item The simplest multiplet is the 
\textit{hyper-multiplet}\index{hyper multiplet}
 $\lam=1/2$ with:
\begin{center}
\renewcommand{\arraystretch}{1.1}
\begin{tabular}{lrl}
helicity:&$1/2$&1 \maj spinor\\
&$0$&2 real scalars\\
&$-1/2$&1 \maj spinor
\end{tabular}
\end{center}
\item
To built a \textit{vector multiplet}\index{vector multiplet} 
we need two \cliff vacua with $\lam=1$ and $\lam^\pr=0$
so that it contains
\begin{center}
\renewcommand{\arraystretch}{1.1}
\begin{tabular}{lrl}
helicity:&$1$&1 gauge field\\
&$1/2$&2 \maj spinors\\
&$0$&2 real scalars\\
&$-1/2$&2 \maj spinors\\
&$-1$&1 gauge field
\end{tabular}
\end{center}
The two \maj spinors combine to a
\dirac spinor. This multiplet has been considered
in the \textsc{Seiberg-Witten} solution to $\cN=2$ supersymmetric 
gauge theories \cite{SW}.\\
\end{itemize}
\item
\textbf{$\cN=4$ supersymmetry:}
An irreducible supermultiplet with $4$ supercharges 
consists of $1+4+6+4+1=16$ states. 
\begin{itemize}
\item The unique multiplet 
giving rise to a renormalizable field theory
in flat spacetime is the \textit{vector multiplet}.
It  has $\lam=1$ and contains
\begin{center}
\renewcommand{\arraystretch}{1.1}
\begin{tabular}{lrl}
helicity:&$1$&1 gauge field\\
&$1/2$&4 \maj spinors\\
&$0$&6 real scalars\\
&$-1/2$&4 \maj spinors\\
&$-1$&1 gauge field
\end{tabular}
\end{center}
This multiplet enters the celebrated AdS/CFT correspondence
\cite{Maldacena}.
\end{itemize}
\end{enumerate}
\subsection{Non-zero central charges}
\index{representations!with central charge}
In this case massive supermultiplets can be as short
as massless ones. Under a $U(N)$ transformation
of the super charges,
\eqnn{
\cQ^i_\al\to U^i_{\;j}\cQ^j_\al\mtxt{and}
\bar \cQ^i_{\dal}\to \bar \cQ^j_{\dal}(U^\dagger)_j^{\;i}}
the antisymmetric central charge matrix $Z=(Z^{ij})$
transforms as
\eqnn{
Z\longrightarrow UZU^T,\quad
\bar Z\longrightarrow \bar U\bar Z U^\dagger.}
In can be 
shown that there exists
a unitary $U$ such that $Z$ becomes block diagonal\footnote{Consider even
$\cN$.}, 
\eqnl{
\left(\matrix{0&Z_1& 0&0 & &\dots & \cr -Z_1&0&0&0&&\dots&\cr
0&0&0&Z_2&&\dots&\cr 0&0&-Z_2&0&&\dots&\cr
&\dots&\dots&\dots&\dots&\dots&\cr &\dots&&&&0&Z_{\cN/2}\cr
&\dots&&&&-Z_{\cN/2}&0\cr}\right)\,.}{nzcc1}
We have labeled the \emph{real positive} eigenvalues by $Z_m$, 
$m=1,2,\ldots,\cN/2$.
We will split the index $i\to (a,m)$: $a=1,2$ labels positions inside
the $2\times 2$ blocks while $m$ labels the blocks.
Then
\eqnl{
\{\cQ_{\al}^{am},\bar \cQ_{\dal}^{bn}\}=2M\delta_{\al\dal}\delta^{ab}\delta^{mn}
\;\;\;,\;\;\;\{\cQ_{\al}^{am},\cQ_{\beta}^{bn}\}=
2Z_n\veps_{\al\beta}\veps^{ab}\delta^{mn}
\,.}{nzcc3}
Now we define the following fermionic oscillators
\eqnl{
A^{m}_{\al}={1\over
\sqrt{2}}\left(
\cQ^{1m}_{\al}+\veps_{\al\beta}\cQ^{\dagger 2m}_{\beta}\right)\quad,\quad
B^{m}_{\al}={1\over
\sqrt{2}}\left(
\cQ^{1m}_{\al}-\eps_{\al\beta}\cQ^{\dagger 2m}_{\beta}\right) \,,}{nzcc5}
and similarly for the conjugate operators.
Their anticommutators read
\eqngrrl{
\{A^m_{\al},A^n_{\beta}\}&=&\{A^m_{\al},B^n_{\beta}\}=
\{B^m_{\al},B^n_{\beta}\}=0\,,}
{\{A^{m}_{\al},A^{\dagger n}_{\beta}\}
&=&2\delta_{\al\beta}\delta^{mn}(M+Z_m)}
{\{B^{m}_{\al},B^{\dagger n}_{\beta}\}&=&2\delta_{\al\beta}\delta^{mn}(M-Z_m)
\,.}{nzcc7}
Unitarity requires that the right-hand sides in \refs{nzcc7}
be non-negative. This in turn implies the well-known 
\textsc{Bogomol'nyi bound}\index{Bogomol'nye bound}
\eqnl{
M\geq {\rm max}\left| Z_{m}\right|\,.}{nzcc9}
It is required by the unitarity of the underlying 
supersymmetric theory.

Assume, for example, that $Z_m=M$ for all $m$. Then
\eqnl{
\{B_\al^m,B^{\dagger m}_\al\}=0,}{nzcc11}
implies that the $B^m_\al$ vanish identically and we are left with
the following creation and annihilation operators
\eqnn{
A^m_\al,\,A^{\dagger m}_\al,\quad
i=m,\dots,\h\cN,\quad \al=1,2.}
They generate a multiplet with $2^\cN$ states and not
$4^\cN$ states as one gets without central charges. More generally,
consider $0\leq r\leq \cN/2$ of the $Z_m$'s to be equal 
to $M$. Then $2r$ of the $B$-oscillators vanish identically 
and we are left with $2\cN-2r$ creation and annihilation 
operators. The representation has $2^{2\cN-2r}$ states.

The maximal case $r=\cN/2$ gives rise to the \emph{short BPS multiplet} 
\index{short multiplet}
which has the same number of states as the massless multiplet.
The other multiplets with $0<r<\cN/2$ are known as \emph{intermediate BPS
multiplets.}
\index{intermediate BPS multiplet}

BPS states are important probes of non-perturbative physics in
theories with extended ($\cN\geq 2$) supersymmetry.
The BPS states are special for various reasons:
\begin{itemize}
\item Although
they are massive, they form multiplets under extended SUSY which are
shorter than the massive multiplets.
Their mass is given in terms of their charges and \textsc{Higgs} 
(moduli) expectation values.
\item They are the only multiplets that can become massless 
when we vary coupling constants and Higgs expectation values
without breaking supersymmetry.
\item They describe solitonic excitations which (at rest)
exert no force on each other.
\item Their mass-formula is supposed to be exact if one uses
renormalized values for the charges and moduli. If
quantum corrections would spoil the relation of
mass and charges without breaking supersymmetry,
then the number of states in the supermultiplet 
would jump as function of the moduli parameter.
\end{itemize}

\chapter{Supersymmetric Yang-Mills Theories}\label{chap:SYM}
\index{supersymmetric YM Theories}
The simplest $\cN=1$ supersymmetric gauge theory 
is \abel and contains a 'photon' 
and its massless superpartner, the 'photino'.
The neutral photino does not couple to the photon
and the resulting theory has no
interactions. Although the \abel theory has a free
dynamics it shares many \emph{algebraic 
properties} with interacting non-\abel gauge theories.
The transition from the
free \abel to the interacting non-\abel theories
is achieved by replacing ordinary by covariant
derivatives.
The non-\abel models contain
charged 'gluons' and their superpartners, the
massless charged 'gluinos'.
These supersymmetric \ym theories are 
introduced in the second part of 
this chapter. Supersymmetry was applied first to \abel gauge
theories without using the superfield formalism
by \textsc{J. Wess} and \textsc{B. Zumino} 
\cite{wesszumino74}. It was then extended to
non-\abel \ym theories by \textsc{S. Ferrara} and
\textsc{B. Zumino} \cite{ferrara74} and \textsc{A. Salam} and
\textsc{J. Strathdee} \cite{salam74}
\section{$\cN=1$ Abelian gauge theories}
\index{supersymmetric gauge theories!Abelian}
We have seen that a CPT-doubled $\cN=1$ vector 
multiplet contains one gauge field and one massless
\maj spinor. In an off-shell version (and the \wz gauge)
we also need an uncharged pseudo-scalar
field which later may be eliminated. 
Since a \maj particle is its own antiparticle and
thus uncharged, the \lagan density takes the simple form
\index{Lagrangean!of $\cN=1$ SYM}
\eqnl{
\cL=-\ft14 F_{\mu\nu}F^{\mu\nu}+\ft{i}{2}\psib\fdi\psi+\ft12 \cG^2,
\qquad F_{\mu\nu}=\pa_\mu A_\nu-\pa_\nu A_\mu.}{ag1}
The dimensions of the fields are
\eqnl{
[A_\mu]=L^{-1},\quad[\psi]=L^{-3/2},\quad
[\cG]=L^{-2}.}{ag3}
In order to find the supersymmetry transformations
of the fields it is useful to recall the hermiticity properties
\refs{herm13} for fermionic bilinears of 
\maj spinors and the results \refs{fierz9}, were we
calculated the sign in $\vepsb \gam^{(n)}\psi=
\pm\psib\gam^{(n)}\veps$ for two \maj spinors $\veps$ 
and $\psi$.  We use the hermitian $\gamf=-i\gam_0\gam_1\gam_2\gam_3$.

Taking the dimensions and hermiticity properties
into account we could guess the following supersymmetry 
transformations,
\index{susy transformations!for Abelian SYM}
\begin{eqnarray}
\delta_\veps A_\mu&=&i\vepsb\gam_\mu\psi\label{ag5}\\
\delta_\veps\psi&=&i F^{\mu\nu}\Sigma_{\mu\nu}\veps +iq\cG\gamf\veps\label{ag6}\\
\delta_\veps \cG&=&q\vepsb\gamf\fdi\psi,\quad q\in\R.\label{ag7}
\end{eqnarray}
The supersymmetry parameter $\veps$ is a constant
\maj spinor anti-commuting with itself and with
$\psi$. Up to a surface term the \lagan \refs{ag1} 
is invariant under these supersymmetry transformation,
\eqnl{
\delta_\veps \cL=\vepsb \pa_\mu V^\mu,\quad V^\mu=
\ft12\left(^*\!F^{\mu\nu}\gamf-iF^{\mu\nu}\right)\gam_\nu\psi
+\h q\cG\gamf\gam^\mu\psi.}{ag9}
In the proof, which I leave as an exercise, one needs 
the formulas \refs{ap1} and \refs{ap51} in the appendix and 
the \textsc{Bianchi} \emph{identity}\index{Bianchi identity} 
$^*\!F^{\mu\nu}_{\,\;\;,\nu}=0$ for the \emph{dual field strength tensor}
\index{Bianchi identity}\index{dual field strength}
\eqnl{
 ^*\!F_{\mu\nu}=\h \veps_{\mu\nu\rho\sigma} F^{\rho\sigma}.}{ag11}
At this point the real parameter $q$ is not fixed. 
Only when we want to recover the superalgebra are
we forced to choose $q=\pm 1$. 
\subsection{The closing of the algebra}\label{closalg}
\index{supersymmetry algebra!for Abelian gauge model}
We repeat what we have done
for the \wz model and calculate the
commutators of two supersymmetry transformations.
The commutator acting on the bosonic fields 
is easily computed,
\eqngrl{
[\delta_{\veps_1},\delta_{\veps_2}]A_\mu&=&
-F^{\rho\sigma}\left(\vepsb_2\gam_\mu\Sigma_{\rho\sigma}\veps_1
-\vepsb_1\gam_\mu\Sigma_{\rho\sigma}\veps_2\right)
\stackrel{\refs{ap51}}{=}2i\vepsb_2\gam^\nu\veps_1F_{\mu\nu}}
{&=&-2i\left(\vepsb_2\gam^\rho\veps_1\right)\,\pa_\rho A_\mu+
2i\pa_\mu\left(\vepsb_2\As\veps_1\right).}{ag13}
The first term in the second line is the expected infinitesimal
translation of the vector field. The last term we did
not encounter in the \wz model. It
is just a field dependent \emph{gauge transformation} with
gauge parameter
\eqnl{
\lam=2i\vepsb_2\As\veps_1.}{ag15}
Similarly, using \refs{ap1} and the \textsc{Bianchi}
identity one finds
\eqnn{
[\delta_{\veps_1},\delta_{\veps_2}]\cG=
-2iq^2\vepsb_2\gam^\rho\veps_1\pa_\rho \cG.}
For the choices $q=\pm 1$ we find the expected
commutator and hence we shall assume that 
$q\in\{-1,1\}$ in what follows.
To calculate the commutator of two transformations
on the photino field is a bit more tricky. First one
obtains
\eqnn{
[\delta_{\veps_1},\delta_{\veps_2}]\psi=
\left(i\left(\vepsb_1\gamf\gam^\rho\pa_\rho\psi\right)\gamf\veps_2
-2\left(\vepsb_1\gam^\nu\pa^\mu\psi\right)\Sigma_{\mu\nu}\veps_2
\right)
-(\veps_1\leftrightarrow \veps_2).}
With the help of \refs{ap53}, the (anti)commutators of 
the \cliff and \lor algebras and the relation 
\refs{ap1} in the appendix one ends up with the expected result
\eqnl{
[\delta_{\veps_1},\delta_{\veps_2}]\psi=
-2i(\vepsb_2\gam^\rho\veps_1)\,\pa_\rho\psi.}{ag17}
As in chapter \ref{chapWZ} we introduce
the \emph{supercharge} via $\delta_\veps(..)=i[\vepsb \cQ,..]$
and arrive at
\eqnl{
\{\cQ_\al,\bar \cQ^\beta\}=2i\left(\gam^\mu\right)_\al^{\;\,\beta}\pa_\mu
-G_\al^{\;\,\beta}(A),}{ag19}
where $G_\al^{\;\,\beta}$ is the gauge transformation
with gauge parameter $\lam_\al^{\;\,\beta}=2i(A\!\!\!\slash)_\al^{\;\,\beta}$.
On gauge invariant fields the last term on the right 
hand side vanishes and we recover the familiar superalgebra. 
\subsection{Noether charge}
\index{Noether charge!for $\cN=1$ SYM}
To find the \noeth charge we must first calculate
\eqnl{
{\delta\cL\ov\delta\pa_\mu A_\nu}\delta_\veps A_\nu
+{\delta \cL\ov \delta \pa_\mu\psi}\delta_\veps \psi
=-F^{\mu\nu}\delta_\veps A_\nu-\ft{i}{2}\delta_\veps\psib\gam^\mu\psi.}{ag21}
Using the above expressions for the supersymmetry 
transformations this can be written as (we take $q=1$)
\eqnl{
-\ft{3i}{2}F^{\mu\nu}\vepsb\gam_\nu\psi
+\ft12\,^*\!F^{\mu\nu}\vepsb\gam_\nu\gamf\psi
+\ft12 \cG\vepsb\gamf\gam^\mu\psi.}{ag23}
Subtracting $V^\mu$ in \refs{ag9} we find the 
\noeth current\index{Noether current!for $\cN=1$ SYM}
$J^\mu=-i\delta_\veps\bar\psi\gammu\psi\vert_{\cG=0}$.
Inserting the $3+1$ decompositions of the field strength
tensor and its dual,
\eqnl{
(F_{0i},F_{ij})=(E_i,-\eps_{ijk}B_k)\mtxt{and}
(^*\!F_{0i},^*\!F_{ij})=(-B_i,-\eps_{ijk}E_k),
}{ag25}
the corresponding conserved and gauge invariant \noeth charge 
takes the form\index{Noether charge!for Abelian gauge model}
\eqnl{
\cQ=\int d\mbx\left(i\pi_i-\gamf B_i\right)\gam_i\psi,\mtxt{where}
\mbpi=\mbE}{ag27}
is the momentum field conjugate to $\mbA$.
\section{$\cN=1$ SYM theories}
\index{supersymmetric gauge theories!non-Abelian}
\index{$\cN=1$ SYM}
We consider $SU(N)$ gauge theories without matter,
sometimes called SUSY \emph{gluodynamics},
with 'gluons' and massless 'gluinos'. 
The former are described by a vector field $A_\mu$ 
and the latter by a \maj spinor field $\psi$. 
Off-shell we also need an auxiliary
field $\cG$. All fields take their values in the \lie
algebra of the gauge group,
\eqnl{
A_\mu=A^a_\mu T_a,\quad \psi=\psi^a T_a\mtxt{and}
\cG=\cG^aT_a.}{nab1}
The real \emph{structure constants}
\index{structure constants!of Lie algebra} in
\eqnl{
[T_a,T_b]=f_{ab}^{\;\;\,c} T_c}{nab3}
are totally \emph{antisymmetric} and we normalize the
hermitian generators by $\tr T_aT_b=\delta_{ab}$.

The gauge and matter fields transform under gauge transformations
as
\eqngrrl{
A&\to& UAU^{-1}+iUdU^{-1}}
{\psi&\to& U\psi U^ {-1}}
{\cG&\to& U\cG U^{-1}}{nab5}
with group elements $U(x)$. With $U\sim \id+i\lam$
the infinitesimal transformations read
\eqnl{
\delta_\lam A_\mu=D_\mu \lam,\quad \delta_\lam \psi=i[\lam,\psi]\mtxt{and}
\delta_\lam \cG=i[\lam,\cG].}{nab7}
The gauge invariant \lagan density is the expected 
generalization of \refs{ag1},\index{action!of $\cN=1$ SYM}
\index{Lagrangean!of $\cN=1$ SYM}
\eqnl{
\cL=-\ft14 \tr F_{\mu\nu}F^{\mu\nu}
+\ft{i}{2}\tr(\psib\di\psi)+\ft12 \tr\cG^2.}{nab13}
It contains the \emph{covariant derivative}\index{covariant derivative} 
of the spinor field
$D_\mu\psi=\pa_\mu\psi-i[A_\mu,\psi]$
and the gauge covariant \emph{field strength tensor}\index{field strength} 
$F_{\mu\nu}=\pa_\mu A_\nu-\pa_\nu A_\mu-i[A_\mu,A_\nu]$.
Both transform according to the adjoint representation.

The susy transformations are gotten from those
of the \abel model if we only replace ordinary
by covariant derivatives,\index{susy transformations!for $\cN=1$ SYM}
\eqngrrl{
\delta_\veps A_\mu\!&=&i\vepsb\gam_\mu\psi}
{\delta_\veps\psi&=&i F^{\mu\nu}\Sigma_{\mu\nu}\veps +i\cG\gamf\veps}
{\delta_\veps \cG&=&\vepsb\gamf\di\psi.}{nab15}
To calculate the variation of the
\lagan density one needs the formula $\delta F_{\mu\nu}=
D_\mu\delta A_\nu-D_\nu\delta A_\mu$ which yields
\eqnn{
\delta_\veps F_{\mu\nu}=
i\vepsb\left(\gam_\nu D_\mu-\gam_\mu D_\nu\right)\psi,}
together with the identities
\eqnl{
\delta (D_\mu\psi)=D_\mu\delta\psi-i[\delta A_\mu,\psi]
\mtxt{and}
\delta_\veps\psib=-i \vepsb F^{\mu\nu}\Sigma_{\mu\nu} +i\vepsb\cG\gamf
.}{nab21}
As for the \abel model one uses the
\textsc{Bianchi} identity $D_{[\rho}F_{\mu\nu]}=0$ 
and obtains\index{Bianchi identity!in YM theories}
\begin{eqnarray}
\delta_\veps\cL&=&\bar\veps\pa_\mu V^\mu+\ft{i}{2}
\tr\left(\psib \gam^\mu[\vepsb\gam_\mu\psi,\psi]\right)
\qquad\hbox{with}
\label{nab23a}\\
V^\mu&=& \ft12\tr\left\{\left(^*\!F^{\mu\nu}\gamf
-iF^{\mu\nu}\right)\gam_\nu\psi\right\}
+\ft12 \tr\left(\cG\gamf\gam^\mu\psi\right).\label{nab23b}
\end{eqnarray}
If we can show that the last term in \refs{nab23a}
vanishes, then we have proved the invariance of
the action. It is not straightforward
to show this, so let me indicate the proof.
First we expand $\psi=\psi^a T_a,\;\psib=\psib^a T_a$ and
rewrite this term as
\eqnl{
\ft{i}{2}\tr\left(\psib \gam^\mu[\vepsb\gam_\mu\psi,\psi]\right)=
\ft12 f_{abc}(\psib^a\gam^\mu\psi^b)(\vepsb\gam_\mu\psi^c).}{nab25}
Then we insert the general \textsc{Fierz} identity \refs{fierz5}
for $\psi^c\psib^a$ in the right hand side of
\eqnl{
(\psib^a\gam^\mu\psi^b)(\vepsb\gam_\mu\psi^c)=
\vepsb\gam_\mu\big(\psi^c\psib^a\big)\gam^\mu\psi^b,}{nab29}
and use the relations \refs{ap3} in the appendix to arrive at the 
useful identity
\eqngrl{
(\psib^a\gam^\mu\psi^b)(\vepsb\gam_\mu\psi^c)&=&
-(\vepsb\psi^b)(\psib^a\psi^c)
+\ft12(\vepsb\gam_\rho\psi^b)(\psib^a\gam^\rho\psi^c)}
{&&-\ft12 (\vepsb\gam_\rho\gamf\psi^b)(\psib^a\gamf\gam^\rho\psi^c)
+(\vepsb\gamf\psi^b)(\psib^a\gamf\psi^c).}{nab33}
Note that all but the second term on the right hand side
are symmetric in $a$ and $c$ such that
\eqnl{
f_{abc}(\psib^a\gam^\mu\psi^b)(\vepsb\gam_\mu\psi^c)=
\ft12 f_{abc}(\psib^a\gam^\mu\psi^c)(\vepsb\gam_\mu\psi^b)
=-\ft12 f_{abc}(\psib^a\gam^\mu\psi^b)(\vepsb\gam_\mu\psi^c)}{nab39}
which implies, that the expression on the left is zero.
We conclude that the last term in \refs{nab23a} is absent
and that the susy variation of the \lagan density is just
a divergence.

The conserved \noeth current has the familiar form
$J^\mu=-i\tr \delta\psib\gammu\psi\vert_{\cG=0}$
such that the conserved supercharge reads
\index{supercharge!for $\cN=1$ SYM}
\eqnl{
\cQ=\int d\mbx\,\tr\left\{(i\pi_i-B_i\gamf)\gam_i\psi\right\}.}{nab41}
The commutator of two supersymmetry transformations
on the fields is a translation plus a field
dependent infinitesimal \emph{gauge transformation}.
After some lengthy but straightforward manipulations 
in which one uses (\ref{ap5},\ref{ap57}), 
the \textsc{Bianchi} identity, (\ref{ap5},\ref{ap53})
and $2i\Sigma_{\mu\nu}\gam^\nu=3\gam_\mu$ one ends
up with the expected commutators
\eqngrrl{
\,[\delta_{\veps_1},\delta_{\veps_2}]A_\mu&=&
-2i\left(\vepsb_2\gam^\nu\veps_1\right)\pa_\nu A_\mu
+D_\mu\lam,}
{\,[\delta_{\veps_1},\delta_{\veps_2}]\cG&=&
-2i\left(\vepsb_2\gam^\rho\veps_1\right)\pa_\rho\cG+i[\lam,\cG],}
{\,[\delta_{\veps_1},\delta_{\veps_2}]\psi&=&
-2i(\vepsb_2\gam^\rho\veps_1)\pa_\rho\psi+i[\lam,\psi].}
{nab43}
The gauge parameter
$\lam=2i(\vepsb_2\As\veps_1)$ depends on the gauge field
and the susy parameters $\veps_i$.
As in the \abel case the superalgebra closes only
on \emph{gauge invariant fields}.
\section{$\cN=2$ SYM theories}
\index{$\cN=2$ SYM}
In this section we consider theories with
two supersymmetries. Realistic (i.e. chiral) models of 
particle interactions have at most one  supersymmetry. 
However, there are good reasons for discussing models
with extended susy. Their dynamics is under much better control 
which may lead to precise statements about their 
spectra, in perturbation theory and beyond. Some models 
possess \emph{central charges} which are saturated 
by BPS-configurations.\index{central charge} Magnetic 
monopoles of BPS type\footnote{See \cite{bruck} for a 
recent review on topological objects in
gauge theories.} are particular BPS-configurations. 
In \cite{adda} these monopoles were first discussed 
in the context of SYM with two supercharges.
The rather simple $\cN=2$ model has played an important
role in recent developments about confinement in
asymptotically free gauge theories. \textsc{Seiberg} and 
\wit derived an analytic expression for the low energy 
effective action (the leading two terms in a derivative expansion)
of this theory \cite{SW}.

On shell the model contains 
one vector field $A_\mu$, two \maj spinors $\lam_1$ and $\lam_2$, 
one scalar field  $\phi_1$ and one pseudo-scalar field $\phi_2$. 
All fields transform according to the adjoint representation 
of the gauge group. The two \maj spinors define a \dirac spinor as
$\kappa\psi=\lam_1+i\lam_2$, where $\kappa=\sqrt{2}$,
with inverse relation
$\kappa\lam_1=\psi+\psi_c$.
Also the two \maj supersymmtry parameters $\veps_1$
and $\veps_2$ are combined to a \dirac parameter  
$\kappa\al=\veps_1+i\veps_2$. It follows, that
\eqngrl{
i\psib\di\psi&=&\ft{i}{2}\sum \left(\bar\lam_i\di\lam_i\right)-\ft12
\pa_\rho\left(\bar\lam_1\gam^\rho\lam_2\right)}
{\alb M\psi\!\!&\!+\!&\!\!\alb_c M\psi_c=
\sum \vepsb_i M\lam_i}{seib1}
hold true for any matrix $M$.
Now we could return to the $\cN=1$ vector multiplet
and obtain the following transformation rule for the
gauge potential,\index{susy transformations!for $\cN=2$ SYM}
\eqnl{
\delta_\al A_\mu=i\sum \vepsb_i\gam_\mu\lam_i=
i\alb\gam_\mu\psi+i\alb_c\gam_\mu\psi_c.}{seib3}
By dimensional arguments there are no other terms
which are linear combinations of the $\psi$ and $\psi_c$ 
contracted with the supersymmetry parameters.
\subsection{Action of Seiberg-Witten model}
The SYM theory with two supercharges has the same particles 
as the SYM theory with one supercharge combined with the 
massless \wz model. But since all fields are in
one supermultiplet they all must transform according
to the adjoint representation of the gauge group.
To construct the supersymmetric action of the $\cN=2$ 
model we could try to add the \wz \lagan \refs{wzi23}
with \emph{massless} and \lie algebra valued fields 
and $\pa_\mu$ replaced by $D_\mu$ to the gauge model 
\lagan \refs{nab13}. 
If $\phi_1$ and $\phi_2$ are in the adjoint representation, then 
the self interaction term in \refs{wzi23}, namely
\eqnl{
-g\psib \phi^\dagger\psi-\ft12 g^2\,
\left(\phi_1^2+\phi_2^2\right)^2,\quad \phi=\phi_1+i\gamf\phi_2,}{seib5}
should be replaced by something like
\eqnl{
-g\tr \psib [\phi^\dagger,\psi]+
\ft12 g^2\,\tr \left([\phi_1,\phi_2]^2\right),\quad
\phi=\phi_1+i\gamf\phi_2.}{seib7}
Hence we would guess that the on-shell \lagan density
of the $\cN=2$ vector multiplet has the 
form\index{Lagrangean!of $\cN=2$ SYM}
\index{action!of $\cN=2$ SYM}
\eqngrl{
\cL&=&
-\ft14 \tr F_{\mu\nu}F^{\mu\nu}
+\ft12 \sum\tr(D_\mu \phi_i)^2
+\ft12 g^2\,\tr \big([\phi_1,\phi_2]^2\big)}
{&&+i\tr\psib\di\psi-g\tr\psib[\phi^\dagger,\psi]}{seib9}
where $\psi$ is a \dirac spinor.
Expanding the \lie algebra valued fields $A_\mu,\phi_i,\psi$
in a trace-orthonormal basis $\{T_a\}$, 
for example $\psi=\psi^a T_a$, such that
\eqngrl{
F^a_{\mu\nu}&=&\pa_\mu A^a_\nu-\pa_\nu A^a_\mu+g\fabc A^b_\mu A^c_\nu}
{(D_\mu \phi)^a&=&\pa_\mu \phi^a+g\fabc A_\mu^b \phi^c,}{seib11}
we obtain for the component fields
\eqngrl{
\cL&=&
-\ft14 F^a_{\mu\nu}F^{\mu\nu}_a
+\ft12 (D_\mu \phi_i)^a (D^\mu \phi_i)_a
-\ft12 g^2 \fabc f_{apq}\big(\phi_1^b\phi_1^p\phi_2^c\phi_2^q)}
{&+&i\psib^a\gam^\mu
\left(\pa_\mu\psi_a\!+\!gf_{abc} A_\mu^b\psi^c\right)
+igf_{abc} \phi_1^a\psib^b\psi^c+gf_{abc} \phi_2^a\psib^b\gamf\psi^c,}{seib13}
where one sums over all indexes.
The model contains only \emph{one coupling constant} $g$ and the
self-coupling of the scalar fields and \textsc{Yukawa} couplings 
are determined by $g$.
There is a potential term in the \lagan but it has
\emph{flat directions}\index{flat directions!for $\cN=2$ SYM} 
whenever $[\phi_1,\phi_2]=0$. Classically the model is
scale invariant but this invariance is spontaneously
broken by a non-zero expectation value of the scalar fields. 

For the \textsc{Legendre} transformation from the
\lagan to the \haman formulation we need
the momenta conjugate to the vector, scalar and
\dirac fields,
\eqnl{
\pi_i=F_{0i}=E_i,\quad\pi_{\phi_i}=D_0\phi_i\mtxt{and}
\pi_\psi=i\psi^\dagger.}{seib15}
In the convenient \weyl or temporal gauge $A_0=0$\index{Weyl gauge}
these relations simplify to $\pi_i=\dot A_i$ and $\pi_{\phi_i}=\dot \phi_i$.
The \haman density $\cH$ takes the form
\eqngrl{
\cH&=&
\tr\Big\{\ft12 \left(\mbpi^2+\mbB^2\right) +
\ft12\sum \left(\pi_{\phi_i}^2+(\mbD \phi_i)^2\right)
-i\psi^\dagger \gam^0\gam^iD_i\psi}
{&&\qquad\qquad\quad
+g\psib [\phi_1,\psi]-ig \psib\gamf [\phi_2,\psi]-\ft12 g^2 
\left[\phi_1,\phi_2\right]^2\Big\}.}{seib17}
This density is (formally) hermitian and the bosonic
part is non-negative. The latter follows from the
fact that the square of the antihermitian matrix 
$[\phi_1,\phi_2]$ is non-positive.
\subsection{Susy transformations and invariance of $S$}
\index{susy transformations!for $\cN=2$ SYM}
Now we shall fix the variations of the remaining fields
$\phi_i$ and $\psi$ such that $\cL$ is invariant up to
surface terms. The calculations parallel those
of the $\cN=1$ models and I need not give
many details here. With
\eqnl{
\delta D_\mu \phi_i=D_\mu\delta \phi_i-ig[\delta A_\mu,\phi_i]\quad,\quad
\delta\di\psi=\di\delta\psi-ig[\delta A_\mu,\gam^\mu\psi]}{seib19}
and $\delta F_{\mu\nu}=D_\mu\delta A_\nu-D_\mu\delta A_\mu$
one can express the variations of the
terms in the \lagan density \refs{seib9} as functions of
$\delta A_\mu,\delta\phi_i$ and $\delta\psi$.
To fix the transformation rules for the scalar fields
one considers the terms in $\delta \cL$ which are
\emph{cubic} in the \dirac field. After using
the transformation rules \refs{seib3} for the
vector fields and a suitable \textsc{Fierz} identity
one obtains the transformation rules for the
scalar fields such that the terms trilinear in $\psi$ 
cancel in $\delta \cL$,\index{susy transformations!for $\cN=2$ SYM}
\eqnl{
\delta_\al \phi_1=i(\alb\psi-\alb_c\psi_c)\mtxt{and}
\delta_\al \phi_2=\alb\gamf\psi-\alb_c\gamf\psi_c.}{seib21}
Again, the variations of $\phi_1$ and $\phi_2$ can 
only be linear combinations of $\psi$ and $\psi_c$ contracted 
with the supersymmetry parameters. Next one fixes
the transformation rules for $\psi$ such that the 
terms in $\delta \cL$ which are cubic in 
the bosonic fields $\phi_i$ cancel. This happens for
\index{susy transformations!for $\cN=2$ SYM}
\eqnl{
\delta_\al\psi=iF^{\mu\nu}\Sigma_{\mu\nu}\al
-D_\mu\phi\,\gam^\mu\al+g\,[\phi_1,\phi_2]\gamf \al}{seib23}
Let us summarize our findings. The action
\refs{seib9} can only be invariant under susy
transformations if they have the forms (\ref{seib3},\ref{seib21}) 
and \refs{seib23}. Note that a constant 
background field $\phi_1$ is left invariant by an 
arbitrary symmetry transformation. It does not 
'break supersymmetry'.

Actually one can show that the action is invariant
under the above transformation. The corresponding
\noeth currents reads\index{Noether current!for $\cN=2$ SYM}
\eqngrl{
J^\mu&=&
\alb\left\{\,^*\!F^{\mu\nu}\gam_\nu\gamf-iF^{\mu\nu}\gam_\nu
+i\gam^\nu D_\nu\phi\gam^\mu+ig[\phi_1,\phi_2]\gam^\mu\gamf\right\}\psi}
{&+&\psib\left\{\,^*\!F^{\mu\nu}\gam_\nu\gamf+iF^{\mu\nu}\gam_\nu
-i\gam^\mu D_\nu \phi\gam^\nu
+ig[\phi_1,\phi_2]\gam^\mu\gamf \right\}\al}{seib25}
and the \noeth charge takes the simple form
\index{supercharge!for $\cN=2$ SYM}
\eqnl{
\cQ=\alb\int d\mbx\left(R+S\right)\psi
+\int d\mbx\,\psib\left(R-S\right)\al,}{seib27}
where we have defined the fields
\eqngrl{
R&=&\gamf\gam^i\pi_i-i\gam^0 D_i\phi\gam^i+i\gam^0\gamf[\phi_1,\phi_2]}
{S&=&-i\gam^i\pi_i+i\pi_{\phi_1}+\pi_{\phi_2}\gamf.}{seib29}
The conserved supercharge $\cQ$ can be decomposed into
two \maj charges. These \maj charges satisfy the $\cN=2$
superalgebra.\\
\emph{Exercise:} Check, that the commutator of two
susy transformations on any field is
\eqnl{
[\delta_{\al_1},\delta_{\al_2}]\Phi=a^\rho\pa_\rho \Phi
+\delta_\lam \Phi}{seib31}
with parameter $a^\rho=-2i\left(\alb_2\gam^\rho\al_1
-\alb_1\gam^\rho\al_2\right)$ for infinitesimal
translation.  $\delta_\lam$ is a small gauge transformation
with field dependent gauge parameter
\eqnl{
\lam=-a^\rho A_\rho
-2i\alb_2\phi^\dagger\al_1+2i\alb_1\phi^\dagger\al_2,\quad
\phi=\phi_1+i\gamf\phi_2.}{seib33}
\section{Chiral basis}
\index{chiral spinors}
We use our conventions in section \ref{sect:spinors}
to rewrite the action \refs{seib9} and supersymmetry
transformations of the $\cN=2$ model in terms of \weyl 
fermions $\varphi$ and $\chib$ in
\eqnl{
\psi=\pmatrix{\varphid\cr\chibu}\mtxt{and}
\psib=\pmatrix{\chiu,\varphibd}}{seibc1}
Also we combine the two scalar fields $\phi_1$ and
$\phi_2$ to a scalar field with values in the complexified
\lie algebra,
\eqnl{
\kappa\phi=\phi_1+i\phi_2 \mtxt{with}\kappa=\sqrt{2}.}{seibc3}
For example, the \textsc{Yukawa} term in \refs{seib9} takes the form
\eqnl{
\psib[\phi_1-i\gamf\phi_2,\psi]=
\kappa\chi [\phi,\varphi]+\kappa\varphib [\phi^\dagger,\chib].}{seibc5}
In terms of the \weyl spinors and complex scalar 
the action reads
\eqngrl{
\cL&=&
-\ft14 \tr F_{\mu\nu}F^{\mu\nu}
+\tr(D_\mu \phi D^\mu\phid)
-\ft12 g^2\,\tr\big([\phi,\phid]^2\big)}
{&&+\,\tr \left\{i\chi(\sigma D)\chib+i\varphib(\tsigma D)\varphi
-\kappa g\chi[\phi,\varphi]-\kappa\varphib \big[\phid,\chib\big]\right\}}{seibc7}
To rewrite the susy-transformations we also use a
chiral basis for the susy parameter,
\eqnl{
\al=\pmatrix{\thetad\cr\zetabu}\mtxt{and}
\alb=\pmatrix{\zetau,\thetabd}.}{seibc9}
Next we insert the relations \refs{ap101} in the appendix
into the susy transformations (\ref{seib3},\ref{seib21},\ref{seib23})
and obtain
\begin{eqnarray}
\delta_\al A_\mu\!\!&=&i\left(\theta\sigma_\mu\varphib+\thetab\tsigma_\mu\varphi
+\zeta\sigma_\mu\chib+\zetab\tsigma_\mu\chi\right),\label{seibc11}\\
\kappa\delta_\al\phi&=&i\theta\chi+i\thetab\chib\label{seibc12}\\
\kappa\delta_\al\phi^\dagger\!\!&=&i\zeta\varphi+i\zetab\varphib\label{seibc13}\\
\delta_\al\varphi&=&
\left(\ft12 F^{\mu\nu}\sigma_{\mu\nu}-ig\big[\phi,\phid\big]\right)\theta
-\kappa\big(\sigma_\mu D^\mu\phid\big)\,\zetab,\label{seibc14}\\
\delta_\al\chib&=&
\left(\ft12 F^{\mu\nu}\tsigma_{\mu\nu}+ig\big[\phi,\phid\big]\right)\zetab
-\kappa\left(\tsigma_\mu D^\mu \phi\right)\theta\label{seibc15}
\end{eqnarray}
This finishes our introduction into the $\cN=2$ supersymmetric
\ym theory.

\chapter{Supersymmetry, Solitons and Fluctuations}
\index{soliton}
In this chapter we construct solitonic type
solutions of the $\cN=2$ supersymmetric gauge theory
considered in chapter \ref{chap:SYM}.
In \mink spacetime these are static solutions 
of the field equations with quantized magnetic
charge \cite{hooft}.  At a distance these magnetically charged
solutions look like \dirac monopoles \cite{diracm}. For the
\textsc{Euclid}ean field theory we consider 
solutions which are localized in space and 'time',
socalled instantons. Both types of solutions are
left invariant by half of the supersymmetries.
The other supersymmetries transform the
solitonic solutions into fermionic zero-modes,
the \textsc{Jackiw-Rebbi} modes \cite{jackiw}.

\section{Field equations}
\index{field equations for $\cN=2$ SYM}
The \textsc{Euler-Lagrange} equations for the
action \refs{seib9} are the following hyperbolic 
field equations:\pan
$\bullet$
The variation with respect to the gauge potential yields the
\ym equations\index{Yang-Mills equations!for $\cN=2$ SYM}
\eqnl{
D_\mu F^{\mu\nu}=
ig[\phi_1,D^\nu \phi_1]+ig[\phi_2,D^\nu \phi_2]-g[\psib,\gam^\nu\psi],}{sol1}
where $[\psib,M\psi]\equiv i\fabc \psib^bM\psi^c T_a.$

$\bullet$ The variations with respect to the scalar and pseudo-scalar
fields yield the \textsc{Klein-Gordon}-type equations
\index{Klein-Gordon equation!for $\cN=2$ SYM}
\eqngrl{
D_\mu D^\mu \phi_1&=&-g[\psib,\psi]\;+\;g^2[\phi_2,[\phi_2,\phi_1]]}
{D_\mu D^\mu \phi_2&=&-ig[\psib,\gamf\psi]+g^2[\phi_1,[\phi_1,\phi_2]],}{sol3}
$\bullet$ The variation with respect to the spinor field yields the 
\dirac equation\index{Dirac equation!for $\cN=2$ SYM}
\eqnl{
\di\psi=g[\phi_1-i\gamf \phi_2,\psi].}{sol5}
A particular class of solutions is obtained if we set
\eqnl{
\psi=0\mtxt{and}\phi_2=0,}{sol7}
in which case the field equations for the remaining 
fields reduce to
\eqnl{
D_\mu F^{\mu\nu}=ig[\phi_1,D^\nu \phi_1]\mtxt{and}
D_\mu D^\mu \phi_1=0.}{sol9}
If we further set $\phi_1=0$ then 
we obtain the pure \ym equations
$D_\mu F^{\mu\nu}=0$,
which in the \textsc{Euclid}ean sector possess (anti)selfdual
\emph{instanton solutions}.\index{instantons}

Now we have prepared the ground for discussing the intriguing
relation between supersymmetry and solutions of the
field equations. Consider a classical background
configuration with vanishing $\phi_2$ and $\psi$. 
Since the variation $\delta_\al\phi_2$ vanishes 
identically for $\psi=0$ such a configuration is left 
invariant by the susy transformations if
\eqnl{
\delta_\al\psi=0\stackrel{\refs{seib23}}{\Longleftrightarrow}
iF^{\mu\nu}\Sigma_{\mu\nu}\al=\di \phi_1\al.}{sol11}
To see what second order equation is implied by
this first order equation we act with $\di$ on it. 
Using \refs{ap1} and the \textsc{Bianchi} identity 
the left hand side yields  
\eqnn{
i\di\,F^{\mu\nu}\Sigma_{\mu\nu}\al=D_\mu F^{\mu\nu}\gam_\nu\al}
and the right hand side becomes
\eqnn{
\di  \di \phi_1\al=D_\mu D^\mu\phi_1\al 
+g[F_{\mu\nu},\phi_1]\Sigma^{\mu\nu}\al.}
Now we may use the first order equations \refs{sol11} once 
more to rewrite the last term as $ig[\phi_1,D_\mu\phi_1]\gammu\al$.
Thus we have shown that the first order equation
$\delta_\al\psi=0$ implies the \emph{second order equation}
\eqnl{
\left(D_\mu F^{\mu\nu}-ig[\phi_1,D^\nu \phi_1]\right)
\gam_\nu\al=D_\mu D^\mu\phi_1\,\al.}{sol13}
If $\al$ is an arbitrary \textsc{Dirac}, 
\weyl or \maj spinor, then both sides must vanish separately.
To see this we note that $\alb\gam_\nu\al$ 
vanishes for \maj spinors and $\alb\al$ for \weyl spinors.
The resulting two equations are just the \emph{field equations} 
\refs{sol9} for the scalar and gauge field. This then proves that
for backgrounds with vanishing $\phi_2$ and $\psi$ 
the first order equation $\delta_\al\psi=0$ implies the second 
order field equations \refs{sol9} for $\phi_1$ and $A_\mu$.

The opposite is of course not true. If \refs{sol11} would 
hold for all susy parameters $\al$ then 
$F_{\mu\nu}$ would vanish and $\phi_1$ would be
constant, up to a gauge transformation. Thus only 
\emph{trivial background fields} respect all supersymmetries. 
The idea is to impose the first order condition $\delta_\al\psi=0$
only for part of supersymmetry, that is for 
a restricted set of supersymmetry parameters $\al$.

\section{Bogomol'nyi bound and monopoles}
\index{magnetic monopole}
\index{Bogomol'nyi bound}
Let us consider magnetic monopole solutions
in more detail. Hence we assume that the only non-vanishing
fields $(A_\mu,\phi_1)$ are static and choose
the \weyl or temporal gauge $A_0=0$. With $2\Sigma_{ij}=-\eps_{ijk}\sigma_0\otimes\tau_k$
the first order equation \refs{sol11} reads 
\eqnl{
0=\delta_\al\psi=
i\pmatrix{\mbtau \mbB&0\cr 0&\mbtau\mbB}\al-
\pmatrix{0&\mbtau\mbD \phi_1\cr -\mbtau \mbD \phi_1&0}\al.}{mon1}
This system of equations is satisfied if
\eqnl{
\al=\pmatrix{\theta\cr \bar\zeta}\mtxt{with}
\theta\pm i\bar\zeta=0
\mtxt{and}
\mbB=\pm\mbD \phi_1}{mon3}
hold true. The first condition means that only an $\cN=1$ susy
is left intact and the second condition is the first order \bog
\emph{monopole equation}.\index{Bogomol'nyi equation}
Solutions of this equation describe magnetic monopoles.
To see this we rewrite the energy or mass \refs{seib31} 
for a static configuration $(A_i,\phi_1)$ as follows, 
\eqngrl{
M&=&\ft12 \int d\mbx\,\tr\left(\mbB^2+(\mbD \phi_1)^2\right)}
{&=&
\ft12 \int d\mbx\,\tr\left(\mbB\pm \mbD \phi_1\right)^2\mp
\int d\mbx\tr \left(\mbB\cdot\mbD \phi_1\right),}{mon5}
from which the celebrated \bog \emph{bound}\index{Bogomol'nyi bound}
follows at once,
\eqnl{
M\geq \Big| \int \tr (\mbB\cdot\mbD \phi_1)\Big|.}{mon7}
Using the \textsc{Bianchi} identity for the 'chromo-magnetic'
field, $\mbD\cdot\mbB=0$, the integrand can be written as
a divergence,
\eqnn{
\tr (\mbB\cdot\mbD \phi_1)=\nabla\,\tr(\mbB\, \phi_1).}
Inserting this into the \bog bound and using
the \textsc{Gauss} theorem yields
\eqnl{
M\geq \Big|\oint \tr(\mbB\,\phi_1)d\mbS\,\Big|.}{mon9}
So far we did not specify the gauge group. 
In what follows we assume that it is the group 
$SU(2)$ with generators
\eqnn{
T_a={1\ov \sqrt{2}}\tau_a,\mtxt{and}
f_{abc}=\eps_{abc}.}
A smooth configuration has finite energy \refs{mon5} if 
both $\mbB$ and $\mbD \phi_1$ vanish sufficiently
fast at large distances $r=\vert\mbx\vert$ from the monopole core.
If we further assume that the length of the scalar field 
tends to a constant value $v$,
\eqnl{
v^2=\lim_{\vert\mbx\vert \to\infty}\tr \phi_1^2(\mbx),}{mon11}
then the last surface integral in \refs{mon9} is to be interpreted as
\emph{magnetic flux}\index{magnetic flux} of the magnetic
field $\mbb$ belonging to unbroken
$U(1)$ gauge group. More precisely, far away from the monopole
the 'chromo-magnetic' field becomes \textsc{Abel}ian,
\eqnl{
\mbB\stackrel{\vert\mbx\vert\to\infty}{\longrightarrow}{1\ov v^2}\phi_1\tr(\phi_1\mbB)=
\hat\phi_1\mbb,}{mon13}
and the non-\abel monopole becomes an \abel \dirac monopole \cite{diracm}.
As a consequence the \bog bound simplifies to
\eqnl{
M\geq v\big\vert\Phi\big\vert,\mtxt{where} \Phi=\oint\mbb d\mbS}{mon15}
is the magnetic charge of the monopole.\index{magnetic charge!of monopole}
The last surface integral is over a large $2$-sphere
surrounding the pole. Since $\vert \mbD \phi_1\vert$ tends to
zero at spatial infinity it follows that
\eqnl{
v^2 A\stackrel{r\to\infty}{\longrightarrow}
\frac{i}{g}[\phi_1,d \phi_1]+\phi_1 a,\quad
a=\tr (\phi_1 A).}{mon17}
If we now compute the leading order behavior of the non-\abel field
strength we find the result \refs{mon13}, that is $F=\hat \phi_1 f$ with
\abel field strength
\eqnl{
f=\ha f_{ij}dx^i\wedge dx^j={i\ov gv^3}\tr (\phi_1 d\phi_1\wedge d\phi_1)+da.}{mon19}
One can show that far away from the monopole core the \textsc{Yang-Mills} 
equations for $F$ simplify to the \textsc{Maxwell} equations for $f$
\cite{monopoles}.

With the help of \refs{mon19} the surface integral
in \refs{mon17} defining the magnetic charge \index{magnetic charge} 
can be rewritten as
\eqnl{
\Phi=\oint f={1\ov g}
\oint \tr (i\hat\phi_1 d\hat\phi_1\wedge d\hat\phi_1).}{mon21}
The field $\hat\phi_1$ has a constant length $1$ 
in the asymptotic region and the map
\eqnl{
\hat x\longrightarrow \phi_1(\hat x)=\lim_{r\to\infty}
\phi_1(r\hat x)}{mon23}
is from the asymptotic sphere $S^2$ surrounding the
monopole to the unit sphere $S^2$ in the
internal space. The last surface integral in \refs{mon21}
is just $4\pi N/g$, where $N$ denotes the 
\emph{winding number}\index{winding number} of  
this map \cite{monopoles}. Thus we obtain the celebrated 
\dirac \emph{quantization condition}
for the magnetic charge of a monopole,\index{Dirac quantization condition}
\eqnl{
\Phi={4\pi\ov g} N.}{mon25}
It relates the gauge coupling to the charges
of the monopoles. 

Now the \bog bound \refs{mon15} for the mass can be written as
\eqnl{
M\geq {4\pi N\ov g}v.}{mon27}
This inequality becomes an equality for
\eqnl{
\mbB=\pm \mbD\phi_1,}{mon29}
that is for BPS-monopoles which leave half of the
supersymmetries intact. The monopole solutions of the 
\bog equations \refs{mon29} have been constructed in \cite{prasad}.
Two such monopoles neither repel nor attract each other.
This must be so since the \bog bound is attained 
independent of the distance between the constituent monopoles. 
When one changes the collective coordinates, and in particular
the distance between BPS monopoles, then the energy does not 
change and this implies that there is no interaction
between the constituents. This
behavior is typical for BPS states which saturate
a \bog type bound.
\section{Jackiw-Rebbi modes}
\index{Jackiw-Rebbi modes}
We have characterized BPS configurations as those fields
which are left invariant by part of the original supersymmetry.
Now we shall see that the other supersymmetry transformations, 
which do not leave the BPS configurations invariant, generate
the so-called fermionic \textsc{Jackiw-Rebbi} zero-modes \cite{jackiw}. 
These are time-independent solutions of the \dirac equation in 
the BPS background fields. Note that the \dirac equation in 
the \weyl gauge $A_0=0$ reads\index{Dirac equation!in monopole background}
\eqnl{
i\dot\psi=-i\al^iD_i\psi+g\gam^0[\phi_1,\psi]\mtxt{with} \al^i=\gam^0\gam^i.}{jr1}
For the static monopole background ($A_i,\phi_1)$
we may factorize the time dependence and arrive
at the time-independent \dirac equation
\eqnl{
E\psi=
-i\mbal \mbD\psi+g\gam^0[\phi_1,\psi]=H\psi.}{jr3}
Now we shall prove that $\delta_\al\psi$ is a zero-mode
of $H$ if $(A_i,\phi_1)$ satisfy the \bog equation.
For $\mbB=\mbD \phi_1$ the susy variation of the
\dirac spinor simplifies to
\eqnl{
\delta_\al\psi=\delta_\kappa\psi=\mbD\phi_1 \pmatrix{i\mbtau\,\kappa\cr
\mbtau\,\kappa},\mtxt{where} \alpha=\pmatrix{\theta\cr\bar\zeta},\quad
\kappa =\theta+i\bar\zeta.}{jr5}
Since $D_iD^i\phi_1=0$ we conclude that
\eqngrl{
\gam^j D_j\delta\psi&=&
D_jD_i \phi_1\pmatrix{\tau_j\tau_i\kappa\cr -i\tau_j\tau_i\kappa}
=\h \eps_{jik}[D_j,D_i]\phi_1\pmatrix{i\tau_k\kappa\cr \tau_k\kappa}}
{&=&
ig[B_k,\phi_1]\pmatrix{i\tau_k\,\kappa\cr
\tau_k\,\kappa}=-ig[\phi_1,\delta\psi],}{jr7}
which is equivalent to the \dirac equation \refs{jr3} with zero energy,
\eqnl{
H\delta_\kappa\psi=0.}{jr9}
Thus we have found an explicit 
expression for the \textsc{Jackiw-Rebbi} zero-modes
in the vicinity of magnetic BPS-monopoles. Supersymmetry not only
opens up an elegant way to characterize and construct
BPS-monopoles, it also gives the associated 
\textsc{Jackiw-Rebbi} modes. For the physical consequences
of these modes I refer to \cite{bruck}.
\section{$\cN=2$-SYM in Euclidean spacetime}
\index{$\cN=2$ SYM!in Euclidean spacetime}
Any supersymmetric theory contains both bosonic and fermionic fields and
the transition from \textsc{Lorentz}ian to \textsc{Euclid}ean
signature is not as straightforward as it is for purely
bosonic models. In spacetimes with \textsc{Euclid}ean
signature the gamma matrices must be hermitian and thus we choose
\eqnl{
(\gam_0,\gam_i)_M=(\gam_0,-i\gam_i)_E,}{cont1}
where the index $M$ refers to \textsc{Minkowski}an and
$E$ to \textsc{Euclid}ean spacetime.
From
\eqnn{
\cL_M\sim\ft12 \tr(F_{0i})^2-\ft14 \tr(F_{ij})^2+\dots}
we infer that the \lagan becomes negative definite,
irrespective whether we multiply the time coordinate
or the space coordinates with $i$. Because
\eqnn{
\cL_M=\ft12 \tr(D_0 \phi_1)^2-\ft12 \tr(D_i\phi_1)^2+\dots}
we prefer to continue the time coordinate 
for these terms to become negative definite as well. 
Hence we choose
\eqnl{
(\pa_0,\pa_i)_M=(i\pa_0,\pa_i)_E,}{cont3}
such that
\eqnn{
(A_0,A_i,F_{0i},F_{ij})_M=(iA_0,A_i,iF_{0i},F_{ij})_E\mtxt{and}
\di_M=i\di_E.}
In \textsc{Euclid}ean spacetime the \dirac term must have the form
\eqnn{
\pm i\psida\di\psi}
in order to be hermitian and $SO(4)$ invariant. This tells
us, that
\eqnl{
(\psi,\psib)_M=(\psi,i\psida)_E\,,\mtxt{such that}
(i\psib\di\psi)_M=-i(\psida\di\psi)_E}{cont5}
holds true. For the choices
\eqnl{(\phi_1)_M=(\phi_1)_E\mtxt{and}(\phi_2)_M=(\phi_2)_E}{cont7}
the \yuk interaction term
\eqnn{
-g\tr(\psib[\phi_1,\psi])=-ig\tr(\psida[\phi_1,\psi])}
becomes antihermitean in \textsc{Euclid}ean spacetime.
More generally, one can show that there are no consistent 
transformation rules of the coordinates, 
$\gamma$-matrices and fields such that the resulting 
\textsc{Euclid}ean action is hermitean and bounded from below
\cite{zuminoeucl}.

This is not as bad as it seems, since in the path
integral formulation of finite
temperature field theories the mass term in the 
\textsc{Euclid}ean \lagan $\cL_E=\psida(i\di+im)\psi+\dots$
is not hermitean either. Although 
the eigenvalues of $i\di+im$ are not real, they come in pairs 
$im\pm\lam$ such that the partition function stays real, 
since $(im+\lam)(im-\lam)$ is real. Actually, in a spontaneous 
broken phase the \yuk term may induce a mass term for the
fermions and we even expect an \emph{antihermitean} \yuk interaction
in \textsc{Euclid}ean spacetime.
Hence we may accept the transformation rules (\ref{cont1}-\ref{cont7})
in which case\index{Lagrangean!for Euclidean SYM}
\eqngrl{
\cL_E&=&
\ft14 \tr F_{\mu\nu}F^{\mu\nu}+
\ft12 \tr(D_\mu \phi_1)^2+\ft12 \tr(D_\mu \phi_2)^2
-\ft12 g^2\,\tr \left([\phi_1,\phi_2]^2\right)}
{&&\,+i\tr\psida\di\psi+ig\tr\left(\psida[\phi_1,\psi]\right)
+g \tr\left(\psida\gamf [\phi_2,\psi]\right).}{eucl1}
The indices are raised and lowered with the
\textsc{Euclid}ean metric $\delta_{\mu\nu}$. A problem
with this \lagan will be discussed below. 

In the chiral representation the hermitean $\gam$-matrices 
have the form
\eqnl{\gam_\mu=\pmatrix{0&\sigma_\mu\cr \sigma^\dagger_\mu&0}
\mtxt{with} (\sigma_\mu)=(\sigma_0,i\tau_i)=(\sigma^\mu),}{eucl3}
and the hermitean $\gamf$ is block diagonal $
\gamf=\gam_0\gam_1\gam_2\gam_3=\tau_3\otimes\sigma_0$. 
Note that the $\sigma_\mu$ in the \textsc{Euclid}ean
space differ from the one in \textsc{Minkowski}an 
spacetime. Since
\eqnl{
\gam_{i0}=i\tau_3\otimes\tau_i\mtxt{and}
\gam_{ij}=i\eps_{ijk}\sigma_0\otimes\tau_k}{eucl5}
the anti-hermitean generators of  \textsc{Euclid}ean 
spin rotations
\eqnl{
\gam_{\mu\nu}=\pmatrix{\sigma_{\mu\nu}&0\cr 0&\tsigma_{\mu\nu}}}{eucl7}
contain selfdual and antiselfdual objects,
\eqnl{
\tsigma_{\mu\nu}=\h\eps_{\mu\nu\al\beta}\tsigma_{\al\beta}\mtxt{and}
\sigma_{\mu\nu}=-\h\eps_{\mu\nu\al\beta}\sigma_{\al\beta}.}{eucl9}
In the \textsc{Euclid}ean space
there exists a \ym theory with $\cN=2$ extended 
supersymmetry. The field transformations 
follow from the transformations in the \textsc{Lorentz}ian model
with the appropriate replacements (\ref{cont1}-\ref{cont7}),
\eqngrrl{
\delta_\al A_\mu&=&i(\alda\gam_\mu\psi-\psida\gam_\mu\al)}
{\delta_\al \phi_1&=&-\alda\psi +\psida\al}
{\delta_\al \phi_2&=&i(\alda\gamf\psi-\psida\gamf\al).}{eucl11}
Here we encounter a first problem: The second formula shows
that $\phi_1$ should be \emph{antihermitean}. But with an 
antihermitean $\phi_1$ the term
$\tr (D\phi_1)^2$ and the potential term quartic in
the scalar fields become \emph{unbounded from below}.
The corresponding model is unstable, as has been
pointed out in \cite{zuminoeucl}.
For a recent work on this annoying stability and
hermiticity problem you may consult 
Belitsky et al. \cite{vandoren}.

The susy transformation of the \dirac spinor reads
\eqnl{
\delta_\al\psi=-iF^{\mu\nu}\Sigma_{\mu\nu}\al
-iD_\mu\phi\gam^\mu\al+g\,[\phi_1,\phi_2]\gamf \al,}
{eucl13}
where we have set $\phi_1+i\gamf\phi_2=\phi$.
To prove the invariance of $S$
one uses the general \textsc{Euclid}ean \textsc{Fierz} identity
\eqngrl{
4\psi\chida&=&-(\chida\psi)-\gam_{\mu}(\chida\gam^\mu\psi)
+\ft12\gam_{\mu\nu}(\chida\gam^{\mu\nu}\psi)}
{&&+\,\gamf\gam_\mu(\chida\gamf\gam^\mu\psi)
-\gamf(\chida\gamf\psi).}{eucl15}
The corresponding \noeth current takes the form
\eqngrl{
J^\mu&=&\alda\left\{i\gam_\nu(\,^*\!F^{\mu\nu}\gamf+F^{\mu\nu})
+ig[\phi_1,\phi_2]\gam^\mu\gamf-\di\phi\gam^\mu\right\}\psi}
{&+&\!\!\psida\left\{i\gam_\nu(\,^*\!F^{\mu\nu}\gamf-F^{\mu\nu})
+ig[\phi_1,\phi_2]\gam^\mu\gamf+\gam^\mu D_\nu\phi\gam^\nu\right\}\al,}{eucl17}
and is just the continuation of the current \refs{seib25}
in the \textsc{Seiberg-Witten} model.
\subsection{Instantons as BPS configurations}
\index{instantons}
Here we are interested in background
configurations which preserve half of the
\textsc{Euclid}ean supersymmetries. We assume that all 
fields with the exception of the gauge potential vanish.
This condition is preserved by supersymmetry
transformations if
\eqnl{
0=\delta_\al\psi=-iF^{\mu\nu}\Sigma_{\mu\nu}\al,\qquad
\al=\pmatrix{\theta\cr\zetab}}{inst1}
Decomposing this condition into its chiral parts yields
\eqnl{
F^{\mu\nu}\sigma_{\mu\nu}\theta=F^{\mu\nu}\tsigma_{\mu\nu}\zetab=0.}{inst3}
There are two ways to fulfill these conditions:
\begin{eqnarray}
i)\quad \theta=0&,&F^{\mu\nu}\tsigma_{\mu\nu}=0\label{inst5a}\\
ii)\quad \zetab=0&,&F^{\mu\nu}\sigma_{\mu\nu}=0.\label{inst5b}
\end{eqnarray}
In the first case only the $\zetab$-supersymmetry
survives and $F^{\mu\nu}$ is \emph{anti-selfdual} whereas in the
second case the $\theta$-supersymmetry survives
and $F^{\mu\nu}$ is \emph{selfdual}. As expected on general
ground these selfdual 
and anti-selfdual  gauge fields are \emph{solutions} of
the classical field equations which in the present case
reduce to the \textsc{Euclid}ean \textsc{Yang-Mills} equation
$D_\nu F^{\mu\nu}=0.$ As for the monopoles solutions we 
can interprete the instantons as BPS-states which
preserve half of the \textsc{Euclid}ean supersymmetry.

Now we pick a selfdual instanton configuration $\bar A_\mu$ 
and consider its supersymmetry variation. For a pure
gauge field background the only nontrivial variation is
\eqnl{
\delta_\al\psi=-iF^{\mu\nu}\Sigma_{\mu\nu}\al,}{inst7}
and since $\bar A_\mu$ is a classical solution we have
\eqnn{
S[\bar A_\mu]=S[\bar A_\mu,\delta_\al\psi]\sim
S[\bar A_\mu]+\big(\delta_\al\psida,i\di\,\delta_\al\psi\big)}
which indicates (only indicates since $i\di$ is indefinite) 
that $\delta_\al\psi$ is a zero-mode
of the \dirac operator. This is easily proven by acting
with $i\di$ on $\delta_\al\psi$,
\eqnl{
i\di\,\delta_\al\psi=i\gam_\nu D_\mu\left(\,^*\!F^{\mu\nu}\gamf
-F^{\mu\nu}\right)\al=0,}{inst9}
where we used the \textsc{Bianchi} identity and the
\ym equation. This shows that the susy variation $\delta_\al\psi$
is a zeromode of the \dirac operator. For a selfdual gauge field
the zero-mode is chiral
\eqnl{
\delta_\al\psi=-\ft12\pmatrix{0&0\cr 0&\tsigma^{\mu\nu} F_{\mu\nu}\zetab},}{inst11}
and for a anti-selfdual gauge field the mode has opposite
chirality.

Actually, from the \emph{index theorem} we infer that
in an $SU(2)$-instanton background with instanton number
\eqnn{
q={1\ov 8\pi^2}\int d^4x\, \tr (F\wedge F)}
the operator $i\di$ possesses at least
\eqnl{
\ft23 (2j+1)(j+1)j\cdot q}{inst13}
righthanded zero-modes in the spin-$j$ representation. 
For $\cN=2$ SYM the fermions transform according to the adjoint 
representation such that there are at least $4q$ zero modes.

\subsection{Small fluctuations about instantons}
Now we wish to relate the various fluctuation fields
about selfdual instantons \cite{Vecchia}. We start with 
perturbing an arbitrary background field by small fluctuations
\eqnl{
A_\mu=\bar A_\mu+a_\mu,\quad
\phi_1=\bar \phi_1+\delta\phi_1,\quad
\phi_2=\bar \phi_2+\delta\phi_2.}{fluct1}
The \dirac field is considered as a fluctuation itself.
We perform the Taylor expansion of the (unstable) 
\textsc{Euclid}ean action about the bosonic background,
\eqnl{
S=S_0+S_1+S_2+\dots,}{fluct3}
where $S_i$ is of order $i$ in the fluctuation fields.
The first term $S_1$ yields the \textsc{Euler-Lagrange}
equations and vanishes for backgrounds which solve the 
field equations. For general backgrounds the expression for $S_2$ 
is rather complicated. But for vanishing $\bar\phi_i$ 
and without gauge fixing (see below) it 
becomes rather simple
\eqnl{
S_2=\int d^ 4x\left(a_\mu M_{\mu\nu}a_\nu
+\delta\phi_1 M \delta\phi_1+\delta\phi_2 M \delta\phi_2+\psida M_\psi\psi\right),}{fluct5}
with fluctuation operators
\eqngrl{
M_{\mu\nu}&=&-D^2\delta_{\mu\nu}+D_\mu D_\nu+2ig\; \hbox{ad}\,(F_{\mu\nu}),}
{M\!&=&\!-D^2\mtxt{and}M_\psi=i\di.}{fluct7}
All covariant derivatives act on fields in the
adjoint representation. 

In a one-loop approximation
to the partition function one needs the eigenvalues
of these fluctuation operators,
\begin{eqnarray}
M_{\mu\nu}a_\nu&=&\lam^2 a_\nu\label{fluct9a}\\
M_i \delta\phi_i&=&\lam^2 \delta\phi_i\label{fluct9b}\\
M_\psi \psi&=&\lam\psi.\label{fluct9c}
\end{eqnarray}
Let us assume that $\psi$ is an eigenmode of the
\dirac operator, $M_\psi\psi=i\di\psi=\lam\psi$.
For selfdual gauge fields the squared \dirac operator simplifies 
as follows,
\eqnl{
\di^2=
D^2+\Sigma^{\mu\nu}\,\hbox{ad}(F_{\mu\nu})=
D^2-\ft{i}{2}\pmatrix{0&0\cr 0&\tsigma^{\mu\nu} \hbox{ad}(F_{\mu\nu})}.}{fluct13}
We see that the chiral spinor $\varphi$ in the decomposition
\eqnl{
\psi=\pmatrix{\varphi\cr\chib}}{fluct15}
fulfills a \weyl type equation without \textsc{Pauli} term,
\eqnn{
-D^2\varphi=\lam^2\varphi.}
This equation is identical to the eigenvalue equation for the
fluctuations $\delta\phi_i$ of the scalar and pseudoscalar 
fields. Hence every left-handed eigenmode $\varphi$ of 
the squared \dirac operator transforms into an scalar and
pseudoscalar eigenmode of $D^2$ with the same eigenvalue,
\eqnl{
\delta_\al\phi_1
=-\theta^\dagger\varphi+\varphi^\dagger\theta\mtxt{and}
\delta_\al\varphi_2=i\theta^\dagger\varphi-i\varphi^\dagger\theta
.}{fluct17}
As in \refs{inst1} we denoted the chiral parts 
of $\al$ by $\theta$ and $\zetab$.

The same procedure applies to the eigenvalue
equation of the vector fluctuations in an selfdual
instanton background. 
We start with a righthanded fermionic eigenmode of 
$-\di^2$ which fulfills
\eqnn{
\left(-D^2+\ft{i}{2}\tsigma^{\mu\nu}
\hbox{ad}(F_{\mu\nu})\right)\chib=\lam^2\chib.}
We multiply this equation from the left with $\sigma_\nu$
and use the identity
\eqnl{
\sigma_\rho \tsigma_{\mu\nu}=\delta_{\rho\mu}\sigma_\nu
-\delta_{\rho\nu}\sigma_\mu
+\eps_{\mu\nu\rho\sigma}\sigma_\sigma,}{fluct19}
together with the selfduality condition $^*\!F_{\nu\mu}=F_{\mu\nu}$.
This leads to
\eqnl{
\left(-D^2\delta_{\mu\nu}+2i\,\hbox{ad}(F_{\mu\nu})\right)
\,\sigma_\nu\chib=\lam^2\sigma_\mu\chib.}{fluct21}
Acting with $D_\mu$ on this equation and using $[D_\mu,D_\nu]=-i\hbox{ad}(F_{\mu\nu})$ we find
\eqnl{
-D^2(\sigma^\mu D_\mu\chib)=\lam^2(\sigma^\mu D_\mu\chib).}{fluct23}
Now we take the vector field which is gotten as 
supersymmetry transformation of a righthanded eigenmode,
\eqnl{
a_\mu=i\theta^\dagger\sigma_\mu\chib+\hbox{h.c}.}{fluct25}
For eigenvalue $\lam\neq 0$ this vector field has a non-vanishing
divergence,
\eqnl{
D_\mu a_\mu=
\lam\theta^\dagger\varphi+\hbox{h.c.}
.}{fluct27}
It can be used to construct the following
'source free' vector field
\eqnl{
b_\mu=a_\mu+{1\ov\lam^2}D_\mu (D_\nu a_\nu)}{fluct29}
which satisfies the so-called \emph{background gauge condition}
\index{background gauge condition}
\eqnl{
D_\mu b_\mu=0.}{fluct31}
The difference between $b_\mu$ and $a_\mu$ is
an infinitesimal gauge transformation. Actually
the proof of \refs{fluct31} is simple when one recalls 
\refs{fluct23},
\eqngr{
D_\mu b_\mu&=&D_\mu a_\mu+{1\ov\lam^2}
D^2\left(i\theta^\dagger (\sigma D\chib)+\hbox{h.c.}\right)}
{&=&D_\mu a_\mu-i\left(\theta^\dagger (\sigma D\chib)
+\hbox{h.c}\right)=D_\mu a_\mu-D_\mu a_\mu=0.}
We return to the fluctuation operator $M_{\mu\nu}$ for the
gauge bosons in \refs{fluct7}. It is just the
sum of the operator on the left in \refs{fluct21},
plus $D_\mu D_\nu$ which annihilates $b_\nu$.
We conclude that $b_\nu$ is an eigenmode of the fluctuation 
operator with eigenvalues $\lam^2$,
\eqnl{
M_{\mu\nu}b_\nu=\lam^2 b_\mu,\qquad D_\mu b_\mu=0.}{fluct33}
Thus we have diagonalized $M_{\mu\nu}$ on fluctuations 
fulfilling the background gauge condition \refs{fluct31}.

We summarize our findings: the fluctuation
operators $M,M_{\mu\nu}$ for the scalars and
gauge field fluctuations (in the background gauge)
have the same spectra as the fluctuation operator 
$-\di^2$ for the \dirac field.
The eigenmodes are related by supersymmetry. 
This important result can be used to calculate the
one-loop partition function of the \textsc{Euclid}ean
\ym theory with $\cN=2$ supersymmetry.

\section{One-loop $\beta$-function}
In this section we shall calculate the 
one-loop $\beta$-function and anomalous dimension 
of $A_\mu$ for the $\cN=2$ SYM. We shall use the
results in \cite{wies} which relate the finite size 
scaling of the partition function to the $\beta$-functions
and anomalous dimensions.  After our previous fluctuation
analysis it is natural to use the background gauge fixing. 
Hence we add a gauge fixing term $\ft12 (D_\mu a_\mu)^2$
to the \textsc{Lagrange}an such that the eigenvalue 
equation for the fluctuations $a_\mu$ in \refs{fluct9a} is 
modified to
\eqnl{
K_{\mu\nu}a_\nu=
(M_{\mu\nu}-D_\mu D_\nu)a_\nu=\lam^2 a_\mu.}{pf1}
The general formula in \cite{wies,osborn} for the \emph{one-loop} 
generating functional in the $q$-\emph{instanton sector} and spacetime
volume $V$ simplifies for the $\cN=2$ SYM as follows,
\eqngrr{
Z_q(V,g,\eta)&=&e^{W_q(V,g)}=
{1\ov \cal{N}}\int {\cal D} (a_\mu,\psida,\psi,\phi_1,\phi_2)
\,e^{-S+\int (\psida\eta+\eta^\dagger\psi)}}
{&=&e^{-S(\bar A_\mu)}\left({2g^2\pi\ov V}\right)^{d_H/2}
{1\ov V_H}\int\prod_1^p d\gam_r (\det J)^{1/2}
{{\detpr}M_\psi\, \detpr M_{gh} \ov
\detpr M\det'^{1/2}(-K_{\mu\nu})}}
{&&\hskip 3.3cm\times \prod _n(\eta^\dagger,\psi_n)(\psi_n^\dagger,\eta)
\exp\left(-\int \eta^\dagger G^\pr\eta\right).}
We expanded the action about a classical gauge
field background such that the determinants depend
on this background field.
In the non-perturbative sectors with non-vanishing
instanton number the fluctuation operators, and in
particular $M_\psi$, may admit zero-modes. 
These zeromodes must be omitted when one computes the 
determinants and the product of the non-zero eigenvalues 
is denoted by $\det'$ in the above formula.
Since the fermions are of \dirac type we get the determinant of the 
\dirac operator. For \maj fermions we would get the square root
of this determinant.
For the background gauge fixing the fluctuation operator
for the ghosts is $M_{gh}=-D^2$ and coincides with
the operator $M$ for the scalar fields. Each of the
two scalar fields yields a square root of $\det'M$ in
the denominator. Also we have used the dimension $d_H$ and 
volume $V_H$ of the stability group $H$ which commutes 
with the $su(2)$-algebra defined by the instanton \cite{osborn}. 
For the gauge group $SU(2)$ we have $d_H=0$. The fluctuation operator
$K_{\mu\nu}$ for the gauge bosons may possess $p$ additional zero-modes
arising from the variation of the collective parameters $\{\gam_r\}$.
$J$ denotes the Jacobian when one converts the $p$ expansion
parameters (in the expansion of the gauge field) into
collective parameters.

In \cite{wies} it was shown that the $\beta$-functions
and anomalous dimensions are the same in 
\emph{all instanton sectors} 
and hence it suffices to consider the 
perturbative $q=0$ sector and $\eta=0$. Hence
we skip the index $q$ in what follows. 
To extract the $\beta$-function for the gauge
coupling we keep an arbitrary background gauge
field.
We use the \emph{zeta-function}\index{zeta function} 
regularization to 'calculate' the determinant of an 
selfadjoint and non-negative operator $A$,
\eqnl{
\log \det A=-{d\ov ds}\,\zeta_A(s)\big\vert_{s=0},\quad
\zeta_A(s)=\sum_{\lam_n>0} \lam_n^{-s}.}{pf3}
The so defined determinant has a simple scaling
property,
\eqnl{
\log\det \left({1\ov\lam}A\right)=\log\det A-\log\lam\cdot \zeta_A(0).}{pf5}
Now we rescale the quantization volume and background field as
\eqnl{
V\longrightarrow \tilde V=\lam V\mtxt{and}
A_\mu(x)\longrightarrow \tilde A_\mu(\tilde x)
=\lam^{-1}A_\mu(x),}{pf7}
where $\tilde x=\lam x$, and compute how the 
\textsc{Schwinger} \emph{functional}
\index{Schwinger functional!of $\cN=2$ SYM}
changes. We take the general results in \cite{wies}, applied
to the present situation, and find
\eqnl{
W[\lam V,\tilde A,g]=W(V,A,g)+{\log\lam\ov 16\pi^2}\int 
\tr\left(a^{A_\mu}_2(x)-a^\psi_2(x)\right),}{pf9}
where the second \textsc{Seeley-deWitt} coefficients
\index{Seeley-deWitt coefficient} of the
various fluctuation operators appeared. 
We used that the scalar and ghost operators
are equal such that their contributions cancel. Setting
\eqnn{
X=g^2\,\tr_{\!\rm A}F^{\mu\nu}F_{\mu\nu},}
where $\tr_{\!\rm A}$ is
the trace in the adjoint representation, the needed
coefficients are \cite{gilkey}
\eqnl{
a_2^{A_\mu}(x)=-\ft53 X,\quad a_2^{gh}={1\ov 12}X\mtxt{and}
a_2^{\psi}(x)=-\ft23  X.}{pf11}
Thus we obtain
\eqnl{
W[\lam V,\tilde A,g]=
W[V,A,g]-{\log\lam\ov 16\pi^2}\,C_{\rm A}\,g^2\int
F_{\mu\nu}^a F^{\mu\nu}_a,}{pf13}
where $C_{\rm A}$ is the second-order \textsc{Casimir} of the
adjoint representation. This then implies the following
scaling law for the \emph{effective action}, that is the
generating functional of the one-particle irreducible
\textsc{Feynman} graphs \cite{wies},
\index{effective action of $\cN=2$ SYM}
\eqnl{
\Gamma[\lam V,\tilde A,g]={1\ov 4g^2}\,Z_3
\int F^{\mu\nu}_aF_{\mu\nu}^a\mtxt{with} Z_3=1-{\log\lam \ov 4\pi^2}
\,C_{\rm A}\,g^2.}{pf15}
The effective action stays invariant if the 
background field and gauge coupling scale in
a non-canonical way,
\eqnl{
\Gamma\left[\lam V,\tilde A,g\right]=
\Gamma\left[V,Z_3^{1/2}A,Z_3^{-1/2}g\right]}{pf17}
For the gauge group $SU(2)$ the second order \textsc{Casimir}
in the adjoint representation is $2$ and the coupling runs with
the inverse size $\mu$ of the quantization volume as
\eqnl{
g^2(\mu)={2\pi^2 g^2\ov 2\pi^2+g^2\log\mu},\mtxt{where}
\mu={1\ov\lam}.}{pf19}
Hence the $\beta$\emph{-function} and \emph{anomalous dimension} 
of the gauge fields are
\index{$\beta$-function! of $\cN=2$ SYM}
\index{anomalous dim. of $\cN=2$ SYM}
\begin{eqnarray}
\beta(g)&=&\mu{\pa\ov\pa\mu}g(\mu)=-{1\ov 4\pi^2}g^3(\mu)\label{pf21a}\\
\gam_A(g)&=&\mu{\pa\ov\pa\mu}\log Z_3={1\ov 2\pi^2}g^2(\mu)\label{pf21b}.
\end{eqnarray}
The theory is asymptotically free, similarly as $QCD$.
Actually, it has been shown in \cite{seibertbeta} that there
is only the one-loop contributions \refs{pf21a}
to the perturbative $\beta$-function 
of $\cN=2$ SYM. The exact non-perturbative $\beta$-function
has been calculated in \cite{bonelli}.

I leave it as an exercise to calculate the $\beta$-function
and anomalous dimension for the \ym theory with $\cN=1$
supersymmetry. Again you should find that the model is asymptotically
free.

\chapter{$N=4$ Super-Yang-Mills Theory}
\index{$\cN=4$ SYM}
The largest supersymmetry that can be represented on
a multiplet with spins $\leq 1$ is the one with
four \maj supercharges and for this reason the $\cN=4$ model 
\cite{brink77} is called \textit{maximally extended}.
The renewed interest in this model is twofold. On
one hand, it is expected to be $S$-dual and the \emph{complete effective 
action} (including all instanton and
anti-instanton effects) should organize into an 
$SL(2,\Z)$ invariant expression. On the other hand, not 
unrelated to the previous, it appears in the 
celebrated AdS/CFT correspondence
\cite{Maldacena,aharony00}.

On-shell the theory contains one gauge field $A_\mu$,
four \maj spinors\footnote{or equivalently $4$ \textsc{Weyl} 
spinors which may be grouped into two \dirac spinors.} and
six scalar fields.  All fields transform according
to the adjoint representation of the gauge group.
\section{Scale invariance in one-loop}
Without knowing the action we can already
calculate the one-loop $\beta$-function of this theory.
Under scale transformations the Schwinger functional
changes as
\eqngr{
W[\lam V,\tilde A,g]&=&W(V,A,g)+{\log\lam\ov 16\pi^2}
\int \tr\left(
6a^{\phi}_2(x)-2a^{gh}_2(x)-2a^\psi_2(x)+a^{A_\mu}_2(x)\right)}
{&=&W(V,A,g)+{\log\lam\ov 16\pi^2}\int 
\tr\left(4a_2^\phi(x)-2a^\psi_2(x)+a^{A_\mu}_2(x)\right),}
where the Seeley-deWitt coefficients have been given
in \refs{pf11}. We have used the field content of the
theory and in particular that all fields transform under 
the adjoint representation of the gauge group. 
We have also used that in the background gauge
the scalars and ghosts have the same fluctuation
operators. The sum of the \textsc{Seeley-deWitt}
coefficients is zero,
\eqnn{
\left(\frac{4}{12}+\frac{4}{3}-\frac{5}{3}\right)X=0,\quad
X=g^2\tr_{\rm A}F^{\mu\nu}F_{\mu\nu},}
and the $1$-loop action
is scale invariant. Hence the $\beta$-function and
wave function renormalization are trivial,
\index{$\beta$-function!of $\cN=4$ SYM}
\eqnl{
\beta(g)=0\mtxt{and}Z_3(g)=1.}{nfourbeta1}
In \cite{caswell} it was found that the
$\beta$-function remains zero up to three loops
and that therefore there were no divergent graphs at all
to that order. This led everyone to suspect that the theory
may be a finite field theoretical model in four space-time
dimensions and arguments were put forward to proof the
finiteness to all orders in perturbation theory.
Some arguments are based on the relation between the trace
anomaly in the energy-momentum tensor, the $\beta$-function
and conformal invariance, other arguments used the explicit matching
of bosonic and fermionic counting in the light-cone gauge and
yet other arguments were based on the non-renormalisation theorem
and the background gauge. The vanishing of the $\beta$-function
to all orders has been shown in \cite{Nfbeta}.
\section{Kaluza-Klein reduction}
\index{Kaluza-Klein reduction}
Models with extended supersymmetry are intimately
related to \lagan field theories in higher dimensions.
The idea of employing higher dimensions to construct
$4$-dimensional models with extended supersymmetry has
been pioneered by \textsc{J. Scherk} and co-workers \cite{brink77}.
We shall see how the maximally extended gauge model fits 
into a world in higher dimensions. We derive this 
theory by a sort of \textsc{Kaluza-Klein} reduction
of a $\cN=1$ super-\ym theory in higher dimensions.
After compactification all but four components of
the vector potential in higher dimensions become 
scalar fields. Since the $\cN=4$ SYM theory has $6$ 
scalar fields and a gauge field with $4$ components
we must reduce a gauge theory with a $4+6=10$-component 
gauge potential, that is a gauge theory in $10$
dimensions.

$\cN=1$ SYM theory in $10$ dimensions has the gauge
invariant action
\eqnl{
S=\int d^{10}x\left(-\ft{1}{4} \tr F_{mn}F^{mn}+\ft{i}{2}
\tr\Psib\di\Psi\right)}{nfein1}
and is invariant under the \emph{on-shell} susy transformations 
\eqnl{
\delta_\al A_m=i\alb\Gam_m\Psi\mtxt{and}
\delta_\al\Psi=iF^{mn}\Sigma_{mn}\al.}{nfein3}
The only difference to the $4$\emph{-dimensional}
transformations $(\ref{ag5},\ref{ag6})$ (with
$\cG=0$) is that $\psi$ and $\al$ are
\textsc{Majorana-Weyl} spinors.

The proof that the $10$-dimensional action
is invariant is very similar to the corresponding
proof for the $\cN=1$ theory in $4$ dimensions. 
On the way  one needs to show that the quartic term
\eqnl{
\ft{i}{2}\tr\big(\Psib\Gam^m [(\alb\Gam_m\Psi),\Psi]\big)=
\ft12 f_{abc}(\Psib^a\Gam^m\Psi^b)(\alb\Gam_m\Psi^c)}{nfein5}
vanishes. Again one applies a suitable \textsc{Fierz} 
identity. In $10$ dimensions the general identity is 
more complicated than in $4$ dimensions and is given 
in \refs{ap71} in the appendix. After a lengthy calculation
on finds the following \textsc{Noether} current
\eqnl{
J^m=
-i\tr\left(F^{mn}\Gam_n\Psi\right)
+\ft{i}{2}\tr\left(F_{pq}\Gam^{pqm}\Psi\right)}{nfein7}

\subsection{Reduction of Yang-Mills term}
In $10$ spacetime dimensions a gauge potential $A_m$ and gauge 
coupling constant $\tilde g$ have length-dimensions
\eqnl{
[A_m]=L^{-4},\quad \tilde g=L^3
\Longrightarrow[\tilde g A_m]=L^{-1}.}{nfour1}
First we perform the \kk reduction of the \ym action
\eqnl{
S_{YM}=-\ft{1}{4}\int d^{10}x\, \tr F_{mn}F^{mn}}{nfour3}
on $\R^4\times T^6$. As internal space we choose the 
torus $T^6$ with volume $\tilde V$ and set
\eqnl{
\tilde V^{\ha}A_m=(A_\mu,\Phi_a),\qquad m=0,\dots,9,\quad \mu=0,\dots,4,
\quad a=1,\dots,6.}{nfour5}
We assume that \emph{all fields} are independent of the 
\emph{internal coordinates} $x^4,\dots,x^9$. This assumption may be 
justified on dynamical grounds for internal spaces with tiny
volumes. On length scales much larger then the typical 
size of the internal space the modes which depend
on the internal coordinates are not excited.
With this crucial assumption 
the $10$-dimensional \ym action reduces to
the action of a $4$-dimensional 
\textsc{Yang-Mills-Higgs} model
\eqngrl{
S_{YM}&\longrightarrow &S_{YMH}=\int d^4x\, \cL_{YMH}}
{\cL_{YMH}\hskip -3mm&=&\hskip-3mm -\ft14 \tr F_{\mu\nu}F^{\mu\nu}
+\ft12\sum_a \tr D^\mu \Phi_a D_\mu \Phi_a
+\ft14 g^2
\sum_{ab}\tr [\Phi_a,\Phi_b]^2.}{nfour9}
We used $\Phi_a=-\Phi^a$ and that the dimensionful 
coupling constant $\tilde g$ in $10$ dimensions and 
the \emph{dimensionless coupling constant} $g$ in $4$ dimensions
are related as
\eqnl{
\tilde g^2=\tilde Vg^2.}{nfour11}
Not unexpected we got the action of a four-dimensional
\textsc{Yang-Mills-Higgs} theory with $6$ \textsc{Higgs} 
fields in the adjoint representation. Typical
is the quartic potential $\sim \sum \tr [\Phi_a,\Phi_b]^2$
with many flat directions.

\subsection{Spinors in $10$ dimensions}
In $10$ dimensions a \dirac spinor has $32$ complex components.
But the $\cN=4$ SYM theory in $4$ dimensions has $4$ \maj
spinors which together have only $16$ real components. Hence we must assume 
that the spinors in $10$ dimensions are both \weyl and 
\textsc{Majorana}. Such spinors do exist, see chapter \ref{chap:3}.

We start with Dirac matrices $\gammu$ in $4$-dimensional
\mink space-time and give an explicit realization for the matrices
$\Gamma^m, m=0,\dots,9$ in $10$ dimensions. We make the ansatz
\eqnl{
\Gamma_\mu=\Delta\ot\gam_\mu,\quad \Gamma_{3+a}=\Delta_a\ot\gamf,\qquad
\mu=0,1,2,3,\quad a=1,\dots,6,}{nfour13}
with $8\times 8$ matrices $\Delta$ and $\Delta_a$. They
must satisfy
\eqnl{
\Delta^2=\id_6,\quad [\Delta,\Delta_a]=0\mtxt{and}
\{\Delta_a,\Delta_b\}=-2\delta_{ab}\id_8}{nfour15}
in order for 
\eqnl{
\{\Gamma_m,\Gamma_n\}=2\eta_{mn},\qquad
\eta=\hbox{diag}(1,-1,\dots,-1)}{nfour17} 
 to hold true. Because $\Gamma^0$ is hermitian and 
the $\Gamma^{m>0}$ antihermitian we conclude that
$\Delta$ is hermitian and $\Delta_a$ antihermitian.

Since $\Delta$ commutes with all matrices we may choose
it to be the identity,
\eqnl{
\Delta=\id_8.}{nfour19}
Note that the hermitian $i\Delta_a$ generate the
\textsc{Euclid}ean \cliff algebra in $6$ dimensions and that 
the $[\Delta_a,\Delta_b]$ generate the group $Spin(6)$.
In $6$ \textsc{Euclid}ean dimensions there is a \maj 
representation with \emph{real and antisymmetric} $\Delta_a$. 
An explicit representation is given in the appendix.
 
We choose a \maj representation in $4$-dimensions
such that the $\gam_\mu$ and $\gamf=-i\gam_0\gam_1\gam_2\gam_3$ 
are imaginary. As charge conjugation matrix in $4$ dimensions 
we take $\cC_4=-\gam_0$. The real $\Delta_a$ lead to 
imaginary $\Gamma_m$ such that the charge 
conjugation matrix in $10$ dimensions is simply
\eqnl{
\cC_{10}=-\Gamma_0=\id_8\otimes \cC_4
\mtxt{with}\cC_4=-\gam_0.}{nfour21} 
For the chiral projections we need
$\Gamma_{11}=-\Gamma_0\cdots\Gamma_9=\Gamma_{11}^\dagger$  
which takes the form
\eqnl{
\Gamma_{11}=\Gamma_*\ot\gamf,\qquad \Gam_*=
-i\Delta_1\cdots\Delta_6,\qquad \Gam_*^\dagger=\Gam_*=-\Gam_*^T.}{nfour23}
A spinor $\Psi=\xi\ot\psi$ is \maj in $10$ dimensions
if $\xi$ is real and $\psi$ a \maj spinor in $4$ dimensions.
Hence an arbitrary \maj \emph{spinor} has the expansion
\index{Majorana spinor!in 10 dimensions}
\eqnl{
\Psi=\sum_{r=1}^8 \mbe_r\otimes \psi_r,}{nfour25}
where the $\mbe_r$ form a basis of $\R^8$ and the $\psi_r$
are \maj spinors in $4$ dimensions.
A spinor has positive chirality if
\eqnl{
\Psi=\Gamma_{11}\Psi=(\Gam_*\ot\gamf)\Psi.}{nfour27}
To characterize \weyl spinors we expand the first factor
of $\xi\ot\psi$ in eigenvectors of the imaginary and
hermitian $\Gam_*$. 
Let $\mbg_1,\dots,\mbg_4\in\C^8$ be the orthonormal (and necessarily
complex) eigenvectors of  $\Gamma_*$ with eigenvalue $1$. 
Together with the complex conjugate eigenvectors 
$\mbg_1^*,\dots,\mbg_4^*$ with eigenvalue $-1$ they
form a basis of $\C^8$. A spinor with \emph{positive chirality} 
has the expansion
\eqnl{
\Psi=\sum_{p=1}^4\left(\mbg_p\ot \psi_p^+
+\mbg_p^*\ot \psi_p^-\right),\mtxt{where} \gamf\psi^\pm=\pm\psi^\pm,}
{nfour29}
that is $\psi^\pm$ are the chiral parts of $\psi$. 
A \textsc{Majorana-Weyl} \emph{spinor} in $10$ dimensions has
at the same time the expansion \refs{nfour23} 
and thus has the form
\eqnl{
\Psi=\sum_{p=1}^4 \left(\mbg_p\ot \psi_p^++\mbg^*_p\ot \psi_p^-\right)}{nfour31}
with four \maj spinors $\psi_p=\psi_p^++\psi_p^-$ in $4$ dimensions. 
The important formula \refs{nfour31} assigns to each
\textsc{Majorana-Weyl} spinor in $10$ dimensions
four \maj spinors in $4$ dimensions and vice versa.
It is the analog of \refs{nfour5} for spinor
fields.
\subsection{Reduction of the Dirac term}
In ten and four dimensions a spinor field has the dimension
\eqnl{
[\Psi]=L^{-9/2}\mtxt{and}
[\psi]=L^{-3/2},}{redd1}
respectively. In the general expansion \refs{nfour31}
for a \textsc{Majorana-Weyl} spinor in $10$ dimensions 
we rescale the spinor such that the $\psi_p$ in
\eqnl{
\tilde V^{\ha}\Psi=\sum_p \left(\mbg_p\ot \psi_p^++
\mbg_p^*\ot \psi_p^-\right)}{redd3}
have correct length-dimension. In the
reduction from ten to four dimensons 
we assume that the spinor field does 
not depend on the internal coordinates and with 
this assumption and (\ref{nfour5},\ref{nfour11}) 
we obtain
\eqngrl{
\tilde V^{\ha}D_\mu \Psi&=&
\sum_p \left(\mbg_p\ot D_\mu \psi_p^+
+\mbg_p^*\ot D_\mu \psi_p^-\right),}
{\tilde V^{\ha}D_{3+a}\Psi&=&-ig
\sum_p \left(\mbg_p\ot [\Phi_a,\psi_p^+]
+\mbg_p^*\ot [\Phi_a,\psi_p^-]\right),}{redd5}
where the covariant derivative $D_\mu=\pa_\mu-ig[A_\mu,.\,]$ 
contains the gauge potential and
coupling constant in $4$ dimensions.
Now we are ready to dimensionally reduce the \dirac 
term from ten to four dimensions. One obtains the \dirac 
terms for the four 'gluino' fields in $4$ dimensions plus 
a particular \textsc{Yukawa} interaction between 
the scalar fields $\Phi_a$ and the 'gluino' fields $\psi_p$. 
This interaction contains the non-zero matrix elements 
of $\Delta^a\equiv\Delta_a$ in the eigenbasis of $\Gam_*$,
\eqnl{
\Delta^a_{pq}=(\mbg_p^*,\Delta^a \mbg_q)\mtxt{and}
\bar\Delta^a_{pq}=(\mbg_p,\Delta^a \mbg_q^*).}{redd7}
The other matrix elements vanish since the
hermitian $\Gam_*$ anticommutes with $\Delta^a$.
The matrix elements $\Delta^a_{pq}$ define a $4$-dimensional
matrix and completely determine the $8$-dimensional
matrix $\Delta^a$. 
Collecting the various terms we end up with
the \lagan density for the $\cN=4$ 
supersymmetric gauge theory in $4$ dimensions,
\index{Lagrangean!of $\cN=4$ SYM}
\index{action!of $\cN=4$ SYM}
\eqngrl{
\cL&=&-\ft14 \tr F_{\mu\nu}F^{\mu\nu}
+\ft12 \tr D^\mu \Phi_a D_\mu \Phi_a+\ft14 g^2\tr [\Phi_a,\Phi_b]^2}
{&&+\ft12\tr\left(i\psib_p\di\psi_p+g \Delta^a_{pq}\psib_p
[\Phi_a,\psi_q^+]
-g\bar\Delta^{a}_{pq}\psib_p [\Phi_a,\psi_q^-]\right),}{redd9}
where $a=1,\dots,6$ and $p=1,\dots,4$. The \textsc{Clebsch-Gordan}
type coefficients $\Delta^a_{pq}$ are needed for the
density \refs{redd9} to be invariant under
$R$-transformations.
\subsection{$R$-symmetry}
Vector and spinor fields in $10$ dimensions transform under
\lor transformations in $10$ dimensions as
\eqnl{
A(x)\longrightarrow \tilde\Lambda A(\tilde\Lambda^{-1}x)\mtxt{and}
\Psi(x)\longrightarrow \tilde S\Psi(\tilde\Lambda^{-1}x),}{rsym1}
where the spin and \lor transformation are related via
\eqnl{
\tilde S^{-1}\Gamma^m \tilde S=\tilde\Lambda^m_{\;\;n}\Gamma^n.}{rsym3}
The spin rotations are generated by the matrices $\Gamma_{mn}=
\ft12 [\Gamma_m,\Gamma_n]$ and the explicit relation
in \refs{rsym3} reads
\eqnl{
\tilde S=\exp\left(\ft12 \om_{mn}\Gamma^{mn}\right)\Longrightarrow
(\tilde\Lambda)^m_{\;n}=\big(e^\om)^m_{\;\;n}.}{rsym5}
The \lagan density is a scalar field such that the
action is \lor invariant. In the dimensional reduction to $\R^4$ we 
demanded the fields not to depend on the internal coordinates.
This condition is not compatible with all \lor transformations
in ten dimensions. Only \lor transformations
which do not mix coordinates on $\R^4$ with 
internal coordinates are admitted. They have the form
\eqnl{
\tilde \Lambda\longrightarrow \pmatrix{\Lambda&0\cr 0&R}
\in SO(1,3)\times SO(6)\subset SO(1,9),}{rsym6}
where $\Lambda$ is a \lor transformation in $4$ dimensions
and $R\in SO(6)$\footnote{actually, only $\det\Lambda\cdot\det R=1$
is required.}.
With our choice for the $\Gamma_m$ in \refs{nfour13}
the generators of the corresponding spin transformations 
in $10$ dimensions read
\eqnl{
\Gamma_{\mu\nu}=\id_8\ot \gam_{\mu\nu}\quad,\quad
\Gamma_{3+a,3+b}=\Delta_{ab}\ot\id_4}{rsym7}
The $\Delta_{ab}$ generate the $Spin(6)\sim SU(4)$ subgroup 
of $Spin(1,9)$. Since the $\Gamma_{\mu\nu}$ act trivially on the
first factor of $\xi\ot\psi$ and
the other generators in \refs{rsym7} act trivially
on the second factor, the admitted spin rotations act as
\eqnl{
\tilde S(\xi\otimes\psi)=(S_6\,\xi)\ot (S\psi),}{rsym9}
where $S_6$ and $S$ define $SO(6)$ and \lor transformations 
according to
\eqnl{
S_6^{-1}\Delta^aS_6=R^a_{\;b}\Delta^b\mtxt{and} S^{-1}\gam^\mu
S=\Lambda^\mu_{\;\,\nu}\gam^\nu.}{rsym11}
It follows that the $SO(6)$-factor of the admitted \lor 
transformations act as global and compact
\emph{internal} $SU(4)$ $R$-symmetry,
\eqnl{
\Phi_a(x)\longrightarrow R_a^{\;\,b}\Phi_b(x),\quad
\Psi(x)\longrightarrow S_6\xi\ot\psi,\quad A_\mu(x)\longrightarrow A_\mu(x)}
{rsym13}
and the $SO(1,3)$-factor as \lor transformations in $4$ dimensions,
\eqnl{
\Phi_a(x)\longrightarrow \Phi_a(\Lambda^{-1}x) ,\hskip2mm
\psi_p(x)\longrightarrow S\psi_p(\Lambda^{-1}x),\hskip2mm
A_\mu(x)\longrightarrow
\Lambda_\mu^{\;\,\nu}A_\nu(\Lambda^{-1}x).}{rsym15}
In the reduced theory the first factor
in $\xi\ot\psi$ has disappeared and we must reinterpret
the $R$-symmetry $\Psi\to S_6\xi\ot\psi$ as a transformation
of the $4$-dimensional spinor $\psi$. To find this
transformation we note that the real
$S_6$ commutes with $\Gam_*$ such that
\eqnl{
S_6 \mbg_p=U_{qp}\mbg_q\mtxt{and}
S_6 \mbg_p^*=U^*_{qp}\mbg^*_q,\qquad U,U^*\in SU(4).}{rsym17}
These relations define group-isomorphisms 
between $Spin(6)$ and $SU(4)$. Therefore,
under the $R$-symmetry the spinors transform as
\eqnl{
\psi_p^+\longrightarrow \sum_q U_{pq}\psi_q^+
\mtxt{and}
\psi_p^-\longrightarrow \sum_q U^*_{pq}\psi_q^-.}{rsym19}
Note that for \maj spinors $\psi_p$ the $R$-transformed objects
$
\sum_q\left(U_{pq}\psi_q^++U^*_{pq}\psi_p^-\right)$
are \maj spinors as well.

By construction the action \refs{redd9} of the $\cN=4$ 
SYM-theory must be invariant under $R$-transformations.
The invariance is easily seen for the
terms without fermions and the \dirac term.
To prove it directly for the \textsc{Yukawa} terms   
one uses the first relation in \refs{rsym11} and the
definitions in \refs{redd7} giving rise to
the following group homomorphism between $SO(6)$ and $SU(4)$:
\eqnl{
U_{rp}\Delta^a_{rs}\, U_{sq}= R^a_{\;\,b}\Delta^b_{pq}
=\left(S_6^{-1}\Delta^a S_6\right)_{pq}.}{rsym21}
These relations implies that the \textsc{Yukawa}
terms are invariant under $R$-transformations.

\section{Susy transformation of reduced theory}
The supersymmetries of the $\cN=1$ SYM-theory
in ten dimensions reduce to  supersymmetries 
of the $\cN=4$ SYM-theory
in four dimensions. To derive these transformations we 
insert the expansion \refs{nfour31} for a \textsc{Majorana-Weyl}
spinor into the supersymmetry  transformations \refs{nfein1}
(and set $\tilde V=1$) 
\begin{eqnarray}
\delta_\al A_\mu\!&=&\!i\alb\left(\id_8\otimes\gam_\mu\right)\Psi\label{nfours1}\\
\delta_\al\Phi_a\!&=&\!i\alb\left(\Delta_a\otimes \gamf\right)\Psi\label{nfours2}\\
\delta_\al\Psi\!&=&\!\!\!\left(iF_{\mu\nu}\id_8\otimes\Sigma^{\mu\nu}
\!+D_\mu\Phi_a\left(\Delta^a\otimes \gam^\mu\gamf\right)
\!+{g\ov 2i}[\Phi_a,\Phi_b]\Delta^{ab}\otimes \id_4\right)\!\al\label{nfours3}
\end{eqnarray}
as well as the corresponding expansion for the
supersymmetry parameter $\al$:
\eqnl{
\al=\sum\left(\mbg_p\ot\veps_p^++
\mbg_p^*\ot \veps_p^-\right).}{nfours5}
Using \refs{redd1} we obtain the variations
of the vector potential and scalar fields 
\begin{eqnarray}
\delta_\al A_\mu&=&i\sum_p \vepsb_p\gam_\mu\psi_p\label{nfours7}\\
\delta_\al\Phi_a&=&i\sum_{pq} \Delta^a_{pq}\,\vepsb_p\gamf\psi_q+i
\sum_{pq}(\Delta^a\Gam_*)_{pq}\,\vepsb_p\psi_q.\label{nfours9}
\end{eqnarray}
To find the transformations of the \maj spinors
is less simple. We insert into \refs{nfours3} and
\eqnn{
\delta_\al\Psi=\sum\left(\mbg_p\otimes (\delta_\al\psi_p)^+
+\mbg_p^*\otimes(\delta_\al\psi_p)^+\right)}
the expansion \refs{nfours5} for $\al$ and compare coefficients. 
With the definition
\eqnl{
\Delta^{ab}_{pq}\equiv (\mbg_p,\Delta^{ab} \mbg_q)}{nfours11}
the variations of the $4$-dimensional \maj spinors
can be written as
\eqngrl{
\delta\psi_p&=&iF_{\mu\nu}\Sigma^{\mu\nu}\veps_p
+\di\Phi_a\sum_q\left(\Delta^a_{pq}\veps_q^+
-\bar\Delta^a_{pq}\veps_q^-\right)}
{&&\qquad\qquad\;\;+\;{g\ov 2i}\,[\Phi_a,\Phi_b]\sum_q\left(
\Delta^{ab}_{pq}\veps_q^++\bar\Delta^{ab}_{pq}\veps_q^-\right).}
{nfours13}
Also the \noeth current is derived from the current 
\refs{nfein7} by dimensional reduction. The four currents 
depend linearly on the $\psi_p$, the field strength and its 
dual and the covariant derivatives of 
the scalar fields. Their explicit forms read
\index{Noether current!for $\cN=4$ SYM}
\eqngrrl{
J^\mu_p&=&-\tr\left(^*F^{\mu\nu}\gamf+iF^{\mu\nu}\right)\gam_\nu\psi_p}
{&&-i\sum_q\tr\left\{D_\al\Phi_a\left(\Delta^a_{pq}\,\gam^\al\gam^\mu\psi_q^+
-\bar\Delta^a_{pq}\,\gam^\al\gam^\mu\psi_q^-\right)\right\}}
{&&-\ft{g}{2}\sum_q\tr\left\{[\Phi_a,\Phi_b]\left(\Delta^{ab}_{pq}\,
\gam^\mu\psi_q^+
+\bar\Delta^{ab}_{pq}\,\gam^\mu\psi_q^-\right)\right\}.}{nfours15}
From these $4$ \noeth currents one derives the $4$ conserved supercharges
of the $\cN=4$ supersymmetric YM theory
\index{supercharge!for $\cN=4$ SYM}
\eqnl{
\cQ^p=\int d\mbx\,J^0_p,\qquad p=1,\dots,4.}{nfours17}
They fulfill the anticommutation relations 
\eqnl{
\{\cQ_\al^p,\cQ_\beta^q\}=(\cC_4\gam^\mu)_{\al\beta}\,\delta^{pq}P_\mu.}{nfours19}
These charges commute with the $4$-momentum $P_\mu$ and
transform as \maj spinors under \lor transformations.

Actually the $\cN=4$ SYM is classically scale invariant
since all couplings in the \lagan \refs{redd9} are dimensionless.
Since the $\beta$-function vanishes to \emph{all orders} in perturbation 
theory the quantized theory is scale invariant as well. Thus
the supersymmetry algebra can be extended to a $\cN=4$
\emph{superconformal algebra}. This enlarged symmetry 
leads to stringent conditions on the particle spectrum of 
the theory. The particles fall into representations of the
superconformal algebra. When one tries to argue in favor
of the AdS-CFT correspondence one needs these representations.
Unfortunately, at this point I must refer to the literature, 
see for example the excellent and exhaustive review \cite{aharony00}.
\section{Omissions}
In these lectures I had to omit many interesting aspects
of supersymmetric theories. Probably I should have introduced
\emph{superfields} \cite{salamsuperspace} from the very 
beginning to shorten some
of the more lengthy calculations in the component
formalism. In particular when it comes to gauge theories
the superspace formulation is superior. I did not
discuss \textsc{Feynman}-diagram in supersymmetric
theories and cancellations of divergences, although these
chancellation lead to a vanishing $\beta$-function in
$\cN=4$ susy Yang-Mills theory. Of course, for constructing
realistic models for particle physics the
breaking of supersymmetry is of paramount interest.
I did not talk about this aspect of supersymmetric 
field theories either. It just would take another 
series of lectures at the troisieme cycle. But this issue 
is discussed in most of the texts \cite{intwess}-\cite{love}.

One of the most interesting aspects of $\cN=1$ SUSY gluodynamics
is the non-vanishing \emph{gluino condensate}, which
vanishes perturbatively, but does not vanish in the
nonperturbative treatment and can be computed exactly.
I refer to the review article by M. Shifman and 
A. Vainshtein \cite{shifman} for the history of 
dynamical supersymmetry breaking and the role of instantons 
in susy gauge theories. 

Also, the more recent
results of \textsc{Seiberg} and \textsc{Witten} on
the low-energy effective action of $\cN=2$ SYM and
the intriguing AdS-CFT correspondence have not been
dealt with. You may consult the quoted literature,
for example the recommended book of \textsc{Weinberg} 
\cite{intweinberg} or the review of Sachs \cite{intsachs} 
for the \textsc{Seiberg-Witten} solution 
and the review \cite{aharony00} for the correspondence.
I very much regretted to find no time for 
supersymmetric field theories on the lattice 
\cite{latticeold,latticenew,muenster}, a topic
which attracted my attention during the past
years \cite{wipfkirch3}.
\section*{Acknowledgements}
I thank Falk Bruckmann, Andreas Kirchberg, Dominique L{\"a}nge, 
Pablo Pisani and Mikhail Volkov for discussions and
collaborations on supersymmetric systems
and Tobias Kaestner and Ulrich Theis for
reading, commenting and improving these lecture notes.
It is a pleasure to thank Ruth Durrer 
and Peter Wittwer for their kind hospitality at the 
\emph{Troisieme Cycle de la Physique en Suisse Romande}.
The days in Geneva have been interesting and very
nice. 
Last but not least I profited from discussions with 
students from Jena and Geneva who followed these lectures.

\appendix
\chapter{Useful formula}
In this appendix we collected some useful
formula which are used in the main body
of the paper.
\section{Gamma matrices and Fierz identities}\label{app1}
For example, the product of generators
of the spingroup and the gamma-matrices in
$4$ dimensions are reduced according to
\begin{eqnarray}
\Sigma_{\mu\nu}\gam_\rho&=&\ft{i}{2}\eta_{\mu\rho}\gam_\nu
-\ft{i}{2}\eta_{\nu\rho}\gam_\mu
-\ft12 \eps_{\mu\nu\rho\sigma}\gam^\sigma\gamf,\label{ap1}\\
\gam_\rho\Sigma_{\mu\nu}&=&-\ft{i}{2}\eta_{\mu\rho}\gam_\nu
+\ft{i}{2}\eta_{\nu\rho}\gam_\mu
-\ft12 \eps_{\mu\nu\rho\sigma}\gam^\sigma\gamf\label{ap2}
\end{eqnarray}
We also use the conjugation formula
\eqngrrl{
\gam_\mu\gam_\rho \gam^\mu&=&(2-d)\gam_\rho}
{\gam_\mu\gam_{\rho\sigma}\gam^\mu&=&(d-4)\gam_{\rho\sigma}}
{\gam_\mu\gamf\gam_\rho\gam^\mu&=&(2-d)\gam_\rho\gamf,}{ap3}
and the simple relation
\eqnl{
i\gamf[\gam_\mu,\gam_\nu]+\eps_{\mu\nu\rho\sigma}\gam^\rho\gam^\sigma=0,}{ap5}
The following particular \textsc{Fierz} identities
follow from the general identity \refs{fierz5}
\begin{eqnarray}
\psib\gam^\rho\Sigma_{\mu\nu}\veps&=&
\vepsb\Sigma_{\mu\nu}\gam^\rho\psi, \label{ap51}\\
\al_2\alb_1-\al_1\alb_2&=&-\ft12 \gam_{\rho}(\alb_1\gam^\rho\al_2)
+\gam_{\rho\sigma}(\alb_1\gam^{\rho\sigma}\al_2)\label{ap53}
\end{eqnarray}
For \maj spinors the in $(a,c)$ antisymmetric part of 
\refs{nab33} can be written as
\eqnl{
(\psib^a\gam^\mu\psi^b)(\vepsb\gam_\mu\psi^c)
+(\psib^b\gam^\mu\psi^c)(\vepsb\gam_\mu\psi^a)
+(\psib^c\gam^\mu\psi^a)(\vepsb\gam_\mu\psi^b)=0.}{ap55}
Another useful identity is
\eqnl{
\vepsb\gamf\gam^\rho\Sigma_{\mu\nu}\psi=\psib \Sigma_{\mu\nu}\gam^\rho\gamf\al}
{ap57}
Sometimes one needs to switch from the chiral
to the \dirac basis. Then the following identities
are useful:
\eqngrrl{
\alb\psi-\alb_c\psi_c&=&\theta\chi+\thetab\chib
+\zeta\lam+\zetab\lamb}
{\alb\gamf\psi-\alb_c\gamf\psi_c&=&
\theta\chi+\thetab\chib-\zeta\lam-\zetab\lamb}
{\alb\gam_\mu\psi+\alb_c\gam_\mu\psi_c&=&
\theta\sigma_\mu\lamb+\thetab\tsigma_\mu\lam
+\zeta\sigma_\mu\chib+\zetab\tsigma_\mu\chi.}{ap101}
In constructing the \lagan of the $\cN=4$ SYM theory
we used the general \textsc{Fierz}-Identity in $10$ dimensions
\begin{eqnarray}
32\Psi\bar\Phi&=&-\chib\psi-\Gam_m X^m+\ft{1}{2!}
\Gam_{mn}X^{mn}+\ft{1}{3!}\Gam_{mnp}X^{mnp}
-\ft{1}{4!}\Gam_{mnpq}X^{mnpq}\nonumber\\
&&-\ft{1}{5!}\Gam_{mnpqr}X^{mnpqr}
-\ft{1}{4!}\Gam_{11}\Gam_{mnpq}X_{11}^{mnpq}
-\ft{1}{3!}\Gam_{11}\Gam_{mnp}X_{11}^{mnp}\label{ap71}\\
&&+\ft{1}{2!}\Gam_{11}\Gam_{mn}X_{11}^{mn}
+\Gam_{11}\Gam_m X_{11}^m
-\Gam_{11} X_{11}\nonumber
\end{eqnarray}
where we have abbreviated
\eqnn{
X^{m\dots}=\bar\Phi\Gam^{m\dots}\Psi\mtxt{and}
X^{m\dots}_{11}=\bar\Phi\Gam_{11}\Gam^{m\dots}\Psi.}

\section{Majorana representation in 6 Euclidean dimensions}\label{app2}
In our dimensional reduction from $10$ to
$4$ dimensions we used a particular representation
for the \textsc{Euclid}ean $\gam$-matrices belonging
to the internal $6$ dimensional space.
To construct this \maj representation 
we make for the $\Delta_a$ in \refs{nfour13} the 
ansatz
\eqnn{
\Delta_i=i\tau_1\ot\al_i\mtxt{and}\Delta_{3+i}=i\tau_3\ot\tal_i,\qquad
i=1,2,3,}
so that
\eqnn{
\{\al_i,\al_j\}=\{\tal_i,\tal_j\}=2\delta_{ij}\id_4,\quad
[\al_i,\tal_j]=0,\quad \al_i^\dagger=\al_i,\quad \tal_i^\dagger=\tal_i}
must hold. A possible solution for the $\al$-matrices is
\eqngr{
&\al_1=\tau_2\ot \tau_1,\quad 
\al_2=\tau_0\ot\tau_2,\quad \al_3=\tau_2\ot\tau_3}
{&\tal_1=\tau_1\ot \tau_2,\quad 
\tal_2=-\tau_3\ot\tau_2,\quad\tal_3=\tau_2\ot\tau_0.}
These hermitean, imaginary and hence antisymmetric matrices obey
\eqnn{
\al_i\al_j=\delta_{ij}+i\eps_{ijk}\al_k\mtxt{and}
\tal_i\tal_j=\delta_{ij}+i\eps_{ijk}\tal_k,\quad
[\al_i,\tal_j]=0.}
They lead to \textit{real and antisymmetric} matrices
$\Delta_a$ which fulfill the \textsc{Euclid}ean 
\textsc{Clifford} algebra in $6$ dimensions.
\def\np{Nucl. Phys. }
\def\pl{Phys. Lett. }
\def\pr{Phys. Rev. }
\def\an{Annals of Phys. }

\printindex
\end{document}